\documentclass[10pt,a4paper]{article}
 \pdfoutput=1
\usepackage{amsmath,fourier,amssymb,amsthm,graphicx,epstopdf,color,tikz}
\usepackage{float}
\usepackage{chngcntr}
\counterwithout{figure}{section}
\usepackage[T1]{fontenc}
\numberwithin{equation}{section}
\allowdisplaybreaks

 \newtheorem{innercustomthm}{Theorem}
 \newenvironment{customthm}[1]
   {\renewcommand\theinnercustomthm{#1}\innercustomthm}
   {\endinnercustomthm}

 \theoremstyle{plain}
 \newtheorem {hypo}{\bf\hspace{-\parindent}Hypothesis}[section]

 \newtheorem {ass}[hypo]{Assumption}
 \newtheorem {prop}[hypo]{Proposition}
 \newtheorem {lemma}[hypo]{Lemma}
  \newtheorem {theo}[hypo]{Theorem}
 \newtheorem {defin}[hypo]{Definition}
 \newtheorem {cor}[hypo]{Corollary}
 
 \theoremstyle{remark}
 \newtheorem {rmk}[hypo]{Remark}
 \newtheorem {eg}[hypo]{Example}

 \newcommand{\pf}{\begin{bpf}}

 \newcommand{\pfms}{\begin{bpfms}}
 \newcommand{\epf}{\end{bpf}\hfill$\square$\vspace{0.1cm}}
 \newcommand{\epfms}{\end{bpfms}\hfill$\square$\\ }
 \newcommand\ben{\begin{equation*}}
 \newcommand\ebn{\end{equation*}}
 \newcommand\beq{\begin{equation}}
 \newcommand\eeq{\end{equation}}
 \newcommand\ds{\displaystyle}
 \newcommand\lb{\left(}
 \newcommand\rb{\right)} 
 \newcommand\Cb{\mathbb{C}} 
 \newcommand\Zb{\mathbb{Z}}
 \newcommand\Pb{\mathbb{P}} 
 
 \newcommand{\eq}[1]{\begin{equation}\begin{gathered} #1\end{gathered}\end{equation}}
 \newcommand{\eqs}[1]{\begin{equation}\begin{gathered}\begin{split}
 #1\end{split}\end{gathered}\end{equation}}
 
 \newcommand\rf[1]{(\ref{#1})}
 \def\Sum{\sum\limits}
 \def\Prod{\prod\limits}
 \def\mc{\mathcal}

\begin{document}
\LARGE
\noindent
\textbf{Fredholm determinant and Nekrasov sum representations
 \vspace{0.1cm}\\
of isomonodromic  tau functions}
\normalsize
 \vspace{1cm}\\
 \noindent\textit{
 P. Gavrylenko$\,^{a,b,c,}$\footnote{pasha145@gmail.com}, 
 O. Lisovyy$\,^{d,}$\footnote{lisovyi@lmpt.univ-tours.fr}}
 \vspace{0.2cm}\\
 $^a$ National Research University Higher School of Economics,
 International Laboratory of Representation Theory and Mathematical
 Physics, Russian Federation\vspace{0.1cm}\\
 $^b$ Bogolyubov Institute for Theoretical Physics,  03680 Kyiv, Ukraine
  \vspace{0.1cm}\\
 $^c$ Skolkovo Institute of Science and Technology,  143026 Moscow, Russia 
 \vspace{0.1cm}\\
 $^d$ Laboratoire de Math\'ematiques et Physique Th\'eorique CNRS/UMR 7350,  Universit\'e de Tours, Parc de Grandmont,
  37200 Tours, France

\begin{abstract}
\noindent We derive Fredholm determinant representation for isomonodromic tau functions of Fuchsian systems with $n$ regular singular points on the Riemann sphere and generic monodromy in $\mathrm{GL}\lb N,\Cb\rb$. The corresponding operator acts in the direct sum of $N\lb n-3\rb $ copies of $L^2\lb S^1\rb$. Its kernel has a block integrable form and is expressed in terms of fundamental solutions of $n-2$ elementary 3-point Fuchsian systems whose monodromy is determined by monodromy of the relevant $n$-point system via a decomposition of the punctured sphere into pairs of pants. For $N=2$ these building blocks have hypergeometric representations, the kernel becomes completely explicit and has Cauchy type. In this case Fredholm determinant expansion yields multivariate series representation for the tau function of the Garnier system, obtained earlier via its identification with Fourier transform of Liouville conformal block (or a dual Nekrasov-Okounkov partition function). Further specialization to $n=4$ gives a series representation of the general solution to Painlevé VI equation.
\end{abstract}

 \section{Introduction}
 \subsection{Motivation and some results}
 The theory of monodromy preserving deformations plays a prominent role in many areas of modern nonlinear mathematical physics.
 The classical works \cite{WMTB,JMMS,TW1} relate, for instance, various correlation and distribution functions of statistical mechanics and random matrix theory models to special solutions of Painlev\'e equations.  The relevant Painlevé functions are usually written in terms of Fredholm or Toeplitz determinants. Further study of these
 relations has culminated in the development by Tracy and Widom
  \cite{TW2} of an algorithmic procedure of derivation of systems of PDEs satisfied by Fredholm determinants with integrable kernels \cite{IIKS} restricted to a union of intervals; the isomonodromic  origin of Tracy-Widom equations has been elucidated in \cite{PalmerTW} and further studied in \cite{HI}. This raises a natural question: 
 \begin{itemize}
 \item[ \textcircled{\raisebox{-0.9pt}{?}}]
 {\it Can the  general solution  of isomonodromy equations be expressed
 in terms of a Fredholm determinant? }
 \end{itemize}
 
  One of the goals of the present paper is to provide a constructive answer to this question in the Fuchsian setting.
   Let us consider a Fuchsian system with $n$ regular singular points $a:=\left\{a_0,\ldots,a_{n-2},a_{n-1}\equiv\infty\right\}$ on ${\mathbb P^1\equiv\mathbb P^1\lb \mathbb C\rb}$:
  \beq\label{fuchsysintro}
  \partial_z\Phi=\Phi A\lb z\rb,\qquad A\lb z\rb =\sum_{k=0}^{n-2}\frac{A_k}{z-a_k},
  \eeq
  where $A_0,\ldots , A_{n-2}$ are $N\times N$ matrices independent of $z$ and  $\Phi\lb z\rb$ is a fundamental matrix solution, multivalued on
  $\mathbb P^1 \backslash a$. The monodromy of $\Phi\lb z\rb$ realizes a representation of the fundamental group $\pi_1\lb \mathbb P^1\backslash a\rb$ in $\mathrm{GL}\lb N,
  \mathbb C \rb$. When the residue matrices $A_0,\ldots, A_{n-2}$ and
  $A_{n-1}:=-\sum_{k=0}^{n-2} A_k$ are non-resonant, the isomonodromy equations are given by the Schlesinger system,
  \begin{align}
  \label{schlSys}
  \begin{cases}
  \partial_{a_i}A_k=&\ds
  \frac{\left[A_i,A_k\right]}{a_k-a_i},\qquad\qquad  i\ne k,\\
  \partial_{a_i}A_i=&\ds\sum_{k\ne i}\frac{\left[A_i,A_k\right]}{a_i-a_k}.
  \end{cases}
  \end{align}
  Integrating the flows associated to affine transformations, we may set without loss of generality $a_0=0$ and $a_{n-2}=1$, so that there remains $n-3$ nontrivial time variables $a_1,\ldots,a_{n-3}$. In the case ${N=2}$, Schlesinger equations  reduce to the Garnier system 
  $\mathcal G_{n-3}$, see for example \cite[Chapter 3]{IKSY} for the details. Setting further $n=4$, we are left with only one time $t\equiv a_1$ and the latter system becomes equivalent to a nonlinear 2nd order ODE --- the Pain\-levé~VI equation.

  The main object of our interest  is the isomonodromic tau function of Jimbo-Miwa-Ueno \cite{JMU}. It is defined as an exponentiated  primitive of the $1$-form 
  \beq\label{taujmud}
  d_a\ln\tau_{\mathrm{JMU}}:=
  \frac12\sum_{k=0}^{n-2}\operatorname{res}_{z=a_{k}}
  \operatorname{Tr} A^2\lb z\rb\;da_k.
  \eeq
  The definition is consistent since the 1-form  on the right is closed on solutions of the deformation equations~(\ref{schlSys}). It generates the hamiltonians of the Schle\-singer  system. 
    Dealing with the Garnier system, we will assume the standard gauge where $\operatorname{Tr}A\lb z\rb=0$ and denote the eigenvalues of 
  $A_k$ by $\pm \theta_k$ with $k=0,\ldots,n-1$. 
  In the Painlevé~VI case, it is convenient to modify this notation  as $\lb\theta_0,\theta_1,\theta_2,\theta_3\rb\mapsto
  \lb \theta_0,\theta_t,\theta_1,\theta_{\infty}\rb$. The logarithmic derivative $\zeta\lb t\rb:=t\lb t-1\rb \frac{d}{dt}\ln\tau_{\mathrm{VI}}\lb t\rb$ then satisfies the $\sigma$-form of Painlevé VI,
    \beq
   \label{sigmapvi}
   \Bigl(t(t-1)\zeta''\Bigr)^2=-2\;\mathrm{det}\left(\begin{array}{ccc}
   2\theta_0^2 & t\zeta'-\zeta & \zeta'+\theta_0^2+\theta_t^2+\theta_1^2-\theta_{\infty}^2 \\
    t\zeta'-\zeta & 2 \theta_t^2 & (t-1)\zeta'-\zeta \\
    \zeta'+\theta_0^2+\theta_t^2+\theta_1^2-\theta_{\infty}^2 & (t-1)\zeta'-\zeta & 2\theta_1^2
   \end{array}\right).
   \eeq
   Monodromy of the associated linear problems provides a complete set of conserved quantities for Pain\-levé~VI, the Garnier system and Schlesinger equations. By the general solution of deformation equations we mean the solution corresponding to generic monodromy data. The precise genericity conditions will be specified in the main body of the text.
   
   In \cite{Palmer90}, Palmer (developing earlier results of Malgrange \cite{Malgrange} and Sato-Segal-Wilson \cite{Sato,SW}) interpreted the Jimbo-Miwa-Ueno tau function (\ref{taujmud}) as a determinant of a singular Cauchy-Riemann operator acting on functions with prescribed monodromy. The main idea of \cite{Palmer90} is to isolate the singular points $a_0,\ldots,a_{n-1}$ inside a circle  $\mathcal C\subset\mathbb P^1$ and represent the Fuchsian system (\ref{fuchsysintro})
   by a boundary space of functions on $\mathcal C$ that can be analytically continued inside with specified branching. The variation of positions of singularities gives rise to a trajectory of this space in an infinite Grassmannian. The tau function is obtained by comparing two sections of an associated determinant bundle.
   
   The construction suggested in the present paper is essentially a refinement of Palmer's approach, translated into the  Riemann-Hilbert framework. A single circle $\mathcal C$ is replaced by the boundaries of $n-3$ annuli which cut the $n$-punctured sphere $\mathbb P^1\backslash a$ into trinions (pairs of pants), see e.g. Fig.~\ref{figsic}a  below. To each trinion is assigned a Fuchsian system with 3 regular singular points whose monodromy is determined by  monodromy of the original system. We show that the isomonodromic tau function is proportional to a 
   Fredholm determinant:
   \beq\label{fred}
   \tau_{\mathrm{JMU}}\lb a\rb = \Upsilon\lb a\rb \cdot\operatorname{det}\lb
   \mathbb 1-K\rb,
   \eeq
   where the prefactor $\Upsilon\lb a\rb$ is a known elementary  function.
   The integral operator $K$ acts on holomorphic vector functions on the union of annuli and involves projections on certain boundary spaces.    
   
   The pay-off of a more complicated Grassmannian model is that the kernel of $K$ may be written explicitly in terms of $3$-point solutions\footnote{
      We would like to note that somewhat similar refined construction emerged in the analysis of massive Dirac equation with  $U\lb 1\rb$ branching on the Euclidean plane  \cite{Palmer93}. Every branch point was isolated there in a separate strip, which ultimately allowed to derive an explicit Fredholm determinant representation for the tau function of appropriate Dirac operator \cite{SMJ}. In physical terms, the determinant corresponds to a resummed form factor expansion of a correlation function of $U(1)$ twist fields in the massive Dirac theory. The paper \cite{Palmer93} was an important source of inspiration for the present work, although it took us more than 10 years to realize that the strips should be replaced by pairs of pants in the chiral problem.}. In particular, for $N=2$ (i.e. for the Garnier system) the latter have hypergeometric expressions.  The $n=4$ specialization of our result is as follows.
   \begin{customthm}{A} Let the independent variable $t$ of Painlevé VI equation vary inside the real interval $]0,1[$ and let $\mathcal C=\left\{z\in\mathbb C\, :\, |z|=R,t<R<1 \right\}$ be a counter-clockwise oriented circle. Let $\sigma$, $\eta$ be a pair of complex parameters satisfying the conditions
   \begin{samepage}
   \begin{gather*}
   \left|\Re\sigma\right|\le \frac12,\qquad \sigma\ne 0,\pm\frac12,\\
   \theta_0\pm\theta_t+\sigma\notin\mathbb Z,\qquad \theta_0\pm\theta_t-\sigma\notin\mathbb Z,\qquad
   \theta_1\pm\theta_{\infty}+\sigma\notin\mathbb Z,\qquad \theta_1\pm\theta_{\infty}-\sigma\notin\mathbb Z.
   \end{gather*}
   \end{samepage}
   General solution of the Painlevé VI equation (\ref{sigmapvi}) admits the following Fredholm determinant representation:
   \beq\label{AUX10}
   \tau_{\mathrm{VI}}\lb t\rb = \mathrm{const}\cdot
   t^{\sigma^2-\theta_0^2-\theta_t^2}
   \lb 1-t\rb^{-2\theta_t\theta_1}
   \operatorname{det}\lb\mathbb 1-U\rb,
   \qquad U=\lb\begin{array}{cc}0 & \mathsf a \\ \mathsf d & 0 \end{array}\rb, 
   \eeq
   where the operators $\mathsf a,\mathsf d\in \operatorname{End}\lb
       \mathbb C^2\otimes L^2\lb \mathcal C\rb\rb$ act on 
       $g=\left(\begin{array}{c} g_+ \\ g_-
                   \end{array}\right)$ with $g_{\pm}\in L^2\lb \mathcal C\rb$ as
   \begin{align}
      \lb\mathsf a g\rb\lb z\rb=
      \,\frac1{2\pi i}\oint_{\mathcal C}\,
      \mathsf a\lb z,z'\rb g\lb z'\rb  dz'\, ,\qquad
     \lb\mathsf d g\rb\lb z\rb=
           \,\frac{1}{2\pi i}\oint_{\mathcal C}
            \mathsf d\lb z,z'\rb g\lb z'\rb  dz',            
   \end{align}
   and their kernels are explicitly given by
   \beq
   \begin{aligned}
   \mathsf{a}\lb z,z'\rb=&\, \frac{\lb 1-z'\rb^{2\theta_{1}}
         \lb\begin{array}{cc}
         K_{++}\lb z\rb & K_{+-}\lb z\rb \\
         K_{-+}\lb z\rb & K_{--}\lb z\rb
         \end{array}\rb
          \lb\begin{array}{rr}
             K_{--}\lb z'\rb & \!\!\!-K_{+-}\lb z'\rb \\
             \!\!\!-K_{-+}\lb z'\rb & K_{++}\lb z'\rb
             \end{array}\rb
             -\mathbb{1}}{z-z'},\\
 \mathsf{d}\lb z,z'\rb=
  &\,   \frac{
    \mathbb{1}-\lb 1-\frac{t}{z'}\rb^{2\theta_{t}}
          \lb\begin{array}{cc}
          \bar K_{++}\lb z\rb & \bar K_{+-}\lb z\rb \\ 
          \bar K_{-+}\lb z\rb & \bar K_{--}\lb z\rb
          \end{array}\rb
           \lb\begin{array}{cc}
              \bar K_{--}\lb z'\rb & -\bar K_{+-}\lb z'\rb \\
              -\bar K_{-+}\lb z'\rb & \bar K_{++}\lb z'\rb
              \end{array}\rb
       }{z-z'},
   \end{aligned}
   \eeq
   with
   \beq
   \begin{aligned}
   K_{\pm\pm}\lb z\rb=& \,{}_2F_1\biggl[\begin{array}{c}
      \theta_1+\theta_{\infty}\pm\sigma,
      \theta_1-\theta_{\infty}\pm\sigma \\
      \pm2\sigma 
      \end{array};z\biggr],\\
   K_{\pm\mp}\lb z\rb=&\, 
   \pm\frac{\theta_{\infty}^2-\lb \theta_1\pm\sigma\rb^2}{2\sigma\lb
         1\pm 2\sigma\rb}\,z\,{}_2F_1\biggl[\begin{array}{c}
         1+\theta_1+\theta_{\infty}\pm\sigma,
         1+\theta_1-\theta_{\infty}\pm\sigma \\
         2\pm2\sigma 
         \end{array};z\biggr],\\
   \bar K_{\pm\pm}\lb z\rb=& \,{}_2F_1\biggl[\begin{array}{c}
         \theta_t+\theta_0\mp\sigma,
         \theta_t-\theta_0\mp\sigma \\
         \mp 2\sigma 
         \end{array};\frac tz\biggr],\\
   \bar K_{\pm\mp}\lb z\rb=&\,\mp t^{\mp 2\sigma}e^{\mp i\eta} 
   \frac{\theta_{0}^2-\lb \theta_t\mp\sigma\rb^2}{2\sigma\lb
            1\mp 2\sigma\rb}\,
   \frac tz\,{}_2F_1\biggl[\begin{array}{c}
            1+\theta_t+\theta_0\mp\sigma,
            1+\theta_t-\theta_0\mp\sigma \\
            2\mp 2\sigma 
            \end{array};\frac tz\biggr].\\
   \end{aligned}      
   \eeq 
   \end{customthm}
   
   Moreover, we demonstrate that for a special choice of monodromy in the Painlevé VI case,  $U$ becomes equivalent to the hypergeometric kernel of \cite{BorodinAnnals} and thereby reproduces previously known family of Fredholm determinant solutions
   \cite{BorodinDeift}. The hypergeometric kernel is known to produce other random matrix integrable kernels in confluent limits.

   Another part of our motivation comes from isomonodromy/CFT/gauge theory correspondence. It was conjectured in \cite{GIL12} that the tau function associated to the general  Painlevé VI solution coincides with a Fourier transform of $4$-point $c=1$ Virasoro conformal block with respect to its intermediate momentum. Two independent derivations of this conjecture have been already proposed in \cite{ILTe} and \cite{BSh}. The first approach \cite{ILTe} also extends the initial statement to the Garnier system. Its main idea is to consider the operator-valued monodromy of conformal blocks with additional level 2 degenerate insertions. At $c=1$, Fourier transform of such conformal blocks reduces their ``quantum'' monodromy  to ordinary $2\times 2$ matrices. It can therefore be used to construct the fundamental matrix solution of a Fuchsian system with prescribed $\mathrm{SL}\lb 2,\mathbb C\rb$ monodromy. The second approach \cite{BSh} uses an embedding of two copies of the Virasoro algebra into super-Virasoro algebra extended by Majorana fermions to prove certain bilinear differential-difference relations for $4$-point conformal blocks, equivalent to Painlevé~VI equation. An interesting feature of this method is that bilinear relations admit a deformation to generic values of Virasoro central charge.    
   
      Among other developments, let us mention the papers \cite{GIL13,ILT14,Nagoya} where asymptotic expansions of Painlevé V, IV and III tau functions were identified with Fourier transforms of irregular conformal blocks of different types. The study of relations between isomonodromy problems in higher rank and conformal blocks of $W_N$ algebras has been initiated in 
   \cite{Gav,GM1,GM2}.
   
   The AGT conjecture \cite{AGT} (proved in \cite{AFLT}) identifies Virasoro conformal blocks with partition functions of $\mathcal N=2$ 4D supersymmetric gauge theories. There exist combinatorial representations of the latter objects \cite{Nekrasov}, expressing them as sums over tuples of Young diagrams. This fact is of crucial importance for isomonodromy theory, since it gives (contradicting to an established folklore) explicit series representations for the Pain\-levé~VI and Garnier tau functions. Since the very first paper \cite{GIL12} on the subject, there has been a puzzle to understand combinatorial tau function expansions directly within the isomonodromic framework. There have also been attempts to sum up these series to determinant expressions; for example, in \cite{Balogh} truncated infinite series for $c=1$ conformal blocks were shown to coincide with partition functions of certain discrete matrix models.

   In this work, we show that combinatorial series correspond to the principal minor expansion of the Fredholm determinant (\ref{fred}), written in the Fourier basis of the space of functions on annuli of the pants decomposition. Fourier modes which label  the choice of rows for the principal minor are related to Frobenius coordinates of Young diagrams. It should be emphasized that this combinatorial structure is valid also for $N>2$ where CFT/gauge theory counterparts of the tau functions have yet to be defined and understood.    
   
   We prove in particular the following result, originally conjectured
   in \cite{GIL12} (the details of notation concerning Young diagrams are explained in the next subsection):
 \begin{customthm}{B}
 General solution of the Painlevé VI equation (\ref{sigmapvi}) can be  written as 
   \begin{align}\label{AUX11}
   \tau_{\mathrm{VI}}(t)=\operatorname{const}\cdot\sum_{n\in\Zb}e^{in\eta'}
   \mathcal{B}\lb \vec{\theta};\sigma+n;t\rb,
   \end{align}
   where $\mathcal{B}(\vec{\theta},\sigma;t)$ is a double sum over Young diagrams,
   \begin{gather*}
   \mathcal{B}\left(\vec{\theta},\sigma;t\right)=
   \mathcal N_{\theta_{\infty},\sigma}^{\theta_1}
   \mathcal N_{\sigma,\theta_0}^{\theta_t}
   t^{\sigma^2-\theta_0^2-\theta_t^2}(1-t)^{2\theta_t\theta_1}
   \sum_{\lambda,\mu\in\mathbb{Y}}
   \mathcal{B}_{\lambda,\mu}\left(\vec{\theta},\sigma\right)   t^{|\lambda|+|\mu|},\\
   \mathcal{B}_{\lambda,\mu}\left(\vec{\theta},\sigma\right)=\!\!\!\!
   \prod_{(i,j)\in\lambda}\!\!\!\!
   \frac{\left(\left(\theta_t+\sigma+i-j\right)^2-\theta_0^2\right)
   \left(\left(\theta_1+\sigma+i-j\right)^2-\theta_{\infty}^2\right)}{
   h_{\lambda}^2(i,j)\left(\lambda'_j-i+\mu_i-j+1+2\sigma\right)^2}
   \!\!\!\!\prod_{(i,j)\in\mu}\!\!\!\!
   \frac{\left(\left(\theta_t-\sigma+i-j\right)^2-\theta_0^2\right)
   \left(\left(\theta_1-\sigma+i-j\right)^2-\theta_{\infty}^2\right)}{
   h_{\mu}^2(i,j)\left(\mu'_j-i+\lambda_i-j+1-2\sigma\right)^2}\,,\\
    \mathcal N_{\theta_3,\theta_1}^{\theta_2}=\frac{
    \prod_{\epsilon=\pm}G\left(1+\theta_3+\epsilon(\theta_1+\theta_2)\right)G\left(1-\theta_3+\epsilon(\theta_1-\theta_2)\right)}{
    G(1-2\theta_1)G(1-2\theta_2)G(1+2\theta_3)}.
   \end{gather*}
   Here $ \sigma\notin\mathbb Z/2$, $\eta'$ are two arbitrary complex  parameters, and $G\lb z\rb$ denotes the Barnes $G$-function.
 \end{customthm}
 
 The parameters $\sigma$ play exactly the same role in the Fredholm determinant (\ref{AUX10}) and the series representation (\ref{AUX11}), whereas $\eta$ and $\eta'$ are related by a simple transformation. An obvious quasiperiodicity of the second representation with respect to integer shifts of $\sigma$ is by no means manifest in the Fredholm determinant.

    \subsection{Notation}
     The monodromy matrices of Fuchsian systems and the jumps of associated Riemann-Hilbert problems appear on the left of solutions. These somewhat unusual conventions are adopted to avoid even more confusing right action of integral and infinite matrix operators.    
     The indices corresponding to the matrix structure of rank~$N$ Riemann-Hilbert problem are referred to as color indices and are denoted by Greek letters, such as $\alpha,\beta\in\left\{1,\ldots,N\right\}$. Upper indices in square brackets, e.g. $[k]$ in $\mathcal T^{[k]}$, label different trinions in the pants decomposition of a punctured Riemann sphere.    
    We denote by $\mathbb Z':=\mathbb Z+\frac12$ the half-integer lattice, and by $\mathbb Z'_{\pm}=\left\{p\in\mathbb Z'\,|\,p\gtrless 0\right\}$ its positive and negative parts. The elements of $\mathbb Z',\mathbb Z'_{\pm}$ will be generally denoted by the letters $p$ and $q$.
    
        \begin{figure}[h!]
          \centering
          \includegraphics[height=3cm]{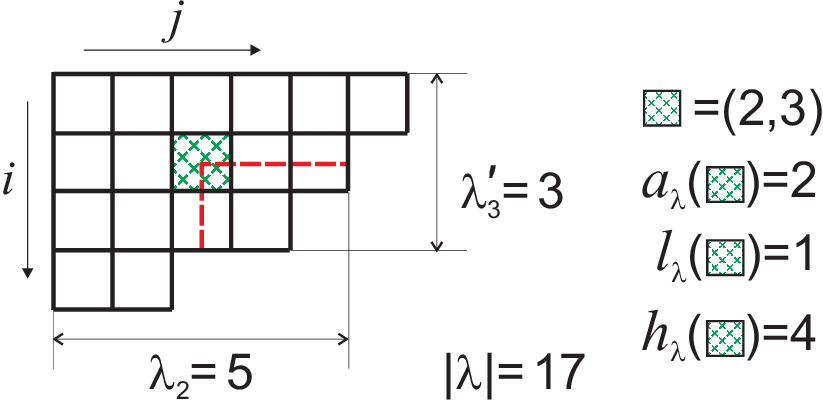}
          \begin{minipage}{0.8\textwidth}
          \caption{\label{youngnot}
          Young diagram associated to the partition
          $\lambda=\left\{6,5,4,2\right\}$.}
         \end{minipage}
          \end{figure}

     The set of all partitions identified with Young diagrams is denoted by $\mathbb Y$. For $\lambda\in\mathbb Y$, we write
     $\lambda'$ for the transposed diagram,  $\lambda_i$ and $\lambda'_j$ for the number of boxes in the $i$th row and $j$th column of $\lambda$, and $|\lambda|$ for the total number of boxes in $\lambda$. Let $\square=\lb i,j\rb$ be the box in the $i$th row and $j$th column of $\lambda\in\mathbb Y$ (see Fig.~\ref{youngnot}). Its arm-length $a_{\lambda}\lb \square\rb
     $ and leg-length $l_{\lambda}\lb\square\rb$ denote the number of boxes on the right and below. This definition is extended to the  case where the box lies outside $\lambda$ by the formulae $a_{\lambda}\lb \square\rb
     =\lambda_i-j$ and $l_{\lambda}\lb\square\rb=\lambda'_j-i$. The
     hook length of the box $\square\in\lambda$ is defined as
     $h_{\lambda}\lb \square\rb=a_{\lambda}\lb \square\rb+l_{\lambda}\lb \square\rb+1$.

  \subsection{Outline of the paper}
  The paper is organized as follows. Section~\ref{sectionfred} is devoted to the derivation of Fredholm determinant representation of the Jimbo-Miwa-Ueno isomonodromic tau function. It starts from a recast of the original rank $N$ Fuchsian system with $n$ regular singular points on $\mathbb P^1$ in terms of a Riemann-Hilbert problem. In Subsection~\ref{subsecauxRHP} we associate to it, via a decomposition of $n$-punctured Riemann sphere into pairs of pants, $n-2$ auxiliary Riemann-Hilbert problems of Fuchsian type having only $3$ regular singular points. Section~\ref{subsecplemelj} introduces Plemelj operators acting on functions holomorphic on the  annuli of the pants decomposition, and deals with their basic properties. The main result of the section is formulated in Theorem~\ref{TauF} of Subsection~\ref{subsectauf}, which relates the tau function of a Fuchsian system with prescribed generic monodromy to a Fredholm determinant whose blocks are expressed in terms of $3$-point Plemelj operators. In Subsection~\ref{subsec4pts}, we consider in more detail the example of $n=4$ points and show that the Fredholm determinant representation can be efficiently used for asymptotic analysis of the tau function. In particular, Theorem~\ref{theojimbo} provides a generalization of the Jimbo asymptotic formula for Painlevé~VI valid in any rank and up to any asymptotic order.
  
  In Section~\ref{seccomb} we explain how the principal minor expansion of the Fredholm determinant leads to a combinatorial structure of the series representations for isomonodromic tau functions. Theorem~\ref{theocauchy} of Subsection~\ref{subseccauchy} shows that  3-point Plemelj operators written in the Fourier basis are given by sums of a finite number of infinite Cauchy type matrices twisted by diagonal factors. Combinatorial labeling of  the minors by $N$-tuples of charged Maya diagrams and partitions is described in Subsection~\ref{subseccomb}.
  
  Section~\ref{secranktwo} deals with rank $N=2$. Hypergeometric representations of the appropriate $3$-point Plemelj operators are listed in Lemma~\ref{abcd2F1} of Subsection~\ref{subsecGC}. Theorem~\ref{theogarnier} provides an explicit combinatorial series representation for the tau function of the Garnier system. In the final subsection, we explain how Fredholm determinant of the  Borodin-Olshanski hypergeometric kernel  arises as a special case of our construction. Appendix contains a proof of a combinatorial identity expressing Nekrasov functions in terms of Maya diagrams instead of partitions.

  \subsection{Perspectives}
  In an effort to keep the paper of reasonable length, we decided to defer the study of several straightforward generalizations of our approach to separate publications. These extensions are outlined below together with a few more directions for future research: 
  \begin{enumerate}
  \item In higher rank $N>2$, it is an open problem to find integral/series representations for general solutions of $3$-point Fuchsian systems and to obtain an explicit description of  the Riemann-Hilbert map. There is however an important exception of rigid systems having two generic regular singularities and one singularity of spectral multiplicity $\lb N-1,1\rb$; these can be solved in terms of generalized hypergeometric functions of type~$_NF_{N-1}$. 
  The spectral condition is exactly what is needed to achieve factorization in Lemma~\ref{lemmafacca}.
  The results of Section~\ref{secranktwo} can therefore be extended to Fuchsian systems with two generic singular points at~$0$ and~$\infty$, and $n-2$ special ones. The corresponding isomonodromy equations (dubbed $\mathcal G_{N,n-3}$ system in \cite{Tsuda}) are the closest  higher rank  relatives  of Painlevé VI and Garnier system. It is natural to expect their tau functions to be related on the 2D CFT and gauge theory side, respectively, to $W_N$ conformal blocks with semi-degenerate fields \cite{FL,Bullimore} and Nekrasov partition functions of 4D linear quiver gauge theories with the gauge
  group ${U\lb N\rb}^{\otimes \lb n-3\rb}$.
  
  In the generic non-rigid case the 3-point solutions depend on $\lb N-1\rb \lb N-2\rb$ accessory parameters and may be interpreted as matrix elements of a general vertex operator for the 
  $W_N$ algebra. They should also be related to the so-called $T_N$ gauge theory without lagrangian description \cite{BMPTY}.
  
  \item Fredholm determinants and series expansions considered in the present work are associated to linear pants decompositions of 
  $\mathbb P^1\left\backslash\left\{n\text{ points}\right\}\right.$, which means that every pair of pants has at least one external boundary component (see Fig.~\ref{figsic}a). Plemelj operators assigned to each trinion act on spaces of functions on internal boundary circles only. To be able to deal with arbitrary decompositions, in addition to 4 operators $\mathsf a^{[k]}$, $\mathsf b^{[k]}$, $\mathsf c^{[k]}$, $\mathsf d^{[k]}$ appearing in (\ref{alphadelta}) one has to introduce 5 more similar operators associated to other possible choices of ordered pairs of boundary components. 
  
    \begin{figure}[h!]
      \centering
      \includegraphics[height=3cm]{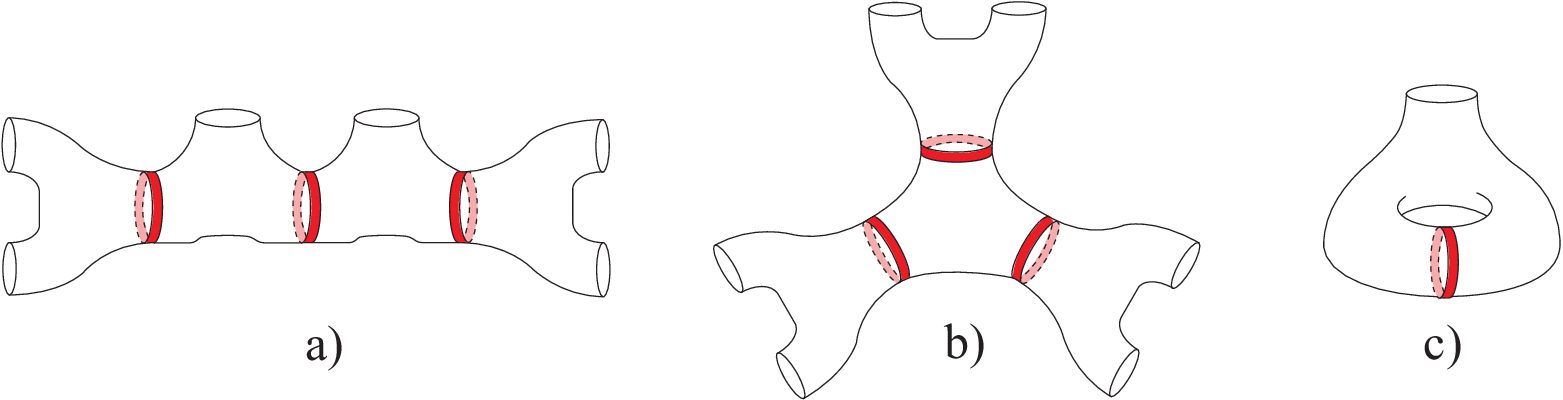}
      \begin{minipage}{0.68\textwidth}
      \caption{\label{figsic}
     (a) Linear and (b) Sicilian pants decomposition 
     of $\mathbb P^1\left\backslash\left\{6\text{ points}\right\}\right.$; 
     (c) gluing 1-punctured torus from a pair of pants.}
     \end{minipage}
      \end{figure}
  
  A (tri)fundamental example where this construction becomes important is known in the gauge theory literature under the name of Sicilian quiver (Fig.~\ref{figsic}b). Already for $N=2$ the monodromies along the triple of internal cicles of this pants decomposition cannot be simultaneously reduced to the form ``$\mathbb 1+$rank~1 matrix'' by factoring out a suitable scalar piece. The analog of expansion
  (\ref{tauGar}) in Theorem~\ref{theogarnier} will therefore be more intricate yet explicitly computable. Since the identification  \cite{ILTe} of the tau function of the Garnier system with a Fourier transform of 
  $c=1$ Virasoro conformal block does not put any constraint on the employed pants decomposition, Sicilian expansion of the Garnier tau function may be used to produce an analog of Nekrasov representation for the corresponding conformal blocks. It might be interesting to compare the results obtained in this way  against instanton counting \cite{HKS}.
  
  Extension of the procedure to higher genus requires introducing additional simple (diagonal in the Fourier basis) operators acting on some of the internal annuli. They give rise to a part of moduli of complex structure of the Riemann surface and correspond to gluing a handle out of two boundary components. Fig.~\ref{figsic}c shows how a 1-punctured torus may be obtained by gluing two boundary circles of a pair of pants. The gluing operator encodes the elliptic modulus, which plays a role of the time variable in the corresponding isomonodromic problem. Elliptic isomonodromic deformations have been studied e.g. in \cite{Korotkin_elliptic}, where the interested reader can find further references.

  \item It is natural to wonder to what extent the approach proposed in the present work may be followed in the presence of irregular singularities, in particular, for Painlevé I--V equations. The contours of appropriate isomonodromic RHPs become more complicated: in addition to circles of formal monodromy, they include anti-Stokes rays, exponential jumps on which account for Stokes phenomenon \cite{FIKN}. We will sketch here a partial answer in rank $N=2$. For this it is useful to recall a geometric representation of the confluence diagram for Painlevé equations recently proposed by Chekhov, Mazzocco and Rubtsov \cite{CM,CMR}, see Fig.~\ref{figCMR}. To each of the equations (or rather associated linear problems) is assigned a Riemann surface with a number of cusped boundary components. They are obtained from Painlevé VI 4-holed sphere using two surgery operations: i) a ``chewing-gum'' move creating from two holes with $k$ and $l$ cusps one hole with $k+l+2$ cusps and ii) a cusp removal reducing the number of cusps at one hole by $1$. The cusps may be thought of as representing the anti-Stokes rays of the Riemann-Hilbert contour.

       \begin{figure}[h!]
          \centering
          \includegraphics[height=7cm]{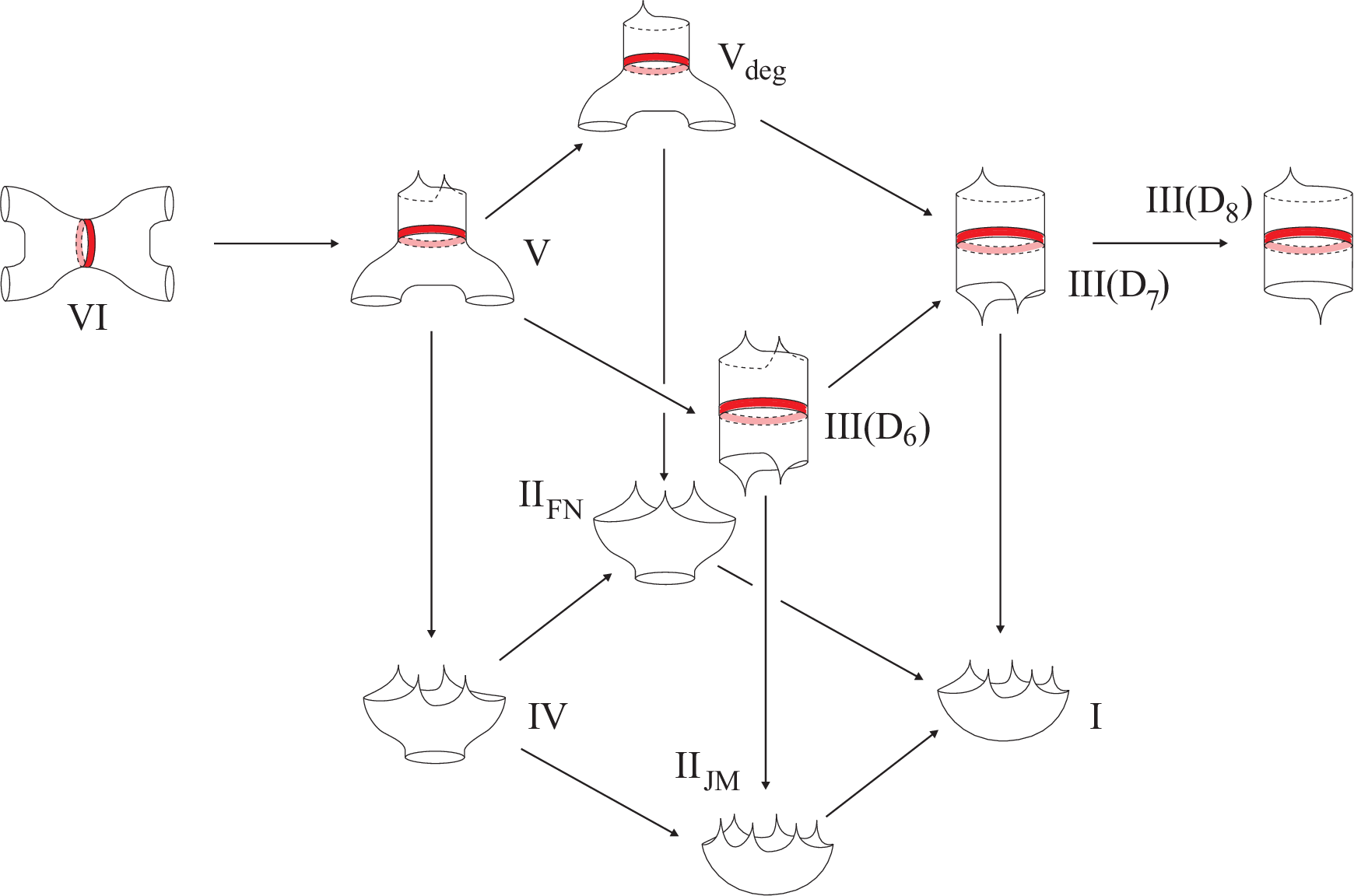}
          \begin{minipage}{0.68\textwidth}
          \caption{CMR confluence diagram for Painlevé equations.\label{figCMR}}
         \end{minipage}
          \end{figure}

  An extension of our approach is straightforward for equations from the upper part of the CMR diagram and, more generally, when the Poincaré ranks of all irregular singular points are either $\frac12$ or $1$. The associated surfaces may be decomposed into irregular pants of three types corresponding to solvable RHPs: Gauss hypergeometric, Whittaker and Bessel systems (Fig. \ref{figsolvable}). They serve to construct local Riemann-Hilbert parametrices which in turn produce the relevant Plemelj operators. 
      \begin{figure}[h!]
          \centering
          \includegraphics[height=2cm]{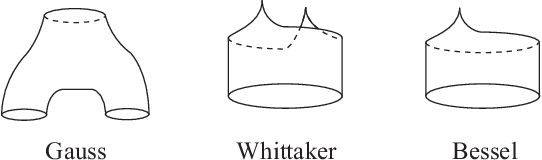}
          \begin{minipage}{0.79\textwidth}
          \caption{Some solvable RHPs in rank $N=2$: Gauss hypergeometric (3 regular punctures), Whittaker (1 regular + 1 of Poincaré rank 1) and Bessel (1 regular + 1 of rank $\frac12$).\label{figsolvable}}
         \end{minipage}
          \end{figure}
          
  The study of higher Poincaré rank  seems to require new ideas. Moreover, even for Painlevé V and Pain\-levé~III Fredholm determinant expansions naturally give series representations of the corresponding tau functions of regular type, first proposed in  \cite{GIL13} and expressed in terms of irregular conformal blocks of \cite{Gaiotto1,Bonelli,GT}.  It is not clear to us how to extract from them irregular (long-distance) asymptotic expansions. Let us mention a recent work \cite{Nagoya} which relates such expansions to irregular conformal blocks of a different type.     
  
  \item Given a matrix $K\in \mathbb C^{\mathfrak X\times \mathfrak X}$ indexed by elements of a discrete set $\mathfrak X$, it is almost a tautology to say that the principal minors $\operatorname{det} K_{\mathfrak Y\in 2^{\mathfrak X}}$ define a determinantal point process on $\mathfrak X$ and a probability measure on $2^{\mathfrak X}$. Fredholm determinant representations and combinatorial expansions of tau functions thus generalize in a natural way various families of measures of random matrix or representation-theoretic origin, such as $Z$- and $ZW$-measures \cite{BorodinAnnals,BO} (the former correspond to the scalar case $N=1$ with $n=4$ regular singular points, and the latter are related to hypergeometric kernel considered in the last subsection). We believe that novel probabilistic models coming from isomonodromy deserve further investigation. 
  \item Perhaps the most intriguing perspective is to extend our setup to $q$-isomonodromy problems, in particular $q$-difference Painlev\'e equations, presumably related to
  the deformed Virasoro algebra \cite{qVir} and 5D gauge theories. Among the results pointing in this direction, let us mention a study of the connection problem for $q$-Painlev\'e VI 
  \cite{Mano} based on asymptotic factorization of the associated linear problem into two systems solved by the Heine basic hypergeometric series $_2\varphi_1$, and critical expansions for 
  solultions of $q$-$P\lb A_1\rb$ equation recently obtained in
  \cite{JR}.
  \end{enumerate}

  \noindent
  { \small \textbf{Acknowledgements}.  We would like to thank F. Balogh, M. Bershtein, M. Bertola, A. Bufetov, M. Cafasso, T. Grava, J. Harnad, G. Helminck, N. Iorgov, A. Its, A. Marshakov, H. Nagoya, V. Poberezhny and V. Rubtsov for their interest to this project and useful discussions. The present work was supported by the CNRS/PICS project ``Isomonodromic deformations and conformal field theory''. P.~G. was supported by the RSF grant No. 16-11-10160 (results of Section 4), the Russian Academic Excellence Project '5-100', and  the ``Young Russian Mathematics'' fellowship. P.~G. would also like to thank  KdV Institute of the University of Amsterdam, where a part of this work was done, and especially G. Helminck, for warm hospitality.
  
  }
  
 \section{Tau functions as Fredholm determinants\label{sectionfred}}
 \subsection{Riemann-Hilbert setup}
  The classical setting of the Riemann-Hilbert problem (RHP) involves two basic ingredients:     
      \begin{tabular}{l l}
      \begin{minipage}{0.7\textwidth}
 \begin{itemize}
 \item A contour $\Gamma$ on a Riemann surface $\Sigma$ of genus $g$ consisting of a finite set of smooth oriented arcs that can intersect transversally.  Orientation of the arcs defines positive and negative side $\Gamma_{\pm}$ of the contour in the usual way, see Fig.~\ref{orient}.
 \item A jump matrix $J:\Gamma\to \mathrm{GL}\lb N,\Cb\rb$ that satisfies suitable smoothness requirements. 
 \end{itemize}
 \end{minipage}
 &
 \begin{minipage}{0.25\textwidth}
 \begin{figure}[H] \centering
        \includegraphics[height=2.2cm]{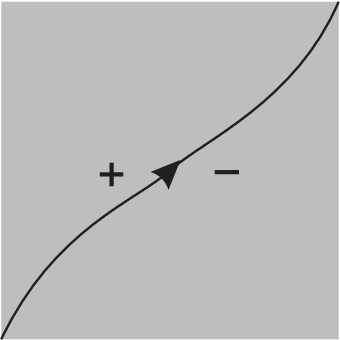}
        \caption{Orientation  and \hspace{\textwidth}labeling of sides
        of $\Gamma$ \label{orient}}
        \end{figure}
 \end{minipage}
 \end{tabular}

 \vspace{0.2cm}The RHP  defined by the pair $\lb \Gamma, J\rb$ consists in finding an analytic invertible matrix function $\Psi:\Sigma\backslash\Gamma\to\mathrm{GL}\lb N,\Cb\rb $ whose boundary values $\Psi_{\pm}$ on $\Gamma_{\pm}$ are related by $\Psi_+=J\Psi_- $. Uniqueness of the solution is ensured  by adding an appropriate normalization condition.

\begin{figure}[h!]
   \centering
 \begin{tikzpicture} 
 \draw(0,0) node{\includegraphics[height=9cm]{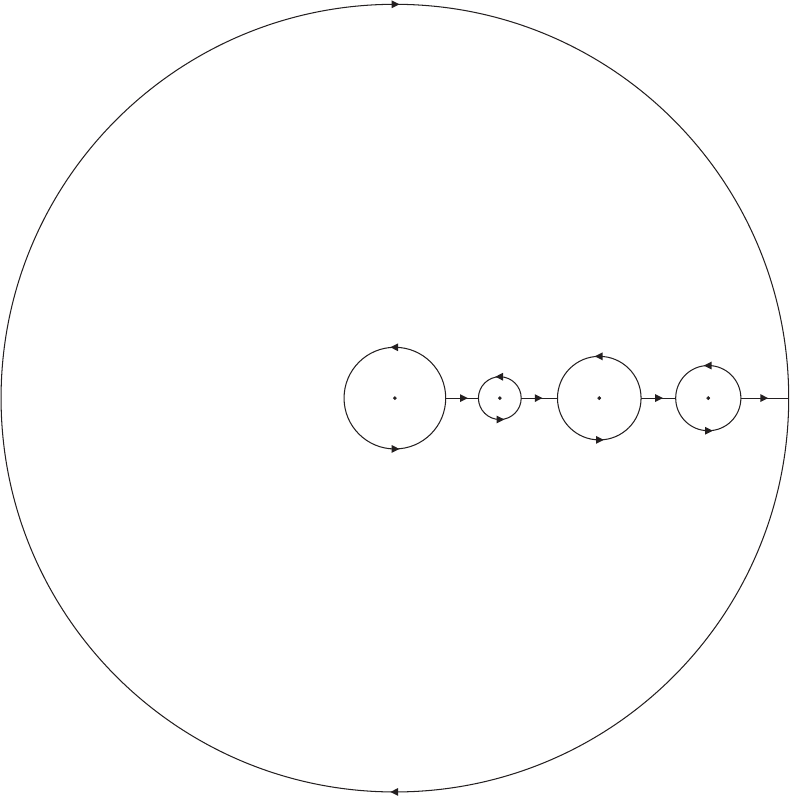}}; 
 \draw(-0.1,-0.15) node {\small $a_0$};  
 \draw(1.15,-0.12) node {\footnotesize $a_1$}; 
 \draw(2.2,-0.15) node {\small $a_2$};  
 \draw(3.5,-0.15) node {\small $a_3$};  
 \draw(0.75,0.25) node {\small $\ell_0$};   
 \draw(1.6,0.25) node {\small $\ell_1$};
 \draw(3,0.25) node {\small $\ell_2$};   
 \draw(4.15,0.25) node {\small $\ell_3$};
 \draw(0,4.2) node {\small $\gamma_4$};  
 \draw(0,0.8) node {\small $\gamma_0$};      
 \draw(1.2,0.5) node {\small $\gamma_1$}; 
 \draw(2.3,0.7) node {\small $\gamma_2$};  
 \draw(3.55,0.6) node {\small $\gamma_3$};  
 \end{tikzpicture}
   \caption{\label{fig1}
   Contour $\Gamma$ for $n=5$\label{figFredRHP}}
   \end{figure}
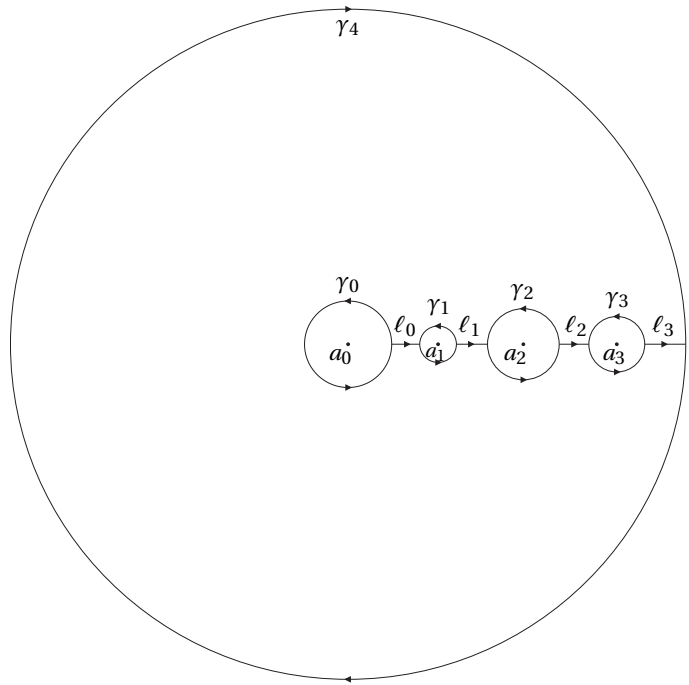

 In the present work we are mainly interested in the genus $0$ case:  $\Sigma=\mathbb P^1$. Let us fix a collection
 \ben
 a:=\lb a_0=0,a_1,\ldots,a_{n-3}, a_{n-2}= 1,a_{n-1}=\infty\rb
 \ebn 
 of $n$ distinct points  on $\Pb^1$ satisfying the condition of radial ordering $0<|a_1|<\ldots <|a_{n-3}|<1$. To reduce the amount of fuss below, it is convenient to assume that
 $a_{1},\ldots, a_{n-3}\in \mathbb R_{>0}$. The contour $\Gamma$ will then be chosen as a collection 
 \ben
 \Gamma=\bigl(\bigcup\nolimits_{k=0}^{n-1}\gamma_k\Bigr)\cup
 \bigl(\bigcup\nolimits_{k=0}^{n-2}\ell_k\Bigr)
 \ebn
 of counter-clockwise oriented circles $\gamma_k$ of sufficiently small radii centered at $a_k$, and the segments $\ell_{k}\subset\mathbb R$   joining the circles $\gamma_k$ and $\gamma_{k+1}$, see Fig.~\ref{figFredRHP}.
 
 The jumps will be defined by the following data:
 \begin{itemize}
 \item An $n$-tuple of diagonal $N\times N$ matrices $\Theta_k=\operatorname{diag}
 \left\{\theta_{k,1},\ldots,\theta_{k,N}\right\}\in\mathbb C^{N\times N}$ (with $k=0,\ldots,n-1$) satisfying Fuchs consistency relation $\sum_{k=0}^{n-1}\operatorname{Tr}\Theta_k=0$ and having non-resonant spectra. The latter condition means that ${\theta_{k,\alpha}-\theta_{k,\beta}\notin
 \mathbb Z\backslash\left\{0\right\}}$. 
 \item A collection of $2n$  matrices $C_{k,\pm}\in \mathrm{GL}\lb N,\Cb\rb$ subject to the constraints 
 \beq\label{connmatrices}
 \begin{gathered}
 M_{0\to k} :=C_{k,-}e^{2\pi i\Theta_k}C_{k,+}^{-1}=
 C_{k+1,-}C_{k+1,+}^{-1},\qquad  k=0,\ldots, n-3,\\
 M_{0\to n-2}:=C_{n-2,-}e^{2\pi i\Theta_{n-2}}C_{n-2,+}^{-1}
 =C_{n-1,-}e^{-2\pi i\Theta_{n-1}}C_{n-1,+}^{-1},\\
 M_{0\to n-1}:=\mathbb1=C_{n-1,-}C_{n-1,+}^{-1}=C_{0,-}C_{0,+}^{-1},
 \end{gathered}
 \eeq 
 which are simultaneously viewed as the definition of $M_{0\to k}\in
 \mathrm{GL}\lb N,\Cb\rb$. Only $n$ of the initial matrices (for example, $C_{k,+}$) are therefore independent. 
  \end{itemize}
 The jump matrix $J$ that we are going to consider is then given by
 \beq\label{jumps}
 \begin{gathered}
 \begin{aligned}
 J\lb z\rb\Bigl|_{\ell_k}=M_{0\to k}^{\;-1},\qquad & 
 k=0,\ldots,n-2,\\
 J\lb z\rb\Bigl|_{\gamma_k}=\lb a_k-z\rb^{-\Theta_k}C_{k,\pm}^{-1},\qquad &
 \Im z\gtrless 0,\quad k=0,\ldots, n-2,\\
 J\lb z\rb\Bigl|_{\gamma_{n-1}}=\lb -z\rb^{\Theta_{n-1}}C_{n-1,\pm}^{-1},\qquad &
  \Im z\gtrless 0.
  \end{aligned}
 \end{gathered}
 \eeq
 Throughout this paper, complex powers will always be understood as $z^{\theta}=e^{\theta\ln z}$, the logarithm being defined on the principal branch. The subscripts $\pm$ of $C_{k,\pm}$ are sometimes omitted to lighten the notation.
 
 A major incentive to study the above RHP comes from its direct connection to systems of linear ODEs with rational coefficients. Indeed, define a new matrix $\Phi$ by
 \beq\label{reltofuchs}
 \Phi\lb z\rb =\begin{cases}
 \Psi\lb z\rb,\qquad &z\text{ outside }\gamma_{0\ldots n-1},\\
 C_k\lb a_k-z\rb^{\Theta_k}\Psi\lb z\rb ,\qquad & 
 z\text{ inside }\gamma_{k},\quad k=0,\ldots,n-2,\\
 C_{n-1}\lb -z\rb^{-\Theta_{n-1}} \Psi\lb z\rb ,\qquad & z\text{ inside }\gamma_{n-1}.
 \end{cases}
 \eeq
 It has only piecewise constant jumps $
  J_{\Phi}\lb z\rb\bigl|_{]a_k,a_{k+1}[}=M_{0\to k}^{-1}
  $ on the positive real axis.
 The matrix $ A\lb z\rb:=\Phi^{-1}\partial_z \Phi$ is therefore meromorphic on
 $\mathbb P^1$ with poles only possible at $a_0,\ldots, a_{n-1}$. It follows immediately that
 \beq\label{fuchsys}
 \partial_z\Phi=\Phi A\lb z\rb,\qquad A\lb z\rb =\sum_{k=0}^{n-2}\frac{A_k}{z-a_k},
 \eeq
 with $A_k=\Psi\lb a_k\rb^{-1}\Theta_k \Psi\lb a_k\rb $. 
 Thus $\Phi\lb z\rb$ is a fundamental matrix solution for a class of Fuchsian systems related by constant gauge transformations. It has prescribed monodromy and singular behavior that are encoded in the connection matrices $C_k$ and local monodromy exponents $\Theta_k$. The freedom in the choice of the gauge reflects the dependence on the normalization of $\Psi$. Let us note that the conditions \eqref{connmatrices} on the Riemann-Hilbert data are in fact designed with the aim to reproduce the monodromy of $\Phi\lb z\rb$. This explains a rather non-obvious fact of the absence of monodromy around the nod points of $\Gamma$, which in its turn ensures that the RHP solution does not have singularities at these points.

\begin{figure}
     \centering
 \begin{tikzpicture} 
 \draw(0,0) node{ \includegraphics[width=10cm]{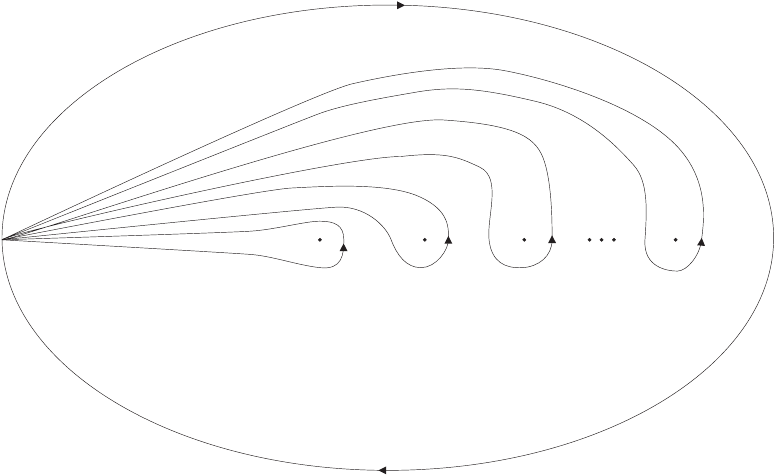}}; 
 \draw(-1.08,-0.1) node {\small $a_0$};  
 \draw(0.3,-0.1) node {\small $a_1$};  
 \draw(1.55,-0.1) node {\small $a_2$};   
 \draw(3.68,-0.15) node {\footnotesize $a_{n-2}$};  
 \draw(-0.45,-0.5) node {\small $\xi_0$};    
 \draw(0.85,-0.5) node {\small $\xi_1$};  
 \draw(2.15,-0.5) node {\small $\xi_2$};  
 \draw(4.3,-0.5) node {\small $\xi_{n-2}$};  
 \draw(0,-2.7) node {\small $\xi_{n-1}$};   
 \end{tikzpicture}      
     \caption{\label{Fredholm_fungr}
     Generators of $\pi_1\lb \Pb^1\backslash a\rb$}
     \end{figure}

   The monodromy representation 
   $ \rho:\pi_1\lb\mathbb P^1
   \backslash a\rb\to \mathrm{GL}\lb N,\Cb\rb $ associated to $\Phi$ is uniquely determined by the jumps. It is generated by the matrices $M_k=\rho\lb\xi_k\rb$ assigned to counter-clockwise loops $\xi_0,\ldots,\xi_{n-1}$ represented in Fig.~\ref{Fredholm_fungr}. They are obtained from 
   $\left[T_{\xi_k}\lb\Phi\rb\right]\lb z\rb=M_k\Phi\lb z\rb$, where $T_{\xi_k}$ denotes the operator of analytic continuation along the loop $\xi_k$. These matrices may be  expressed as
   \ben
   M_0=M_{0\to 0},\qquad M_{k+1}={ M_{0\to k}}^{-1}M_{0\to k+1},
   \ebn
   which means simply that $M_{0\to k}=M_0\ldots M_{k-1} M_k$.
   It is a direct consequence of the definition (\ref{connmatrices}) that the spectra of $M_k$ coincide with those of $e^{2\pi i \Theta_k}$. In fact, $M_k=C_{k,+}e^{2\pi i \Theta_k}C_{k,+}^{-1}$.

  \begin{ass}\label{asssp}
 The matrices $M_{0\to k}$ with $k=1,\ldots,n-3$ are assumed to be diagonalizable:
  \ben
  M_{0\to k}= S_k e^{2\pi i \mathfrak S_k}S_k^{-1},
  \qquad \mathfrak S_k=\operatorname{diag}\left\{
  \sigma_{k,1},\ldots,\sigma_{k,N}\right\}.
  \ebn
  It can then be assumed without loss in generality that $\operatorname{Tr}\mathfrak S_k=\sum_{j=0}^{k}\operatorname{Tr}\,\Theta_j$ and $\left|\Re\lb\sigma_{k,\alpha}-\sigma_{k,\beta}\rb\right|\le1$. We further impose a non-resonancy condition $\sigma_{k,\alpha}-\sigma_{k,\beta}\neq
  \pm 1$.
  \end{ass}
  \noindent 
  In order to have uniform notation, we may also identify $\mathfrak S_0\equiv\Theta_0$, $\mathfrak S_{n-2}\equiv -\Theta_{n-1}$.
  Note that any sufficiently generic monodromy representation can be realized as described above. 
 
 \subsection{Auxiliary $3$-point RHPs\label{subsecauxRHP}} 
\begin{figure}
      \centering
  \begin{tikzpicture} 
  \draw(0,0) node{\includegraphics[width=16cm]{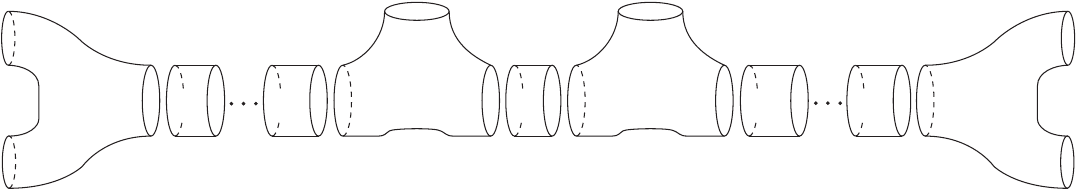}}; 
  \draw(-1.2,-0.2) node {\footnotesize $\mathcal C^{[k]}_{\mathrm{out}}$}; 
  \draw(-2.4,-0.2) node {\footnotesize $\mathcal C^{[k]}_{\mathrm{in}}$};  
  \draw(-3.7,-0.2) node {\footnotesize $\mathcal A_{k-1}$};  
  \draw(-0.2,-0.2) node {\footnotesize $\mathcal A_{k}$};  
  \draw(-5.2,-0.2) node {\footnotesize $\mathcal A_{1}$}; 
  \draw(-6.2,-0.2) node {\footnotesize $\mathcal C^{[1]}_{\mathrm{out}}$};  
  \draw(-7.5,-1.1) node {\footnotesize $\gamma_0$}; 
  \draw(-7.5,0.8) node {\footnotesize $\gamma_1$};  
  \draw(-1.8,0.9) node {\footnotesize $\gamma_k$};   
  \draw(-6.9,-0.1) node {$\mathcal T^{[1]}$};     
  \draw(-1.7,0.3) node {$\mathcal T^{[k]}$};
  \draw(1.7,0.9) node {\footnotesize $\gamma_{k+1}$};
  \draw(1.8,0.3) node {$\mathcal T^{[k+1]}$}; 
  \draw(1.15,-0.2) node {\footnotesize $\mathcal C^{[k+1]}_{\mathrm{in}}$}; 
  \draw(2.2,-0.2) node {\footnotesize $\mathcal C^{[k+1]}_{\mathrm{out}}$};     
  \draw(3.4,-0.2) node {\footnotesize $\mathcal A_{k+1}$};  
  \draw(4.95,-0.2) node {\footnotesize $\mathcal A_{n-3}$};   
  \draw(6.35,0) node {\footnotesize $\mathcal C^{[n-2]}_{\mathrm{in}}$}; 
  \draw(7.4,-1.1) node {\footnotesize $\gamma_{n-1}$}; 
  \draw(7.4,0.8) node {\footnotesize $\gamma_{n-2}$};  
  \draw(6.9,-0.3) node {$\mathcal T^{[n-2]}$};                        
  \end{tikzpicture}      
     \caption{\label{figfredpants}
      Labeling of trinions, annuli and boundary curves}
      \end{figure}
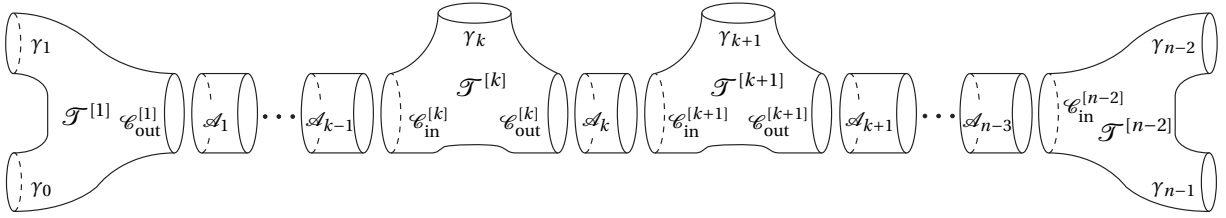

 Consider a decomposition of the original $n$-punctured sphere 
 into $n-2$ pairs of pants $\mathcal T^{[1]},\ldots,\mathcal T^{[n-2]}$ by
 $n-3$ annuli $\mathcal A_1,\ldots, \mathcal A_{n-3}$ represented in Fig.~\ref{figfredpants}. The labeling is designed so that two boundary components of the annulus $\mathcal A_k$ that belong to trinions $\mathcal T^{[k]}$ and $\mathcal T^{[k+1]}$ are denoted by $\mathcal C^{[k]}_{\text{out}}$ and $\mathcal C^{[k+1]}_{\text{in}}$.  We are now going to associate to the $n$-point RHP described above $n-2$ simpler $3$-point RHPs assigned to different trinions and defined by the pairs $\lb \Gamma^{[k]},J^{[k]}\rb$ with $k=1,\ldots,n-2$.
 
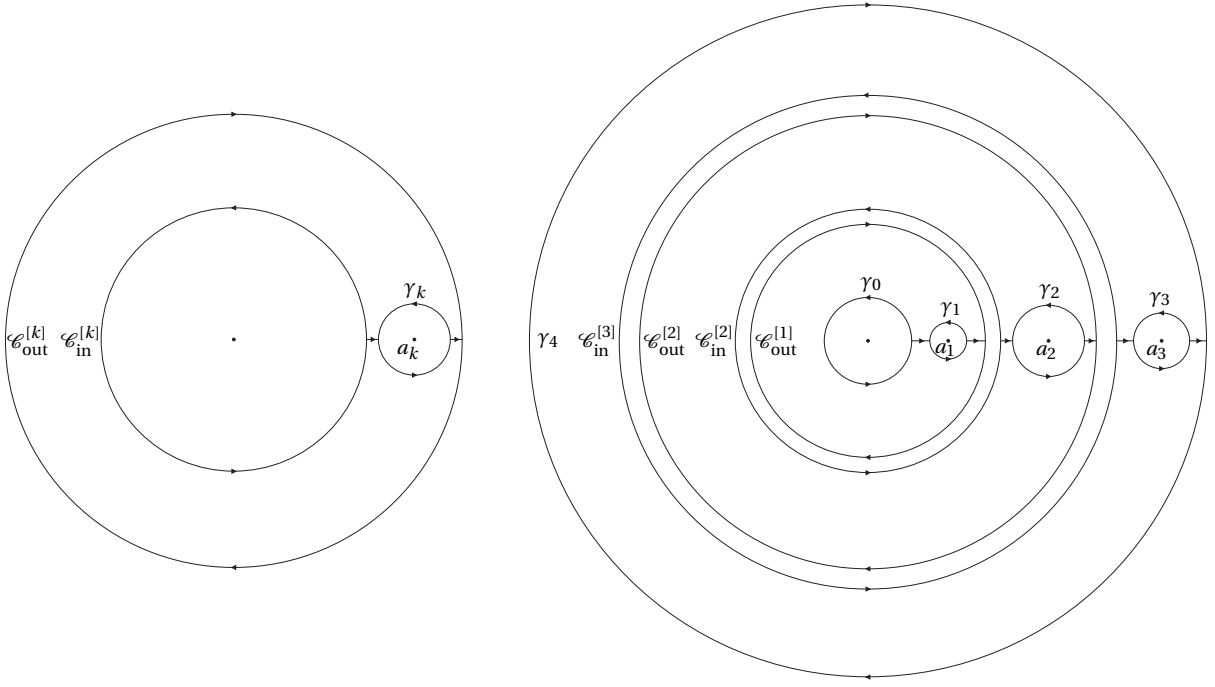
\begin{figure}
     \centering
 \begin{tikzpicture} 
 \draw(0,0) node{\includegraphics[height=9cm]{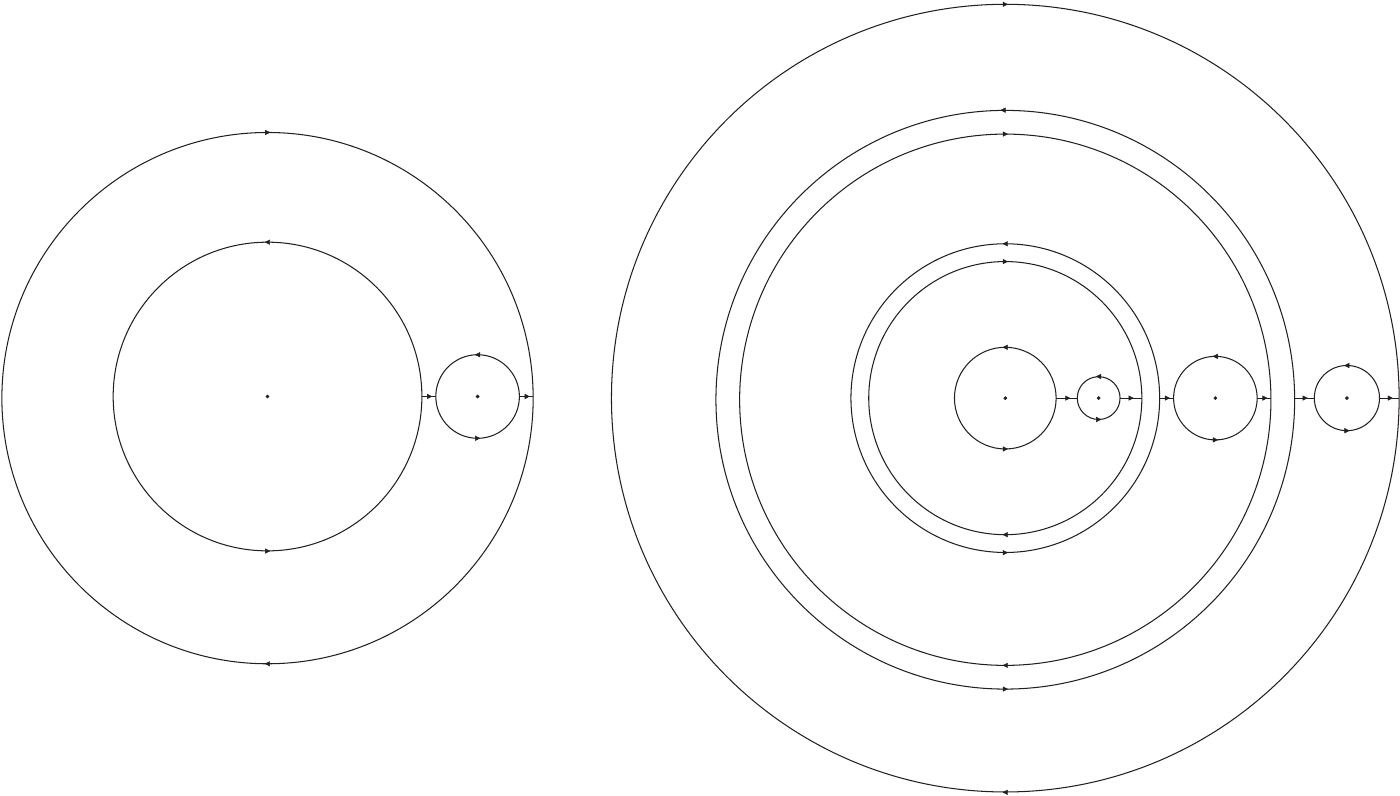}}; 
 \draw(-7.6,0) node {\footnotesize $\mathcal C^{[k]}_{\mathrm{out}}$};  
 \draw(-6.9,0) node {\footnotesize $\mathcal C^{[k]}_{\mathrm{in}}$}; 
 \draw(-2.5,0.67) node {\footnotesize $\gamma_k$};    
 \draw(-2.6,-0.15) node {\footnotesize $a_k$}; 
 \draw(-0.75,0) node {\footnotesize $\gamma_4$}; 
 \draw(-0.1,0) node {\footnotesize $\mathcal C^{[3]}_{\mathrm{in}}$}; 
 \draw(0.78,0) node {\footnotesize $\mathcal C^{[2]}_{\mathrm{out}}$}; 
 \draw(1.44,0) node {\footnotesize $\mathcal C^{[2]}_{\mathrm{in}}$};
 \draw(2.25,0) node {\footnotesize $\mathcal C^{[1]}_{\mathrm{out}}$};   
 \draw(3.5,0.75) node {\footnotesize $\gamma_0$};
 \draw(4.55,0.42) node {\footnotesize $\gamma_1$}; 
 \draw(4.47,-0.12) node {\footnotesize $a_1$}; 
 \draw(5.85,0.65) node {\footnotesize $\gamma_2$}; 
 \draw(5.8,-0.15) node {\footnotesize $a_2$};   
 \draw(7.3,0.55) node {\footnotesize $\gamma_3$};   
 \draw(7.25,-0.15) node {\footnotesize $a_3$};                
 \end{tikzpicture}      
   \caption{\label{figgammahat}
   Contour  $\Gamma^{[k]}$ (left) and  $\hat\Gamma$ for $n=5$ (right)}
     \end{figure}

 The curves $\mathcal C^{[k]}_{\text{in}}$ and $\mathcal C^{[k]}_{\text{out}}$ are represented by circles of positive and negative orientation as shown in Fig.~\ref{figgammahat}. For $k=2,\ldots,n-3$, the contour $\Gamma^{[k]}$ of the RHP assigned to trinion $\mathcal T^{[k]}$ consists of three circles $\mathcal C^{[k]}_{\text{in}}$, $\mathcal C^{[k]}_{\text{out}}$,  $\gamma_{k}$ associated to boundary components, and two segments of the real axis. For leftmost and rightmost trinions $\mathcal T^{[1]}$ and $\mathcal T^{[n-2]}$, the role of $\mathcal C^{[1]}_{\text{in}}$ and $\mathcal C^{[n-2]}_{\text{out}}$ is played respectively by 
 the circles $\gamma_0$ and $\gamma_{n-1}$ around $0$ and $\infty$. 

 The jump matrix $J^{[k]}$ is constructed according to two basic rules:
 \begin{itemize}
 \item The arcs that belong to original contour give rise to the same jumps: $\lb J^{[k]}-J\rb\bigl|_{\Gamma^{[k]}\cap\Gamma}=0$.
 \item The jumps on the boundary circles $\mathcal C^{[k]}_{\text{out}}$, $\mathcal C^{[k+1]}_{\text{in}}$ mimic regular singularities characterized by counter-clockwise monodromy matrices $M_{0\to k}$:
 \beq\label{jump3point}
 J^{[k]}\Bigl|_{\mathcal C_{\text{out}}^{[k]}}=
 \lb -z\rb^{-\mathfrak S_{k}}S_{k}^{-1},\qquad
 J^{[k+1]}\Bigl|_{\mathcal C_{\text{in}}^{[k+1]}}=
 \lb -z\rb^{-\mathfrak S_{k}} S_{k}^{-1},
 \qquad k=1,\ldots,n-3.
 \eeq
 \end{itemize}
 The solution $\Psi^{[k]}$ of the RHP defined by the pair $\lb \Gamma^{[k]}, J^{[k]}\rb$ is thus related in a way analogous to (\ref{reltofuchs}) to the fundamental matrix solution $\Phi^{[k]}$ of a Fuchsian system with 3 regular singular points at $0$, $a_{k}$ and $\infty$ characterized by monodromies $M_{0\to k-1}$, $M_{k}$, $M_{0\to k}^{-1}$:
 \beq\label{FS3point}
 \partial_z \Phi^{[k]}=\Phi^{[k]}A^{[k]}\lb z\rb ,\qquad
 A^{[k]}\lb z\rb=\frac{A^{[k]}_0}{z}+\frac{A^{[k]}_1}{z-a_{k}}.
 \eeq
 We note in passing that the spectra of $A^{[k]}_0$, $A^{[k]}_1$ and $A^{[k]}_{\infty}:=-A^{[k]}_0-A^{[k]}_1$ coincide with the spectra of 
 $\mathfrak S_{k-1}$, $\Theta_{k}$ and $-\mathfrak S_{k}$. We also assume that the monodromy is sufficiently generic so that the $3$-point Fuchsian systems with the local exponents of the fundamental solution leading to the jumps (\ref{jump3point}) in $\Psi^{[k]}$ exist\footnote{An effective characterization of the non-solvable cases for $N\ge 3$ does not seem to be available in the literature  even for irreducible monodromy representations; see \cite{Bolibrukh} for a discussion of related matters. For rank $N=2$, however, if the non-resonance condition in Assumption~\ref{asssp} is satisfied, the $3$-point solutions can be constructed explicitly in terms of hypergeometric functions.} for all $k=1,\ldots,n-2$. 
 
 It will be convenient to replace the $n$-point RHP described in the previous subsection by a slightly modified one. It is defined by a pair $\lb \hat\Gamma,\hat J\rb$ such that (cf right part of Fig.~\ref{figgammahat})
 \beq\label{RHPhat}
 \hat\Gamma=\bigcup_{k=1}^{n-2}\Gamma^{[k]},\qquad \hat J\Bigl|_{\Gamma^{[k]}}=J^{[k]}.
 \eeq
 Constructing the solution $\hat\Psi$ of this RHP is equivalent to  finding $\Psi$: it is plain that
 \beq\label{psipsihat}
 \hat\Psi\lb z\rb=
 \begin{cases}
 \lb -z\rb^{-\mathfrak S_{k}}S_{k}^{-1}\Psi \lb z\rb  , \qquad &
 z\in\mathcal A_{k},\\
 \Psi\lb z\rb,\qquad & z\in\mathbb P^1\backslash \bigcup\limits_{k=1}^{n-3}\mathcal A_k.
 \end{cases}
 \eeq
 Our aim in the next subsections is to construct the isomonodromic tau function in terms of $3$-point {solutions~$\Phi^{[k]}$}. This construction employs in a crucial way integral Plemelj operators acting on spaces of holomorphic functions on $\mathcal A:=\bigcup_{k=1}^{n-3}\mathcal A_k$.
 
 \subsection{Plemelj operators\label{subsecplemelj}}
  Given a positively oriented circle $\mathcal C\subset\mathbb C$ centered at the origin, let us denote by $\mathcal V\lb \mathcal C\rb$ the space of functions holomorphic in an annulus containing $\mathcal C$. Any $f\in
  \mathcal V\lb \mathcal C\rb$ is canonically decomposed as $f=f_++f_-$, where $f_+$ and $f_-$ denote the analytic and principal part of $f$. Let us accordingly write $\mathcal V\lb \mathcal C\rb=
  \mathcal V_+\lb \mathcal C\rb\oplus \mathcal V_-\lb \mathcal C\rb$ and denote by $\Pi_\pm\lb \mathcal C\rb$ the projectors on the corresponding subspaces. Their explicit form is 
  \ben
   \Pi_\pm\lb \mathcal C\rb f\lb z\rb =\frac1{2\pi i}
  \oint_{\mathcal C_\pm,|z'|=|z|\pm 0}\frac{f\lb z'\rb dz'}{z'-z},
  \ebn  
  where the subscript of $\mathcal C_{\pm}$ indicates the orientation of $\mathcal C$. Projectors $\Pi_\pm\lb \mathcal C\rb$
  are the classical Cauchy operators, which are simple instances of the operators extensively used below. 
   
   Let us next associate to every trinion $\mathcal T^{[k]}$ with 
   $k=2,\ldots, n-3$ the spaces of vector-valued functions
   \ben
   \mathcal H^{[k]}=\bigoplus_{\epsilon=\mathrm{in,out}}
   \lb\mathcal H^{[k]}_{\epsilon,+}\oplus 
   \mathcal H^{[k]}_{\epsilon,-}\rb,\qquad 
   \mathcal H^{[k]}_{\epsilon,\pm}=\mathbb C^N\otimes \mathcal V_{\pm}
   \lb \mathcal C_{\epsilon}^{[k]}\rb.
   \ebn 
  With respect to the first decomposition, it is convenient to write the elements $f^{[k]}\in\mathcal H^{[k]}$ as
  \ben
  f^{[k]}=\lb\begin{array}{c}
  f^{[k]}_{\mathrm{in},-}\vspace{0.1cm}\\ 
  f^{[k]}_{\mathrm{out},+}
  \end{array}\rb \oplus \lb\begin{array}{c}
    f^{[k]}_{\mathrm{in},+} \vspace{0.1cm}\\ 
   f^{[k]}_{\mathrm{out},-}
    \end{array}\rb.
  \ebn
  Here $f^{[k]}_{\mathrm{\epsilon},\pm}$ denote $N$-column vectors which represent the restrictions of analytic and principal part of $f^{[k]}$ to  boundary circle $\mathcal C_{\mathrm{\epsilon}}^{[k]}$. 
  Now define a Plemelj operator $\mathcal P^{[k]}:\mathcal H^{[k]}\to
  \mathcal H^{[k]}$ by
  \beq\label{onepointpr}
  \mathcal P^{[k]}f^{[k]}\lb z\rb=\frac{1}{2\pi i}
  \oint_{\mathcal C^{[k]}_{\mathrm{in}}\cup 
  \mathcal C^{[k]}_{\mathrm{out}}}\frac{ \Psi^{[k]}_+\lb z\rb
  {\Psi^{[k]}_+\lb z'\rb}^{-1} f^{[k]}\lb z'\rb dz' }{z-z'}.
  \eeq
  The singular factors $1/\lb z-z'\rb$ for $z,z'\in \mathcal C_{\mathrm{in,out}}^{[k]}$ are interpreted with the following prescription: the contour of integration is deformed to appropriate annulus (e.g. $\mathcal A_{k-1}$ for $\mathcal C_{\mathrm{in}}^{[k]}$
  and $\mathcal A_{k}$  for $\mathcal C_{\mathrm{out}}^{[k]}$) as to avoid the pole at $z'=z$. Matrix function $\Psi^{[k]}\lb z\rb$ is a solution of the $3$-point RHP described in the previous subsection. Its normalization is irrelevant as the corresponding factor cancels out in (\ref{onepointpr}). Let us stress once again the triviality of monodromy at the nods of $\Gamma^{[k]}$, which ensures that $\Psi^{[k]}_+\lb z\rb$ can be analytically continued to small annuli containing $\mathcal C^{[k]}_{\mathrm{out}}$ and $\mathcal C^{[k]}_{\mathrm{in}}$.
 
  \begin{lemma}\label{lemalphagamma} We have $\lb\mathcal P^{[k]}\rb^2=\mathcal P^{[k]}$ and  $\operatorname{ker}\mathcal P^{[k]}=\mathcal H^{[k]}_{\mathrm{in},+}\oplus\mathcal H^{[k]}_{\mathrm{out},-}$. Moreover, $\mathcal P^{[k]}$ can be explicitly written as
  \ben
  \mathcal P^{[k]}: \lb\begin{array}{c}
     f^{[k]}_{\mathrm{in},-}\vspace{0.1cm}\\ 
   f^{[k]}_{\mathrm{out},+}
    \end{array}\rb \oplus \lb\begin{array}{c}
    f^{[k]}_{\mathrm{in},+}  \vspace{0.1cm}\\ 
   f^{[k]}_{\mathrm{out},-}   
      \end{array}\rb\mapsto
  \lb\begin{array}{c}
    f^{[k]}_{\mathrm{in},-}  \vspace{0.1cm}\\ 
    f^{[k]}_{\mathrm{out},+}  
      \end{array}\rb \oplus 
   \lb\begin{array}{rr}
  \mathsf a^{[k]} & \mathsf b^{[k]} \vspace{0.1cm}\\ 
  \mathsf c^{[k]} & \mathsf d^{[k]}
   \end{array}\rb   \lb\begin{array}{c}
      f^{[k]}_{\mathrm{in},-}   \vspace{0.1cm}\\ 
  f^{[k]}_{\mathrm{out},+}       
        \end{array}\rb ,   
  \ebn
  where the operators $\mathsf a^{[k]}$, $\mathsf b^{[k]}$, 
  $\mathsf c^{[k]}$, $\mathsf d^{[k]}$
  are defined by
  \begin{subequations}
  \label{alphadelta}
  \begin{alignat}{3}
  \label{alphadelta01}
  \lb\mathsf a^{[k]}g\rb\lb z\rb=&\,\frac1{2\pi i}\oint_{\mathcal C_{\mathrm{in}}^{[k]}}\frac{\left[ \Psi^{[k]}_+\lb z\rb {\Psi^{[k]}_+\lb z'\rb}^{-1}-\mathbb 1\,\right]g\lb z'\rb dz' }{z-z'},
   \qquad\qquad && z\in \mathcal C_{\mathrm{in}}^{[k]},\\
  \lb\mathsf b^{[k]}g\rb\lb z\rb=&\,\frac1{2\pi i}\oint_{\mathcal C_{\mathrm{out}}^{[k]}}\frac{ \Psi^{[k]}_+\lb z\rb 
   {\Psi^{[k]}_+\lb z'\rb}^{-1}g\lb z'\rb dz' }{z-z'},
    \qquad \qquad && z\in \mathcal C_{\mathrm{in}}^{[k]},\\
  \lb\mathsf c^{[k]}g\rb\lb z\rb=&\,\frac1{2\pi i}\,\oint_{\mathcal C_{\mathrm{in}}^{[k]}}\;\frac{ \Psi^{[k]}_+\lb z\rb
   {\Psi^{[k]}_+\lb z'\rb}^{-1}g\lb z'\rb dz' }{z-z'},
    \qquad\qquad && z\in \mathcal C_{\mathrm{out}}^{[k]},\\
   \label{alphadelta04}
 \lb\mathsf d^{[k]}g\rb\lb z\rb=&\,\frac1{2\pi i}\oint_{\mathcal C_{\mathrm{out}}^{[k]}}\frac{\left[ \Psi^{[k]}_+\lb z\rb {\Psi^{[k]}_+\lb z'\rb}^{-1}-\mathbb 1\,\right]g\lb z'\rb dz' }{z-z'},
    \qquad\qquad && z\in \mathcal C_{\mathrm{out}}^{[k]}.    
  \end{alignat}
  \end{subequations}
  \end{lemma}
  \pf Let us first prove that $\mathcal H^{[k]}_{\mathrm{in},+},\mathcal H^{[k]}_{\mathrm{out},-}\subset \operatorname{ker}\mathcal P^{[k]}$. This statement follows from the fact that $\Psi^{[k]}_+$ holomorphically extends inside $\mathcal C_{\mathrm{in}}^{[k]}$ and
  outside $\mathcal C_{\mathrm{out}}^{[k]}$, so that the integration contours can be shrunk to $0$ and $\infty$. To prove the projection property,  decompose for example
  \ben
  \lb\mathcal P^{[k]}f_{\mathrm{out},+}^{[k]}\rb_{\mathrm{out}}
  \lb z\rb=\frac{1}{2\pi i}
    \oint_{ 
    \mathcal C^{[k]}_{\mathrm{out}}}\frac{\left[ \Psi^{[k]}_+\lb z\rb {\Psi^{[k]}_+\lb z'\rb}^{-1}-\mathbb 1\,\right]
    f_{\mathrm{out},+}^{[k]}\lb z'\rb dz' }{z-z'}+
    \frac{1}{2\pi i}
        \oint_{\mathcal C^{[k]}_{\mathrm{out}},|z'|>|z|}\frac{f_{\mathrm{out},+}^{[k]}\lb z'\rb dz' }{z-z'}.
  \ebn
  The first integral admits holomorphic continuation in $z$  outside $\mathcal C_{\mathrm{out}}^{[k]}$ thanks to nonsingular integral kernel, and leads to (\ref{alphadelta04}), whereas the second term is obviously equal to $f^{[k]}_{\mathrm{out},+}$. The action of  $\mathcal P^{[k]}$ on $ f^{[k]}_{\mathrm{in},-}$ is computed in a similar fashion.
  \epf
  
  The leftmost and rightmost trinions $\mathcal T^{[1]}$ and $\mathcal T^{[n-2]}$ 
  play somewhat distinguished role. Let us assign to them boundary spaces
  \ben
  \mathcal H^{[1]}:=\mathcal H^{[1]}_{\mathrm{out},+}\oplus
  \mathcal H^{[1]}_{\mathrm{out},-},\qquad
    \mathcal H^{[n-2]}:=\mathcal H^{[n-2]}_{\mathrm{in},+}\oplus
    \mathcal H^{[n-2]}_{\mathrm{in},-},
  \ebn
  and the operators $\mathcal P^{[k]}:\mathcal H^{[k]}\to
  \mathcal H^{[k]}$ with $k=1,n-2$  defined by
 \begin{align*}
   \mathcal P^{[1]}f^{[1]}\lb z\rb=&\,\frac{1}{2\pi i}
   \oint_{\mathcal C^{[1]}_{\mathrm{out}}}\frac{ \Psi^{[1]}_+\lb z\rb
   {\Psi^{[1]}_+\lb z'\rb}^{-1}f^{[1]}\lb z'\rb dz' }{z-z'},\\
   \mathcal P^{[n-2]}f^{[n-2]}\lb z\rb=&\,\frac{1}{2\pi i}
      \oint_{\mathcal C^{[n-2]}_{\mathrm{in}}}\frac{ \Psi^{[n-2]}_+\lb z\rb {\Psi^{[n-2]}_+\lb z'\rb}^{-1} f^{[n-2]}\lb z'\rb dz' }{z-z'}.
 \end{align*}
 Analogously to the above, one can show that
 \begin{align*}
 \mathcal P^{[1]}:&\,\;f^{[1]}_{\mathrm{out},+}\oplus f^{[1]}_{\mathrm{out},-}\;\;\mapsto 
 f^{[1]}_{\mathrm{out},+}\;\oplus \mathsf d^{[1]}f^{[1]}_{\mathrm{out},+},
 \\
 \mathcal P^{[n-2]}:&\, f^{[n-2]}_{\mathrm{in},-}\oplus f^{[n-2]}_{\mathrm{in},+}\mapsto 
  f^{[n-2]}_{\mathrm{in},-}\oplus \mathsf a^{[n-2]}f^{[n-2]}_{\mathrm{in},-},
 \end{align*} 
 where the operators $\mathsf d^{[1]}$, $\mathsf a^{[n-2]}$ are given by the same formulae (\ref{alphadelta01}), (\ref{alphadelta04}). Note in particular that $\mathcal P^{[1]}$ and $\mathcal P^{[n-2]}$ are projections along their kernels $\mathcal H^{[1]}_{\mathrm{out},-}$ and $\mathcal H^{[n-2]}_{\mathrm{in},+}$.

 Let us next introduce the total space
 \ben
 \mathcal H:=\bigoplus_{k=1}^{n-2}\mathcal H^{[k]}.
 \ebn
 It admits a splitting that will play an important role  
 below. Namely,
 \beq\label{firstsplitting}
 \begin{gathered}
 \mathcal H=\mathcal H_{+}\oplus\mathcal H_{-},\\
 \mathcal H_{\pm}:=\mathcal H^{[1]}_{\mathrm{out},\pm}
 \oplus\lb \mathcal H^{[2]}_{\mathrm{in},\mp} \oplus \mathcal H^{[2]}_{\mathrm{out},\pm}\rb\oplus\ldots \oplus
 \lb \mathcal H^{[n-3]}_{\mathrm{in},\mp} \oplus \mathcal H^{[n-3]}_{\mathrm{out},\pm}\rb \oplus \mathcal H^{[n-2]}_{\mathrm{in},\mp}.
 \end{gathered}
 \eeq
 Combine the 3-point projections $\mathcal P^{[k]}$ into an operator $\mathcal P_{\oplus}:
 \mathcal H\to\mathcal H$ given by the direct sum
 \ben
 \mathcal P_{\oplus}=\mathcal P^{[1]}\oplus\ldots\oplus
 \mathcal P^{[n-2]}.
 \ebn
 Clearly, we have
 \begin{lemma}
  $\mathcal P_{\oplus}^2=\mathcal P_{\oplus}$ and $ \operatorname{ker} \mathcal P_{\oplus}=\mathcal H_- $.
 \end{lemma}
 
 Another important operator $\mathcal P_{\Sigma}:\mathcal H\to \mathcal H$ is defined using the solution $\hat \Psi\lb z\rb $ (defined by (\ref{RHPhat})) of the $n$-point RHP  in a way similar to construction of the projection (\ref{onepointpr}): 
 \beq\label{projsigma}
  \mathcal P_{\Sigma}f\lb z\rb=\frac{1}{2\pi i}
   \oint_{\mathcal C_{\Sigma}}\frac{\hat\Psi_+\lb z\rb
   {\hat\Psi_+\lb z'\rb}^{-1} f\lb z'\rb dz' }{z-z'},
   \qquad 
 \mathcal C_{\Sigma} := \bigcup\limits_{k=1}^{n-3} \mathcal C^{[k]}_{\mathrm{out}}\cup\mathcal C_{\mathrm{in}}^{[k+1]}.
 \eeq
 We use the same prescription for the contours: whenever it is necessary to interpret the singular factor $1/\lb z-z'\rb$, the contour of integration is slightly deformed into the appropriate annulus.
 
 Let $\mathcal H_{\mathcal A}$ be the space of boundary values on 
 $\mathcal C_{\Sigma}$ of functions holomorphic on 
 $\mathcal A=\bigcup_{k=1}^{n-3}\mathcal A_k$.
 \begin{lemma} 
 \label{lemmaprojsigma}
 $\mathcal P_{\Sigma}^2=\mathcal P_{\Sigma}$ and
 $\ds \mathcal H_{\mathcal A}\subseteq \operatorname{ker} \mathcal P_{\Sigma}$.
 \end{lemma}
 \pf Given $f\in \mathcal H_{\mathcal A}$, the integration contours $\mathcal C_{\mathrm{out}}^{[k]}$ and  $\mathcal C_{\mathrm{in}}^{[k+1]}$ in (\ref{projsigma}) can be merged thanks to the absence of singularities inside $\mathcal A_k$, which proves the second statement. To show the projection property, it suffices to notice that
 \ben
 \mathcal P_{\Sigma}^2f^{[k]}\lb z\rb=
 \frac{1}{\lb 2\pi i\rb^2}
    \oiint_{\mathcal C_{\Sigma}}
    \frac{\hat\Psi_+\lb z\rb {\hat\Psi_+\lb z''\rb}^{-1}
    f^{[k]}\lb z''\rb  dz'dz'' }{\lb z-z'\rb \lb z'-z''\rb}.
 \ebn
 Because of the ordering of contours prescribed above, the only obstacle to merging $\mathcal C_{\mathrm{out}}^{[k]}$
 and $\mathcal C_{\mathrm{in}}^{[k]}$ in the integral with respect to $z'$
 is the pole at $z'=z$. The result follows by residue computation.
 \epf
 
 \begin{lemma}\label{lemmarange}
 $\displaystyle \mathcal P_{\Sigma}\mathcal P_{\oplus}=\mathcal P_{\oplus}$ and $\displaystyle \mathcal P_{\oplus}\mathcal P_{\Sigma}\mathcal =\mathcal P_{\Sigma}$.
 \end{lemma}
 \pf Similar to the proof of Lemma~\ref{lemmaprojsigma}. Use that $\hat\Psi^{-1}\Psi^{[k]} $ has no jumps on $\Gamma^{[k]}$ to compute by residues the intermediate integrals in $ \mathcal P_{\Sigma}\mathcal P_{\oplus}$ and $\mathcal P_{\oplus} \mathcal P_{\Sigma}$. \epf
 
 The above suggests to introduce the notation
 \beq\label{ahtau}
 \mathcal H_{\mathcal T}:=\operatorname{im}\mathcal P_{\oplus}=
 \operatorname{im}\mathcal P_{\Sigma}.
 \eeq
 The space $\mathcal H_{\mathcal T}\subset \mathcal H$ can be thought of as the subspace of functions on the union of boundary circles $\mathcal C^{[k]}_{\mathrm{in}}$, $\mathcal C^{[k]}_{\mathrm{out}}$ that can be continued inside $\bigcup_{k=1}^{n-2}\mathcal T^{[k]}$ with monodromy and singular behavior of the $n$-point fundamental matrix solution $\Phi\lb z\rb$. The only exception is the regular singularity at $\infty$ where the growth is slower.  
 
 The structure of elements of $\mathcal H_{\mathcal T}$  is described by Lemma~\ref{lemalphagamma}.  Varying the positions of singular points, one obtains  a trajectory of  $\mathcal H_{\mathcal T}$ in the infinite-dimensional Grassmannian $\mathrm{Gr}\lb \mathcal H\rb$ defined with respect to the splitting $\mathcal H=\mathcal H_+\oplus\mathcal H_-$. Note that each of the subspaces $\mathcal H_{\pm}$ may be identified with $N\lb n-3\rb$ copies of the space $L^2\lb S^1\rb$ of functions on a circle; the factor $n-3$ corresponds to the number of annuli and $N$ is the rank of the appropriate RHP. 
 
 We can also write
 \beq\label{secondsplitting}
 \mathcal H=\mathcal H_{\mathcal T}\oplus \mathcal H_-.
 \eeq
 The operator  $\mathcal P_{\oplus}$ introduced above gives the projection on $\mathcal H_{\mathcal T}$ along $\mathcal H_{-}$.   
    Similarly, the operator $\mathcal P_{\Sigma}$ is  a projection on $\mathcal H_{\mathcal T}$ along $\operatorname{ker} \mathcal P_\Sigma\supseteq\mathcal H_{\mathcal A}$. We would like to express it in terms of $3$-point projectors. To this end let us regard $f^{[k]}_{\mathrm{in},-}$, $f^{[k]}_{\mathrm{out},+}$ as coordinates on $\mathcal H_{\mathcal T}$. Suppose that $f\in\mathcal H$
 can be decomposed as $f=g+h$ with $g\in \mathcal H_{\mathcal T}$ and $h\in \mathcal H_{\mathcal A}$. The latter condition means that
 \ben
 h^{[k]}_{\mathrm{out},\pm}=h^{[k+1]}_{\mathrm{in},\pm},\qquad k=1,\ldots,n-3,
 \ebn
 which can be equivalently written as a system of equations for components of $g$:
 \beq\label{projdet}
 \begin{gathered}
 \begin{aligned}
  g^{[k]}_{\mathrm{in},-}
   -\mathsf c^{[k-1]}g^{[k-1]}_{\mathrm{in},-}
  -\mathsf d^{[k-1]}g^{[k-1]}_{\mathrm{out},+}
  =&\,
   f^{[k]}_{\mathrm{in},-}-f^{[k-1]}_{\mathrm{out},-},\\
 g^{[k]}_{\mathrm{out},+}
 -\mathsf a^{[k+1]}g^{[k+1]}_{\mathrm{in},-}
 -\mathsf b^{[k+1]}g^{[k+1]}_{\mathrm{out},+}
 =&\,
 f^{[k]}_{\mathrm{out},+}-f^{[k+1]}_{\mathrm{in},+},
 \end{aligned}
 \end{gathered}
 \eeq
 where $g^{[1]}_{\mathrm{in},-}=0$, $g^{[n-2]}_{\mathrm{out},+}=0$.
 The first and second equations are valid in sufficiently narrow annuli containing
 $\mathcal C_{\mathrm{in}}^{[k]}$ and $\mathcal C_{\mathrm{out}}^{[k]}$, respectively. Define 
 \beq\label{UVW}
 \begin{gathered}
 \tilde g_k=\lb\begin{array}{c}
 g^{[k]}_{\mathrm{out},+} \\ g^{[k+1]}_{\mathrm{in},-}
 \end{array}\rb,\qquad  \tilde f_k=\lb\begin{array}{c}
  f^{[k]}_{\mathrm{out},+}-f^{[k+1]}_{\mathrm{in},+} \\  f^{[k+1]}_{\mathrm{in},-}-f^{[k]}_{\mathrm{out},-}
  \end{array}\rb,\qquad 
 U_k=\lb\begin{array}{cc}
 0 & \mathsf a^{[k+1]} \\ \mathsf d^{[k]} & 0
 \end{array}\rb, \qquad k=1,\ldots, n-3,\\
 V_k=\lb\begin{array}{cc}
  \mathsf b^{[k+1]} & 0 \\ 0 & 0
  \end{array}\rb,\qquad 
  W_k=\lb\begin{array}{cc}
   0 & 0 \\ 0 & \mathsf c^{[k+1]}
    \end{array}\rb,\qquad k=1,\ldots,n-4, \\
 K=\lb\begin{array}{cccccc}
 U_1 & V_1 & 0 &  . & 0 \\
 W_1 & U_2 & V_2 & . & 0 \\
 0 & W_2 & U_3 & . & . \\
 . & . & . & . & V_{n-4} \\
 0 & 0 & . & W_{n-4} & U_{n-3}
 \end{array}\rb, \qquad
 \vec g=\lb\begin{array}{c}
 \tilde g_1 \\ \tilde g_2 \\ \vdots \\ \tilde g_{n-3}
 \end{array}\rb, \qquad
  \vec f=\lb\begin{array}{c}
  \tilde f_1 \\ \tilde f_2 \\ \vdots \\ \tilde f_{n-3}
  \end{array}\rb.      
  \end{gathered}
 \eeq
 The system (\ref{projdet}) can then be rewritten in a block-tridiagonal form
 \beq\label{projdet2}
 \lb \mathbb 1 -K\rb \vec g =\vec f.
 \eeq
   The decomposition $\mathcal H=\mathcal H_{\mathcal T}\oplus
  \mathcal H_{\mathcal A}$ thus uniquely exists provided that $\mathbb1 - K$ is invertible. 
  
  Let us prove a converse result and interpret $K$ in a more invariant way. Consider the operators $\mathcal P_{\oplus,+}:\mathcal H_+\to\mathcal H_{\mathcal T}$ and ${\mathcal P_{\Sigma,+}:\mathcal H_+\to\mathcal H_{\mathcal T}}$ defined as restrictions of $\mathcal P_{\oplus}$ and $\mathcal P_{\Sigma}$ to $\mathcal H_+$. The first of them is invertible, with the inverse given by the projection on $\mathcal H_+$ along $\mathcal H_-$. Hence one can consider the composition $L\in\operatorname{End}\lb\mathcal H_+\rb$ defined by
 \beq\label{doperator}
 L:={\mathcal P_{\oplus,+}}^{-1}\mathcal P_{\Sigma,+}.
 \eeq
 We are now going to make an important assumption which is expected to hold generically (more precisely, outside the Malgrange divisor).  It will soon become clear that it is satisfied at least in a sufficiently small finite polydisk $\mathbb D\subset \Cb^{n-3}$ in the variables $a_1,\ldots, a_{n-3}$, centered at the origin. 
 \begin{ass}
  $\mathcal P_{\Sigma,+}$ is invertible.
 \end{ass}
 \begin{prop} For $g\in \mathcal H_+$, let $\tilde g_k$ and $\vec g$ be defined by (\ref{UVW}). In these coordinates, $L^{-1}=1-K$.
 \end{prop}
 \pf Rewrite the equation $L^{-1}f'=f$ as $\mathcal P_{\oplus,+} f'=
 \mathcal P_{\Sigma,+}f$. Setting $f=\mathcal P_{\oplus,+}f'+h$, the latter equation becomes equivalent to $\mathcal P_{\Sigma}h=0$. The solution thus reduces to constructing $h\in\mathcal H_{\mathcal A}$ such that $\lb h+\mathcal P_{\oplus,+}f'\rb_-=0$, where the projection is taken with respect to the splitting $\mathcal H=\mathcal H_+\oplus\mathcal H_-$. This can be achieved by setting  
 \begin{align*}
 h^{[k]}_{\mathrm{out},+}=&\,h^{[k+1]}_{\mathrm{in},+}=-
 \lb \mathcal P_{\oplus,+}f'\rb^{[k+1]}_{\mathrm{in},+},\\
  h^{[k]}_{\mathrm{out},-}=&\,h^{[k+1]}_{\mathrm{in},-}=-
  \lb \mathcal P_{\oplus,+}f'\rb^{[k]}_{\mathrm{out},-}.
 \end{align*}
 It then follows that $f=f'+h_+=\lb\mathbb 1-K\rb f'$.
 \epf
 
 Let us emphasize that the operator $L$ involves the (a priori unknown) solution $\hat{\Psi}\lb z\rb$ of the original RHP, equivalent to the solution of the $n$-point Fuchsian system, whereas $K$ is expressed solely in terms of $3$-point parametrices. It is this fact which ultimately allows us to obtain in the next subsections a Fredholm determinant representation of the tau function involving only these elementary building blocks. 

 \subsection{Tau function\label{subsectauf}}
 \begin{defin} Let $L\in\operatorname{End}\lb \mathcal H_+\rb$ be the operator defined by (\ref{doperator}). We define the tau function associated to the Riemann-Hilbert problem for $\Psi$ as
 \beq\label{taudefdet}
 \tau\lb a\rb:=\operatorname{det}\lb L^{-1}\rb .
 \eeq
 \end{defin}
 
 In order to demonstrate the relation of (\ref{taudefdet}) to conventional definition \cite{JMU} of the isomonodromic tau function and its extension \cite{ILP},
 let us compute the logarithmic derivatives of $\tau$ with respect to isomonodromic times $a_1,\ldots, a_{n-3}$. At this point it is convenient to introduce the notation
 \beq\label{notationfordeltas}
 \Delta_k=\frac12\operatorname{Tr}\Theta_k^2,\qquad 
 \bar\Delta_k=\frac12\operatorname{Tr}\mathfrak S_k^2.
 \eeq
 Recall that $\bar\Delta_0\equiv\Delta_0$ and $\bar\Delta_{n-2}\equiv\Delta_{n-1}$.

 \begin{theo}\label{TauF} We have
 \beq\label{tautau}
 \tau\lb a\rb= {\Upsilon\lb a\rb}^{-1} 
 \tau_{\mathrm{JMU}}\lb a\rb,
 \eeq
 where $\tau_{\mathrm{JMU}}\lb a\rb$ is defined up to a constant independent of $a$ by
 \begin{align}\label{taujmudef}
 d_a\ln \tau_{\mathrm{JMU}}=\sum_{0\leq k<l\leq n-2}
   \operatorname{Tr}A_k A_l\;d\ln\lb a_k-a_l\rb,
 \end{align}
 and the prefactor $\Upsilon\lb a\rb$ is given by
 \beq\label{upsdef} \Upsilon\lb a\rb = \prod_{k=1}^{n-3}a_k^{\bar\Delta_{k}-
   \bar\Delta_{k-1}-\Delta_k}.
   \eeq
 \end{theo}
 \pf We will proceed in several steps.\vspace{0.1cm}\\
 \noindent\underline{\it Step 1}. 
 Choose a collection of points $a^0$ close to $a$ in the sense that the same annuli can be used to define the tau function 
 $\tau\lb a^0\rb$. The collection $a^0 $ will be considered fixed whereas $a$ varies. Let us compute the logarithmic derivatives of the ratio
 $\tau\lb a\rb/\tau\lb a^0\rb$. First of all we can write
 \beq\label{tauratio}
 \frac{\tau\lb a\rb}{\tau\lb a^0\rb}=\operatorname{det}\lb
 {\mathcal P_{\oplus,+}\lb a^0\rb}^{-1}\mathcal P_{\Sigma,+}\lb a^0\rb
 {\mathcal P_{\Sigma,+}\lb a\rb}^{-1}\mathcal P_{\oplus,+}\lb a\rb\rb
 \eeq
 Note that since $\mathcal P_{\Sigma,+}\lb a\rb:\mathcal H_+\to \mathcal H_{\mathcal T}\lb a\rb$ can be viewed as a projection of elements of $\mathcal H$ along $\mathcal H_{\mathcal A}$, the composition 
 \ben
 \mathcal P_{a^0\to a}:=\mathcal P_{\Sigma,+}\lb a\rb
  {\mathcal P_{\Sigma,+}\lb a^0\rb}^{-1}:
  \mathcal H_{\mathcal T}\lb a^0\rb\to  \mathcal H_{\mathcal T}\lb a\rb
  \ebn
  is also a projection along $\mathcal H_{\mathcal A}$. It therefore coincides with the restriction $\mathcal P_{\Sigma}\bigl|_{\mathcal H_{\mathcal T}\lb a^0\rb}$.
One similarly shows that
 \ben
  \mathcal F_{a^0\to a}:=\mathcal P_{\oplus,+}\lb a\rb
   {\mathcal P_{\oplus,+}\lb a^0\rb}^{-1}=
 \mathcal P_{\oplus}\bigl|_{\mathcal H_{\mathcal T}\lb a^0\rb}.  
 \ebn  
 The exterior logarithmic derivative of  (\ref{tauratio}) can now be written as
 \begin{align}
 \nonumber d_a\ln\frac{\tau\lb a\rb}{\tau\lb a^0\rb}=&\,
 -\operatorname{Tr}_{\mathcal H_{\mathcal T}\lb a^0\rb}
 \biggl\{d_a\left(\mathcal F_{a\to a^0}\mathcal P_{a^0\to a}\right)\cdot\mathcal P_{a\to a^0}\mathcal F_{a^0\to a}\biggr\}=\\
 \label{auxdertau0} 
 =&\,-\operatorname{Tr}_{\mathcal H_{\mathcal T}\lb a^0\rb}
  \biggl\{\mathcal F_{a\to a^0}\cdot d_a
  \mathcal P_{a^0\to a}
  \cdot\mathcal P_{a\to a^0}\mathcal F_{a^0\to a}\biggr\}=\\
 \nonumber =&\,-\operatorname{Tr}_{\mathcal H}
    \biggl\{\mathcal P_{\oplus}\lb a^0\rb\cdot d_a
    \mathcal P_{\Sigma}\lb a\rb
    \cdot \mathcal P_{\Sigma}\lb a^0\rb\mathcal P_{\oplus}\lb a\rb\biggr\}.
 \end{align}
 The possibility to extend operator domains as to have the second equality is a consequence of (\ref{ahtau}). Furthermore, using once again the projection properties, one shows that
 \ben
 \mathcal P_{\Sigma}\lb a\rb\Bigl(\mathbb1-
 \mathcal P_{\Sigma}\lb a^0\rb\Bigr)=0,\qquad
 \mathcal P_{\oplus}\lb a\rb\Bigl(\mathbb1-
  \mathcal P_{\oplus}\lb a^0\rb\Bigr)=0.
 \ebn
 which reduces the equation (\ref{auxdertau0}) to
 \beq\label{auxdertau1}
 d_a\ln\tau\lb a\rb=-\operatorname{Tr}_{\mathcal H}\biggl\{
 \mathcal P_{\oplus} d_a\mathcal P_{\Sigma}\biggr\}=
 -\sum_{k=1}^{n-2} 
 \operatorname{Tr}_{\mathcal H^{[k]}}\biggl\{
  \mathcal P_{\oplus}^{[k]} d_a\mathcal P_{\Sigma}\biggr\}.
 \eeq
 
 \noindent\underline{\it Step 2}.
 Let us now proceed to calculation of the right side of (\ref{auxdertau1}). Computations of the same type have already been used in the proofs of 
 Lemmata~\ref{lemmaprojsigma} and \ref{lemmarange}. The idea is that 
 $\Psi^{[k]}$ and $\hat\Psi$ have the same jumps on the
 contour $\Gamma^{[k]}$ which reduces the integrals in (\ref{onepointpr}), (\ref{projsigma}) to residue computation. In particular, for $f^{[k]}\in \mathcal H^{[k]}$ with $k=2,\ldots,n-3$ we have
 \beq\label{auxdertau2}
 \mathcal P_{\oplus}^{[k]} d_a\mathcal P_{\Sigma} f^{[k]}\lb z\rb=
 \frac{1}{\lb 2\pi i\rb^2}\oiint_{\mathcal C_{\mathrm{in}}^{[k]}\cup
  \mathcal C_{\mathrm{out}}^{[k]}}\frac{ 
   \Psi_+^{[k]}\lb z\rb  {\Psi_+^{[k]}\lb z'\rb}^{-1}d_a\lb 
  \hat\Psi_+\lb z'\rb  {\hat\Psi_+\lb z''\rb}^{-1} \rb  
  f^{[k]}\lb z''\rb 
   dz'dz''}{\lb z-z'\rb\lb z'-z''\rb}.
 \eeq
  The integrals are computed with the prescription that $z$ is located inside the contour of $z'$, itself located inside the contour of $z''$, and then passing to boundary values. But since the function
  $\lb z'-z''\rb^{-1}
  d_a\lb \hat\Psi_+\lb z'\rb  {\hat\Psi_+\lb z''\rb}^{-1} \rb $       
   has no singularity at $z''=z'$, the contours of $z'$ and $z''$ can be moved through each other. This identifies the trace of the integral operator on the right of (\ref{auxdertau2}) with
   \begin{align*}
   \operatorname{Tr}\lb
   \mathcal P_{\oplus}^{[k]} d_a\mathcal P_{\Sigma}\rb=&\,
   -\frac{1}{\lb 2\pi i\rb^2}\oiint_{\mathcal C_{\mathrm{in}}^{[k]}\cup
     \mathcal C_{\mathrm{out}}^{[k]}}\frac{
     \operatorname{Tr}\left\{ 
     { \Psi_+^{[k]}\lb z\rb\Psi_+^{[k]}\lb z'\rb}^{-1} 
      d_a\lb \hat\Psi_+\lb z'\rb 
      {\hat\Psi_+\lb z\rb}^{-1} 
            \rb\right\}  dz\,dz'}{\lb z-z'\rb^2}=\\
      =&\,-\frac{1}{\lb 2\pi i\rb^2}\oiint_{\mathcal C_{\mathrm{in}}^{[k]}\cup
           \mathcal C_{\mathrm{out}}^{[k]}}\frac{
           \operatorname{Tr}\left\{{\Psi_+^{[k]}\lb z'\rb}^{-1} d_a 
           \hat\Psi_+\lb z'\rb \cdot 
           {\hat\Psi_+\lb z\rb}^{-1} \Psi_+^{[k]}\lb z\rb\right\}  dz\,dz'}{\lb z-z'\rb^2}\\
        &\, -\frac{1}{\lb 2\pi i\rb^2}\oiint_{\mathcal C_{\mathrm{in}}^{[k]}\cup
             \mathcal C_{\mathrm{out}}^{[k]}}\frac{
             \operatorname{Tr}\left\{ d_a\lb {\hat\Psi_+\lb z\rb}^{-1}\rb\cdot 
            \Psi_+^{[k]}\lb z\rb {\Psi_+^{[k]}\lb z'\rb}^{-1}\hat\Psi_+\lb z'\rb       
              \right\}  dz\,dz'}{\lb z-z'\rb^2}    ,
   \end{align*}
  where $z$ is considered to be inside the contour of $z'$. The first term vanishes since the contours $\mathcal C_{\mathrm{in}}^{[k]}$ and $\mathcal C_{\mathrm{out}}^{[k]}$ in the integral with respect to $z$ can be merged. In the second term the integral with respect to $z'$ is determined by the residue at $z'=z$, which yields
  \ben
  \operatorname{Tr}\lb
     \mathcal P_{\oplus}^{[k]} d_a\mathcal P_{\Sigma}\rb=
     \frac{1}{2\pi i}\oint_{\mathcal C_{\mathrm{in}}^{[k]}\cup
                  \mathcal C_{\mathrm{out}}^{[k]}} 
 \operatorname{Tr}\left\{ d_a\lb {\hat\Psi_+\lb z\rb}^{-1}\rb\cdot 
              \Psi_+^{[k]}\lb z\rb\cdot\partial_z\lb
              {\Psi_+^{[k]}\lb z\rb}^{-1}\hat\Psi_+\lb z\rb \rb
               \right\}  dz.                 
  \ebn
  Recall that $\hat \Psi_+$, $\Psi^{[k]}_+$ are related to fundamental matrix solutions $ \Phi$,  $\Phi^{[k]}$ of $n$-point and $3$-point Fuchsian systems by
  \ben
  \begin{gathered}
  \hat \Psi_+\lb z\rb\Bigl|_{\mathcal C_{\mathrm{in}}^{[k]}}=
  S_{k-1}^{-1} \lb -z\rb^{-\mathfrak S_{k-1}}\Phi\lb z\rb,\qquad
   \hat \Psi_+\lb z\rb\Bigl|_{\mathcal C_{\mathrm{out}}^{[k]}}=
   S_{k}^{-1} \lb -z\rb^{-\mathfrak S_{k}} \Phi\lb z\rb,\\
  \Psi_+^{[k]}\lb z\rb\Bigl|_{\mathcal C_{\mathrm{in}}^{[k]}}=
    S_{k-1}^{-1} \lb -z\rb^{-\mathfrak S_{k-1}} \Phi^{[k]}\lb z\rb ,\qquad
   \Psi_+^{[k]}\lb z\rb\Bigl|_{\mathcal C_{\mathrm{out}}^{[k]}}=
    S_{k}^{-1}\lb -z\rb^{-\mathfrak S_{k}} \Phi^{[k]}\lb z\rb   .
    \end{gathered}
  \ebn
  This leads to
  \begin{align}
  \nonumber\operatorname{Tr}\lb
       \mathcal P_{\oplus}^{[k]} d_a\mathcal P_{\Sigma}\rb=&\,
 \frac{1}{2\pi i}\oint_{\mathcal C_{\mathrm{in}}^{[k]}\cup
                   \mathcal C_{\mathrm{out}}^{[k]}} 
  \operatorname{Tr}\left\{ d_a\lb {\Phi}^{-1}\rb\cdot 
                \Phi^{[k]}\cdot \partial_z\lb
                {\Phi^{[k]}}^{-1}\Phi 
                \rb\right\}  dz  =\\
 \label{auxdertau3}
   =&\,\operatorname{res}_{z=a_{k}} \operatorname{Tr}\left\{
   d_a\Phi\cdot \Phi^{-1}\lb \partial_z\Phi\cdot \Phi^{-1}- \partial_z \Phi^{[k]}\cdot {\Phi^{[k]}}^{-1}\rb\right\} . 
  \end{align}
  The contributions of the subspaces $\mathcal H^{[1]}$ and $\mathcal H^{[n-2]}$ to the trace (\ref{auxdertau1}) can be computed in a similar fashion. The only difference is that instead of merging $\mathcal C_{\mathrm{in}}^{[k]}$ with
  $\mathcal C_{\mathrm{out}}^{[k]}$ one should now shrink the contour $\mathcal C_{\mathrm{out}}^{[1]}$ to $0$ and $\mathcal C_{\mathrm{in}}^{[n-2]}$ to~$\infty$. The result is given by the same formula (\ref{auxdertau3}).
  \vspace{0.1cm}\\
  \noindent\underline{\it Step 3}. To complete the proof, it now remains to compute the residues in (\ref{auxdertau3}). Note that near the regular singularity $z=a_{k}$
  the fundamental matrices $\Phi$, $\Phi^{[k]}$ are characterized by the behavior
  \begin{subequations}\label{localexps}
  \begin{align}
  \label{localexp01}
  \Phi\lb z\to a_{k}\rb=&\,C_{k}\lb a_{k}-z\rb^{\Theta_{k}}\lb \mathbb 1+\sum_{l=1}^{\infty}
  g_{k,l}\lb z-a_{k}\rb^l\rb G_{k},\\
  \label{localexp02}
    \Phi^{[k]}\lb z\to a_{k}\rb=&\,C_{k}\lb a_{k}-z\rb^{\Theta_{k}}\lb \mathbb 1+\sum_{l=1}^{\infty}
    g^{[k]}_{1,l}\lb z-a_{k}\rb^l\rb G^{[k]}_{1}.
  \end{align}
  \end{subequations}
  The coinciding leftmost factors ensure the same local monodromy
  properties. The rightmost coefficients appear in the $n$-point and $3$-point RHPs as $G_k=\Psi\lb a_k\rb$, $G_1^{[k]}=\Psi^{[k]}\lb a_k\rb$. It becomes straightforward to verify that as $z\to
  a_{k}$, one has
  \begin{align*}
  \partial_z\Phi\cdot \Phi^{-1}-\partial_z \Phi^{[k]}\cdot
  {\Phi^{[k]}}^{-1} 
  =&\, C_{k} \lb a_{k}-z\rb^{\Theta_{k}}
  \Bigl[g_{k,1}-g^{[k]}_{1,1}+O\lb z-a_{k}\rb\Bigr] \lb a_{k}-z\rb^{-\Theta_{k}}C_{k}^{-1},\\
 d_a\Phi\cdot  \Phi^{-1}=&\,C_{k} \lb a_{k}-z\rb^{\Theta_{k}}
    \left[-\frac{\Theta_{k}da_{k}}{z-a_{k}}+O\lb 1\rb\right] \lb a_{k}-z\rb^{-\Theta_{k}}C_{k}^{-1}.
  \end{align*}
  In combination with (\ref{auxdertau1}), (\ref{auxdertau3}), this in turn implies that 
  \beq\label{auxdertau6}
  d_a\ln\tau\lb a\rb=\sum_{k=1}^{n-3}\operatorname{Tr}\Theta_k\lb
  g_{k,1}-g^{[k]}_{1,1}\rb da_k.
  \eeq
  
  Substituting local expansion (\ref{localexp01}) into the Fuchsian system (\ref{fuchsys}), we may recursively determine the coefficients $g_{k,l}$. In particular, the first coefficient $g_{k,1}$ satisfies
  \beq\label{auxdertau5}
  g_{k,1}+\left[\Theta_k,g_{k,1}\right]=G_k^{-1}\lb
  \sum^{n-2}_{l=0,l\ne k}\frac{A_l}{a_k-a_l}
  \rb G_k,
  \eeq
  so that
  \beq\label{auxdertau7}
  \sum_{k=1}^{n-3}\operatorname{Tr}\lb\Theta_k g_{k,1}\rb da_k
  =\sum_{k=1}^{n-3}\sum^{n-2}_{l=0,l\ne k}\frac{\operatorname{Tr}A_k A_l}{a_k-a_l}da_k=d_a\ln \tau_{\mathrm{JMU}}.
  \eeq
  The $3$-point analog of the relation (\ref{auxdertau5})
  is
    \ben
    g^{[k]}_{1,1}+\left[\Theta_{k},g^{[k]}_{1,1}\right]=
    {G_1^{[k]}}
    \frac{A_0^{[k]}}{a_{k}}
    {G_1^{[k]}}^{-1},
    \ebn
  which gives 
  \beq\label{auxdertau8}
  \operatorname{Tr}\lb \Theta_{k} g^{[k]}_{1,1}\rb =\frac{\operatorname{Tr} A_0^{[k]}A_1^{[k]}}{a_{k}}=
  \frac{\operatorname{Tr}\lb {A_{\infty}^{[k]}}^2-{A_0^{[k]}}^2-{A_1^{[k]}}^2\rb}{2a_{k}}=
  \frac{\bar\Delta_{k}-\bar\Delta_{k-1}-\Delta_{k}}{a_{k}}.
  \eeq
 Combining (\ref{auxdertau6}) with (\ref{auxdertau7})
 and (\ref{auxdertau8}) finally yields the statement of the theorem.
 \epf
 
 \begin{cor}
 Jimbo-Miwa-Ueno isomonodromic tau function $\tau_{\mathrm{JMU}}\lb a\rb$ admits a block Fredholm determinant representation
 \beq\label{fredholmrep}
 \tau_{\mathrm{JMU}}\lb a\rb=\Upsilon\lb a\rb\cdot \operatorname{det}\lb
 \mathbb 1-K\rb,
 \eeq
 where the operator $K$ is defined by (\ref{UVW}). Its $N\times N$ subblocks (\ref{alphadelta}) are expressed in terms of solutions $\Psi^{[k]}$
 of RHPs associated to $3$-point Fuchsian systems with prescribed monodromy.
 \end{cor}
 
 There arises a natural question: is there a relative of the Fredholm determinant representation (\ref{fredholmrep}) for the RHP solution $\hat{\Psi}\lb z\rb$ itself? In the theory of integral operators with integrable kernels there is a familiar procedure of converting the computation of the resolvent into a Riemann-Hilbert problem. It turns out that our situation is similar, namely we have the following
 \begin{theo}\label{RHPso}
 Inside the annuli $\mathcal A$, the RHP solution $\hat{\Psi}\lb z\rb$ is expressed in terms of the $3$-point solutions by
 \beq\label{fundf}
 \hat{\Psi}\lb z\rb=\mathcal P_{\oplus,+}L\;\mathbb 1_{\mathrm{out}}^{[n-3]}\hat\Psi^{[n-2]}_+\lb z\rb
 +\mathbb 1_{\mathrm{in}}^{[n-2]}\hat\Psi^{[n-2]}_+\lb z\rb,
 \eeq
 where $\mathbb 1_{\mathrm{out}}^{[n-3]}$ and $\mathbb 1_{\mathrm{in}}^{[n-2]}$ are indicator functions of the boundary circles $\mathcal C_{\mathrm{out}}^{[n-3]}$ and $\mathcal C_{\mathrm{in}}^{[n-2]}$, and we assume the normalizations
 $\hat{\Psi}\lb \infty\rb=\hat\Psi^{[n-2]}\lb \infty\rb=\mathbb1$.
 \end{theo}
 \pf First of all notice that the right side of (\ref{fundf}) indeed makes sense. Namely, $\mathbb 1_{\mathrm{out}}^{[n-3]}\hat\Psi^{[n-2]}_+\lb z\rb$ belongs to $
 \mathcal H^{[n-3]}_{\mathrm{out},+}\subset \mathcal H_+$ and therefore
 can be acted upon by $L$ defined by (\ref{doperator}). From this definition it trivially follows that $\mathcal P_{\oplus,+}L=\mathcal P_{\Sigma,+}$, so that
 \ben
 \mathcal P_{\oplus,+}L\;\mathbb 1_{\mathrm{out}}^{[n-3]}\hat\Psi^{[n-2]}_+\lb z\rb=
 \frac1{2\pi i}\oint_{\mathcal C^{[n-3]}_{\mathrm{out}}}
 \frac{\hat\Psi_+\lb z\rb
    {\hat\Psi_+\lb z'\rb}^{-1} \hat\Psi^{[n-2]}_+\lb z'\rb dz' }{z-z'}.
 \ebn
 The product ${\hat\Psi_+}^{-1} \hat\Psi^{[n-2]}_+$ has no jumps on
 $\hat\Gamma^{[n-2]}$. The contour $\mathcal C^{[n-3]}_{\mathrm{out}}$ can therefore be collapsed at infinity so that the integral is given by the sum of residues at $z'=\infty$ and $z'=z$. The former contribution is just $\hat\Psi_+\lb z\rb
     {\hat\Psi_+\lb \infty\rb}^{-1} \hat\Psi^{[n-2]}_+\lb \infty\rb$ while the latter appears only if $z$ belongs to the boundary circle $\mathcal C^{[n-2]}_{\mathrm{in}}$ of the rightmost trinion and is given by 
 $-\mathbb 1_{\mathrm{in}}^{[n-2]}\hat\Psi^{[n-2]}_+\lb z\rb$.
 \epf

 \subsection{Example: $4$-point tau function}\label{subsec4pts}
 In order to illustrate the developments of the previous subsection, let us consider the simplest nontrivial case of Fuchsian systems with $n=4$ regular singular points. Three of them have already  been fixed at $a_0=0$, $a_2=1$, $a_3=\infty$. There remains a single time variable 
  $a_1\equiv t$. To be able to apply previous results, it is assumed that $0<t<1$. 
  
  The monodromy data are given by $4$ diagonal matrices $\Theta_{0,t,1,\infty}$ of local monodromy exponents and connection matrices $C_0$, $C_{t,\pm}$, $C_{1,\pm}$, $C_{\infty}$ satisfying the relations 
  \ben
  M_0\equiv C_0 e^{2\pi i \Theta_0}C_0^{-1}=C_{t,-}C_{t,+}^{-1},\qquad
  e^{2\pi i\mathfrak S}= C_{t,-} e^{2\pi i \Theta_t}C_{t,+}^{-1}=
  C_{1,-}C_{1,+}^{-1}
  \ebn  
  Observe that, in the hope to make the notation more intuitive, it has been slightly changed as compared to the general case. The indices $0,1,2,3$ are replaced by $0,t,1,\infty$. Also, for $n=4$ there is only one nontrivial matrix $M_{0\to k}$ (namely, with $k=1$). Therefore it becomes convenient to work from the very beginning in a distinguished basis where $M_{0\to1}$ is given by a diagonal matrix $e^{2\pi i \mathfrak S}$ with
  $\operatorname{Tr}\mathfrak S=\operatorname{Tr}\lb\Theta_0+\Theta_t\rb=-\operatorname{Tr}\lb \Theta_1+\Theta_{\infty}\rb$. In terms of the previous notation, this corresponds to setting $\mathfrak S_1= \mathfrak S$ and $S_1=\mathbb1$. The eigenvalues of $\mathfrak S$ will be denoted by 
  $\sigma_1,\ldots,\sigma_N$. Recall (cf Assumption~\ref{asssp}) that $\mathfrak S$ is chosen so that these eigenvalues satisfy
  \beq\label{ascondsp}
  \left|\Re\lb \sigma_{\alpha}-\sigma_{\beta}\rb\right|\leq 1,\qquad
  \sigma_{\alpha}-\sigma_{\beta}\neq \pm1.
  \eeq

     \begin{figure}
       \centering
   \begin{tikzpicture} 
   \draw(0,0) node{         \includegraphics[height=10cm]{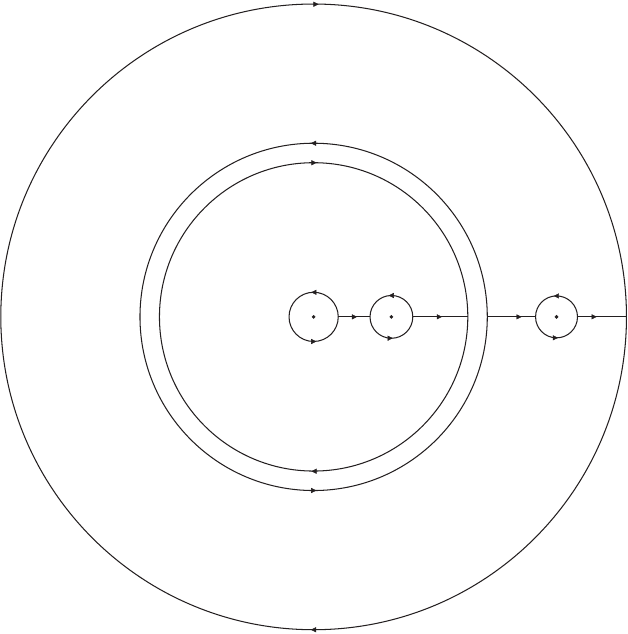}}; 
   \draw(-0.1,-0.15) node {\footnotesize $0$}; 
   \draw(1.15,-0.15) node {\footnotesize $t$};
   \draw(3.75,-0.15) node {\footnotesize $1$};
   \draw(0.7,0.2) node {\scriptsize $M_0^{-1}$}; 
   \draw(2.0,0.25) node {\scriptsize $e^{-2\pi i\mathfrak S}$};  
   \draw(3.2,0.25) node {\scriptsize $e^{-2\pi i\mathfrak S}$};
   \draw(4.5,0.2) node {\scriptsize $M_{\infty}$};  
   \draw(0.2,3.05) node {\scriptsize $\lb -z\rb^{-\mathfrak S}$};   
   \draw(0.2,2.15) node {\scriptsize $\lb -z\rb^{-\mathfrak S}$};        
   \draw(-4.2,0) node {\scriptsize $\lb -z\rb^{\Theta_{\infty}}C_{\infty}^{-1}$};
   \draw(-1.05,0) node {\scriptsize $\lb -z\rb^{-\Theta_{0}}C_{0}^{-1}$};   
   \draw(1.5,0.55) node {\scriptsize $\lb t-z\rb^{-\Theta_{t}}C_{t,+}^{-1}$};  
   \draw(1.5,-0.55) node {\scriptsize $\lb t-z\rb^{-\Theta_{t}}C_{t,-}^{-1}$}; 
   \draw(4,0.55) node {\scriptsize $\lb 1-z\rb^{-\Theta_{1}}C_{1,+}^{-1}$}; 
   \draw(4,-0.55) node {\scriptsize $\lb 1-z\rb^{-\Theta_{1}}C_{1,-}^{-1}$};                            
   \end{tikzpicture}      
         \caption{\label{PD4points}
         Contour $\hat \Gamma$ and jump matrices $\hat J$ for the $4$-punctured sphere}
       \end{figure}
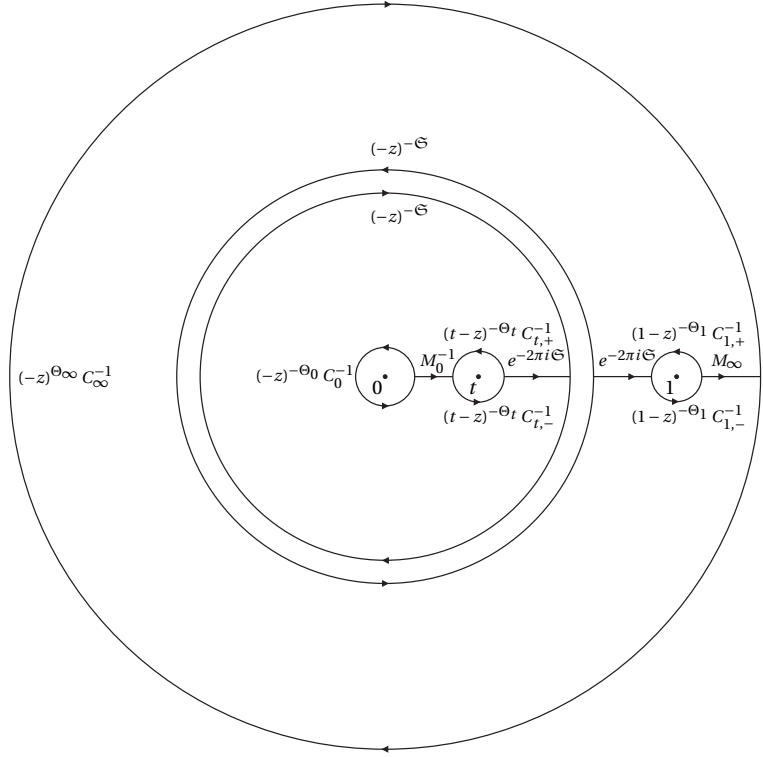

    The $4$-punctured sphere is decomposed into two pairs of pants $\mathcal T^{[L]}$, $\mathcal T^{[R]}$ by one annulus $\mathcal A$ as shown in Fig.~\ref{PD4points}. The space $\mathcal H$ is a sum 
   \beq\label{hplusdec}
   \mathcal H=\mathcal H_+\oplus\mathcal H_-,\qquad
   \mathcal H_{\pm}=\mathcal H_{\text{out},\pm}^{[L]}\oplus\mathcal H_{\text{in},\mp }^{[R]}. 
   \eeq
   Both subspaces $\mathcal H_{\pm}$ may thus be identified with the space
   $\mathcal H_{\mathcal C}:=\Cb^N\otimes L^2\lb \mathcal C\rb$ of vector-valued square integrable functions on a circle $\mathcal C$  centered at the origin and belonging to the annulus $\mathcal A$.   It will be very convenient for us to represent the elements of $\mathcal H_{\mathcal C}$ by their Laurent series inside $\mathcal A$,
      \beq\label{fourierbasis1}
     \qquad \qquad\qquad  f\lb z\rb=\sum_{p\in \Zb'}f^p z^{-\frac12+p},\qquad \qquad f^p\in\Cb^N.
      \eeq
       In particular, the first and second component of $\mathcal H_+$ 
       in (\ref{hplusdec}) consist of  functions with vanishing  negative and positive Fourier coefficients, respectively, i.e. they may be identified with 
   $\Pi_+\mathcal H_{\mathcal C}$ and $\Pi_-
   \mathcal H_{\mathcal C}$. At this point the use of half-integer indices $p\in\mathbb Z'$ for Fourier modes may seem redundant, but its convenience will quickly become clear.
  
   When $n=4$, the representation (\ref{fredholmrep}) reduces to
   \beq\label{det4pttau}
   \tau_{\mathrm{JMU}}\lb t\rb=t^{\frac12\operatorname{Tr}\lb\mathfrak S^2-\Theta_0^2-\Theta_t^2\rb}\operatorname{det}\lb\mathbb 1-U\rb,\qquad
   U=\lb\begin{array}{cc}0 & \mathsf a \\ \mathsf d & 0 \end{array}\rb\in \operatorname{End}\lb\mathcal H_{ \mathcal C}\rb,
   \eeq
   where the operators $\mathsf a\equiv\mathsf a^{[R]}\equiv
   \mathsf a^{[2]}: \Pi_-
      \mathcal H_{\mathcal C}\to \Pi_+
         \mathcal H_{\mathcal C}$ and $\mathsf d\equiv \mathsf d^{[L]}\equiv  \mathsf d^{[1]}:
         \Pi_+
               \mathcal H_{\mathcal C}\to \Pi_-
                  \mathcal H_{\mathcal C}$ are given by
                  \begin{subequations}
   \begin{alignat}{2}\label{intkernelsalphadelta1}
   \lb\mathsf a g\rb\lb z\rb=
   \,\frac1{2\pi i}\oint_{\mathcal C}\,
   \mathsf a\lb z,z'\rb g\lb z'\rb  dz'\, ,\qquad\quad &&\mathsf a\lb z,z'\rb=\frac{\Psi^{[R]}\lb z\rb{\Psi^{[R]}\lb z'\rb}^{-1}-\mathbb 1}{z-z'},\\
   \label{intkernelsalphadelta2}
  \lb\mathsf d g\rb\lb z\rb=
        \,\frac{1}{2\pi i}\oint_{\mathcal C}
         \mathsf d\lb z,z'\rb g\lb z'\rb  dz'\, ,\qquad \quad && \mathsf d\lb z,z'\rb=\frac{\mathbb 1-\Psi^{[L]}\lb z\rb {\Psi^{[L]}\lb z'\rb}^{-1}}{z-z'}.   
   \end{alignat}
   \end{subequations}
   The contour $\mathcal C$ is oriented counterclockwise, which is the origin of sign difference in the expression for $\mathsf d$ as compared to (\ref{alphadelta04}).  In the Fourier basis (\ref{fourierbasis1}), the operators $\mathsf a$ and $\mathsf d$ are given by semi-infinite matrices whose $N\times N$ blocks $\mathsf a^{\; \;\;\,p}_{-q}$, $\mathsf d^{-p}_{\;\,\;\; q}$ are deteremined by 
   \begin{align}\label{alphadeltaME}
   \mathsf a\lb z,z'\rb=\sum_{p,q\in\mathbb Z'_+}\mathsf a^{\; \;\;\,p}_{-q}\,z^{-\frac12+p} z'^{-\frac12+q},\qquad
   \mathsf d\lb z,z'\rb=\sum_{p,q\in\mathbb Z'_+}\mathsf d^{-p}_{\;\,\;\; q}\,z^{-\frac12-p}z'^{-\frac12-q}.
   \end{align}
   It should be emphasized that the indices of $\mathsf a^{\;\; \;p}_{\!-q}$ and $\mathsf d^{-p}_{\;\;\;\, q}$ belong to different ranges, since in both cases $p,q$ are positive half-integers.
   
   \begin{figure}
     \centering
 \begin{tikzpicture} 
 \draw(0,0) node{ \includegraphics[height=6.5cm]{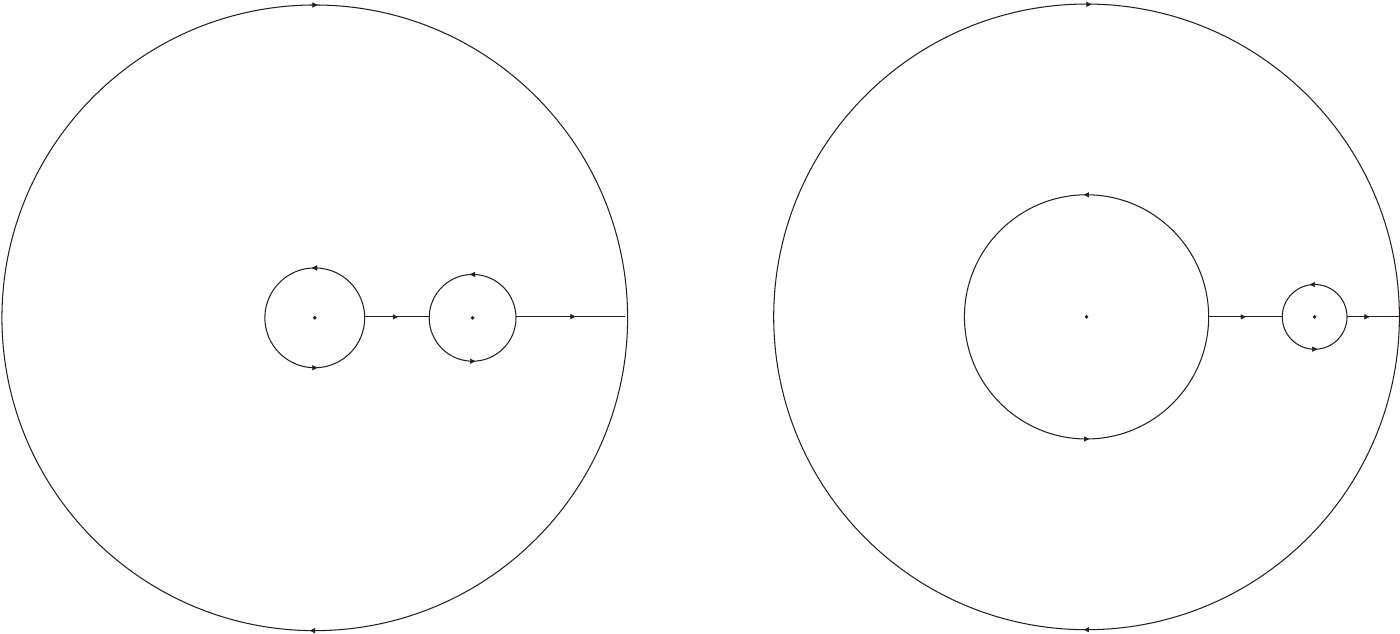}}; 
 \draw(-4.1,-0.1) node {\footnotesize $0$};  
 \draw(-2.45,-0.1) node {\footnotesize $1$};  
 \draw(-6.7,0) node {\scriptsize $\lb -z\rb^{-\mathfrak S}$}; 
 \draw(-5.15,0) node {\scriptsize $\lb -z\rb^{-\Theta_0}C_0^{-1}$}; 
 \draw(-3.1,-0.25) node {\scriptsize $M_0^{-1}$}; 
 \draw(-1.3,0.25) node {\scriptsize $e^{-2\pi i \mathfrak S}$};  
 \draw(-2.1,0.65) node {\scriptsize $\lb 1-z\rb^{-\Theta_t}C_{t,+}^{-1}$}; 
 \draw(-2.1,-0.65) node {\scriptsize $\lb 1-z\rb^{-\Theta_t}C_{t,-}^{-1}$};
 \draw(1.5,0) node {\scriptsize $\lb -z\rb^{\Theta_{\infty}}C_{\infty}^{-1}$};
 \draw(4.1,1.5) node {\scriptsize $\lb -z\rb^{-\mathfrak S}$};           
 \draw(3.85,-0.1) node {\footnotesize $0$};  
 \draw(6.2,-0.1) node {\footnotesize $1$}; 
 \draw(6.9,0.2) node {\scriptsize $M_{\infty}$};  
 \draw(5.68,0.23) node {\scriptsize $e^{-2\pi i \mathfrak S}$}; 
 \draw(6.35,0.55) node {\scriptsize $\lb 1-z\rb^{-\Theta_1}C_{1,+}^{-1}$}; 
 \draw(6.35,-0.55) node {\scriptsize $\lb 1-z\rb^{-\Theta_1}C_{1,-}^{-1}$};       
 \end{tikzpicture}      
            \caption{\label{PD4pointsBis}
            Contours and jump matrices for $\tilde\Psi^{[L]}$ (left) and $\Psi^{[R]}$ (right)}
     \end{figure}
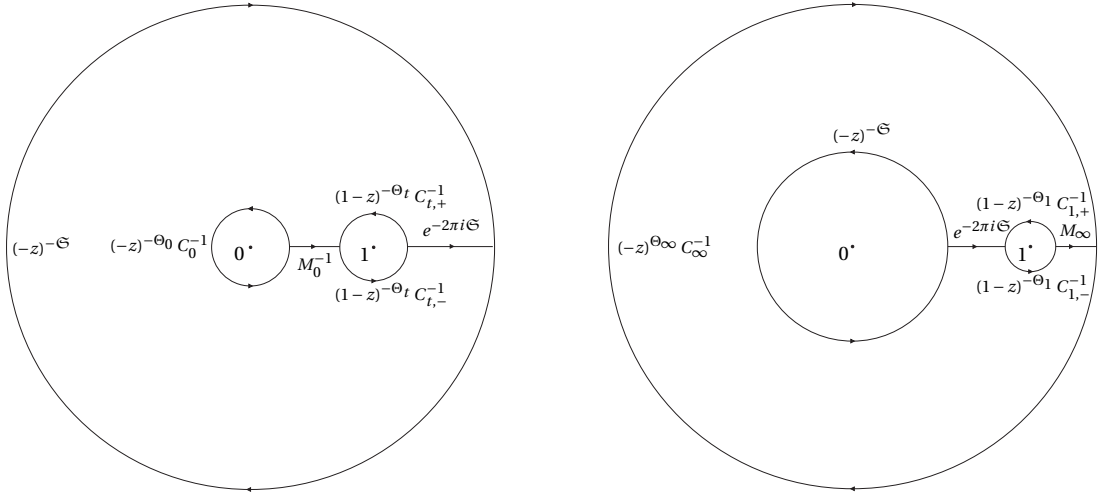

   The matrix functions ${\Psi^{[L]}\lb z\rb}$, ${\Psi^{[R]}\lb z\rb}$ appearing in the integral kernels of $\mathsf a$ and $\mathsf d$ solve the $3$-point RHPs associated to Fuchsian systems with regular singularities at $0,t,\infty$ and $0,1,\infty$, respectively. In order to understand the dependence of  the $4$-point tau function on the time variable~$t$, let us rescale the fundamental solution of the first system by setting 
   \beq\label{rescale}
   {\Phi^{[L]}\lb z\rb}=\tilde\Phi^{[L]}\lb \frac zt\rb .
   \eeq
   The rescaled matrix $\tilde\Phi^{[L]}\lb z\rb$ solves a Fuchsian system characterized by the same monodromy as $\Phi^{[L]}\lb z\rb$ but the corresponding singular points are located at $0,1,\infty$.
   Denote by $\tilde\Psi^{[L]}\lb z\rb$ the solution of the RHP associated to $\tilde\Phi^{[L]}\lb z\rb$. To avoid possible confusion of the reader, we explicitly indicate the contours and jump matrices for RHPs for $\tilde\Psi^{[L]}$ and $\Psi^{[R]}$ in Fig.~\ref{PD4pointsBis}; note the independence of jumps on $t$.  In particular, inside the disk around~$\infty$ we have $\tilde\Phi^{[L]}\lb z\rb=\lb -z\rb^{\mathfrak S}\tilde\Psi^{[L]}\lb z\rb$.
   Since the annulus $\mathcal A$ belongs to the disk 
   around $\infty$ in the RHP for $\Psi^{[L]}$, the formula (\ref{rescale}) 
   yields the following expression for $\Psi^{[L]}$ inside $\mathcal A$:
   \begin{subequations}\label{PsiRL}
   \beq\label{psiLinsideA}
   \Psi^{[L]}\lb z\rb\Bigl|_{\mathcal A}= \lb -z\rb^{-\mathfrak S}\Phi^{[L]}\lb z\rb =t^{-\mathfrak S}
   \tilde\Psi^{[L]}\lb \frac zt\rb 
   =  t^{-\mathfrak S}\left(\mathbb 1+\sum_{k=1}^{\infty}
    g_k^{[L]} t^k z^{-k}\right)G_{\infty}^{[L]},
   \eeq
   where the $N\times N$ matrix coefficients $g_k^{[L]}$ are independent of $t$.
    Analogous expression for $\Psi^{[R]}\lb z\rb$ inside $\mathcal A$  does not contain $t$ at all:
    \beq\label{psiRinsideA}
    \Psi^{[R]}\lb z\rb\Bigl|_{\mathcal A}= \left(\mathbb 1+\sum_{k=1}^{\infty}
        g_k^{[R]} z^{k}\right)G_{0}^{[R]}. 
    \eeq
    \end{subequations}    
    The formulae (\ref{PsiRL}) allow to extract from the determinant representation (\ref{det4pttau}) the asymptotics of $4$-point Jimbo-Miwa-Ueno tau function $\tau_{\mathrm{JMU}}\lb t\rb$ as $t\to0$ to any desired order. We are now going to explain the details of this procedure. 
    
    Rewrite the integral kernel $\mathsf d\lb z,z'\rb$ as
    \ben
    \mathsf d\lb z,z'\rb=t^{-\mathfrak S}
    \frac{\mathbb 1-\tilde\Psi^{[L]}\lb\frac zt\rb
    {\tilde\Psi^{[L]}\lb \frac {z'}t\rb}^{-1}}{z-z'}t^{\mathfrak S}.
    \ebn
    The  block matrix elements of $\mathsf d$ in the Fourier basis are therefore given by
    \beq\label{deltamnt}
    \mathsf d^{-p}_{\;\;\;\, q}=t^{-\mathfrak S}\tilde{\mathsf d}^{-p}_{\;\;\;\, q}t^{\mathfrak S}\cdot t^{p+q},\qquad\qquad p,q\in\mathbb Z'_+,
    \eeq
    where $N\times N$ matrix coefficients $\tilde{\mathsf d}^{-p}_{\;\;\;\, q}$ are independent of $t$. They can be extracted from the Fourier series
    \beq\label{deltatilde}
     \frac{\mathbb 1-\tilde\Psi^{[L]}\lb z\rb{\tilde\Psi^{[L]}\lb z'\rb}^{-1}}{z-z'}=\sum_{p,q\in\mathbb Z'_+}\tilde{\mathsf d}^{-p}_{\;\;\;\, q}z^{-\frac12-p}z'^{-\frac12-q},
    \eeq
    and are therefore expressed in terms of the coefficients of local expansion of  the $3$-point solution $\tilde\Phi^{[L]}\lb z\rb$ around $z=\infty$ by straightforward algebra. For instance, the first few coefficients are given by
    \ben\begin{gathered}
    \tilde{\mathsf d}^{-\frac12}_{\;\;\;\,\frac12}=g_{1}^{[L]},\\ 
    \tilde{\mathsf d}^{-\frac12}_{\;\;\;\,\frac32}=g_2^{[L]}-{g_1^{[L]}}^2,\qquad
    \tilde{\mathsf d}^{-\frac32}_{\;\;\;\,\frac12}=g_2^{[L]},\\
    \tilde{\mathsf d}^{-\frac12}_{\;\;\;\,\frac52}=g_3^{[L]}-g_2^{[L]}g_1^{[L]}
    -g_1^{[L]}g_2^{[L]}+{g_1^{[L]}}^3,\qquad 
     \tilde{\mathsf d}^{-\frac32}_{\;\;\;\,\frac32}=
     g_3^{[L]}-g_2^{[L]}g_1^{[L]},\qquad
     \tilde{\mathsf d}^{-\frac52}_{\;\;\;\,\frac12}=g_3^{[L]},\\
     \ldots\ldots\qquad \ldots\ldots\qquad \ldots\ldots
    \end{gathered}
    \ebn
    Different lines above contain the coefficients of fixed degree $p+q\in\mathbb Z_{>0}$ which appears in the power of $t$ in (\ref{deltamnt}). Very similar formulas are also valid for matrix elements of $\mathsf a$:
    \ben
    \mathsf a^{\;\;\;\frac12}_{\!-\frac12}=g_1^{[R]},\qquad
    \mathsf a^{\;\;\;\frac12}_{\!-\frac32}=g_2^{[R]}-{g_1^{[R]}}^2,\qquad
    \mathsf a^{\;\;\;\frac32}_{\!-\frac12}=g_2^{[R]},\qquad\ldots
    \ebn

    The crucial point for the asymptotic analysis of $\tau\lb t\rb$  is that for small $t$ the operator $\mathsf d$ becomes effectively finite rank. Indeed, fix a positive integer $Q$. To obtain a uniform approximation of $\mathsf d\lb z,z'\rb$ up to order $O\lb t^{Q}\rb$, it suffices to take into account its Fourier coefficients $\mathsf d^{-p}_{\;\;\;\,q}$ with 
    $p+q\leq Q$; recall that the eigenvalues of $\mathfrak S$ are chosen as to satisfy (\ref{ascondsp}). Since here $p,q\in\mathbb Z'_+$, the total number of relevant coefficients is finite and equal to $Q\lb Q-1\rb/2$. It follows that the only terms in the Fourier expansion of $\mathsf a\lb z,z'\rb$ that contribute to the determinant (\ref{det4pttau}) to order $O\lb t^{Q}\rb$ correspond to  monomials $z^{p-\frac12} z'^{q-\frac12}$ with $p+q\le Q$. This is summarized in     
    \begin{theo}\label{theojimbo} Let $Q\in\Zb_{> 0}$. The $4$-point tau function $\tau_{\mathrm{JMU}}\lb t\rb$ has the following asymptotics  as $t\to 0$:
    \beq\label{4pttauas}
    \tau_{\mathrm{JMU}}\lb t\rb\simeq t^{\frac12\operatorname{Tr}\lb\mathfrak S^2-\Theta_0^2-\Theta_t^2\rb}\Bigl[\operatorname{det}\lb\mathbb 1-U_Q\rb+O\lb t^{Q}\rb\Bigr],\qquad
       U_Q=\lb\begin{array}{cc} 0 & \mathsf a_Q \\ \mathsf d_Q & 0 \end{array}\rb.
    \eeq
    Here $U_Q$ denotes a $2NQ\times 2NQ$ finite matrix whose $NQ\times NQ$-dimensional blocks $\mathsf a_Q$ and $\mathsf d_Q$ are  themselves block lower and block upper triangular matrices of the form
    \begin{align*}
    \mathsf a_Q=\lb\begin{array}{cccc}
    \mathsf a^{Q-\frac12}_{\;-\frac12} & 0 & \cdot\cdot\cdot & 0\\
    \vdots & \mathsf a^{Q-\frac32}_{\;-\frac32} & \cdot & \vdots \\
    \mathsf a^{\;\;\;\frac32}_{\!-\frac12} & \cdot & \ddots & 0\vspace{0.1cm}\\    
    \mathsf a^{\;\;\;\frac12}_{\!-\frac12} & \mathsf a^{\;\;\;\frac12}_{\!-\frac32} & \cdot\cdot\cdot & \mathsf a^{\;\;\;\frac12}_{\frac12-Q}
    \end{array}\rb,\qquad
    \mathsf d_Q=t^{-\mathfrak S}\lb\begin{array}{cccc}
       \tilde{\mathsf d}^{\;\;-\frac12}_{Q-\frac12}t^Q & \cdot\cdot\cdot
        & \tilde{\mathsf d}^{-\frac12}_{\;\;\;\,\frac32}t^2 & \tilde{\mathsf d}^{-\frac12}_{\;\;\;\,\frac12}t\\
       0 & \ddots & \cdot & \tilde{\mathsf d}^{-\frac32}_{\;\;\;\,\frac12}t^2 \\
       \vdots & \cdot & \tilde{\mathsf d}^{\frac32-Q}_{\;\;\;\;\frac32}t^Q & \vdots \\
      0 & \cdot\cdot\cdot & 0 & \tilde{\mathsf d}^{\frac12-Q}_{\;\;\;\frac12}t^Q
        \end{array}\rb t^{\mathfrak S},
    \end{align*}
    where $\mathsf a^{\;\;\;\,p}_{-q}$, $\tilde{\mathsf d}^{-p}_{\;\;\;\,q}$ are determined by (\ref{intkernelsalphadelta1}), (\ref{alphadeltaME}), (\ref{deltatilde}), and the conjugation by $t^{\mathfrak S}$ in the expression for 
    $\mathsf d_Q$ is understood to act on each $N\times N$ block of the interior matrix. Moreover, strengthening the condition (\ref{ascondsp}) to strict inequality $\left|\Re\lb \sigma_{\alpha}-\sigma_{\beta}\rb\right|<1$ improves the error estimate in
    (\ref{4pttauas})   to  $o\lb t^Q\rb$.
    \end{theo}
  
   \begin{rmk}The above theorem gives the asymptotics  of  $\tau_{\mathrm{JMU}}\lb t\rb$ to arbitrary finite order $Q$ in terms of solutions $\Phi^{[R]}\lb z\rb$, $\tilde\Phi^{[L]}\lb z\rb$ of two $3$-point Fuchsian systems with prescribed monodromy around regular singular points $0$, $1$, $\infty$. For  $Q=1$ and under assumption $\left|\Re\lb \sigma_{\alpha}-\sigma_{\beta}\rb\right|<1$, its statement  may be rewritten as 
   \beq\label{eqresj}
   \tau_{\mathrm{JMU}}\lb t\rb\simeq t^{\frac12\operatorname{Tr}\lb\mathfrak S^2-\Theta_0^2-\Theta_t^2\rb}\left[\operatorname{det}\lb\mathbb 1-g_1^{[R]}t^{\mathbb 1-\mathfrak S}g_1^{[L]}t^{\mathfrak S}\rb+o\lb t\rb\right].
   \eeq
   A result equivalent to this last formula has been recently obtained  in \cite[Proposition 3.9]{ILP} by a rather involved asymptotic analysis based on the conventional Riemann-Hilbert approach. For 
   $N=2$, the leading term in the expansion of the determinant  appearing in (\ref{eqresj}) gives Jimbo asymptotic formula \cite{Jimbo} for Painlev\'e VI.
    \end{rmk}

  Let us also describe for completeness the $n=4$ specialization of the formulae for the fundamental solution. Fix the bases in $\mathcal H_+$ and $\mathcal H$ with respect to decompositions $\mathcal H_+=
  \mathcal H_{\mathrm{out},+}^{[L]}\oplus\mathcal H_{\mathrm{in},-}^{[R]}$ and $\mathcal H=
    \mathcal H_{\mathrm{out},+}^{[L]}\oplus\mathcal H_{\mathrm{in},-}^{[R]}\oplus  \mathcal H_{\mathrm{out},-}^{[L]}\oplus\mathcal H_{\mathrm{in},+}^{[R]}$. In these bases, different operators appearing in (\ref{fundf})  acquire the form
  \ben
  \begin{gathered}
  \mathcal P_{\oplus,+}=\lb\begin{array}{cc}
  \mathbb 1 & 0 \\ 0 & \mathbb 1 \\ 
  \mathsf d & 0 \\ 0 & \mathsf a
  \end{array}\rb,\qquad
  \mathbb 1^{[L]}_{\mathrm{out}}\Psi^{[R]}\lb z\rb=
  \lb\begin{array}{cccc}
  \Psi^{[R]}\lb z\rb  \\ 0
  \end{array}\rb,\qquad
  \mathbb 1^{[R]}_{\mathrm{in}}\Psi^{[R]}\lb z\rb=
    \lb\begin{array}{cccc}
    0 \\ 0 \\ 0 \\
    \Psi^{[R]}\lb z\rb
    \end{array}\rb,\\
    L=\lb \mathbb 1-U\rb^{-1}=\lb\begin{array}{rr}
    \lb\mathbb 1-\mathsf a\mathsf d\rb^{-1} &
    \mathsf a\lb\mathbb 1-\mathsf d\mathsf a\rb^{-1} \\
    \mathsf d\lb\mathbb 1-\mathsf a\mathsf d\rb^{-1} &
    \lb\mathbb 1-\mathsf d\mathsf a\rb^{-1}
    \end{array}\rb,
    \end{gathered}
  \ebn
 where $\mathbb 1^{[L]}_{\mathrm{out}}\Psi^{[R]}\lb z\rb$ and
 $\mathbb 1^{[R]}_{\mathrm{in}}\Psi^{[R]}\lb z\rb$ are seen as vectors in  $\mathcal H_+$ and $\mathcal H$, respectively. Now using (\ref{fundf}) and representing the answer in the form of a single function defined on the annulus $\mathcal A$, we obtain
 \beq
 \hat\Psi\lb z\rb=\lb \mathbb 1+\mathsf d\rb
 \lb\mathbb 1-\mathsf a\mathsf d\rb^{-1} \Psi^{[R]}\lb z\rb,\qquad z\in\mathcal A.
 \eeq
 
 \section{Fourier basis and combinatorics\label{seccomb}}
   \subsection{Structure of matrix elements\label{subseccauchy}}
   Let us return to the general case of $n$ regular singular points on $\Pb^1$.  We have already seen in the previous subsection
   certain advantages of writing the operators which appear in the Fredholm determinant representation (\ref{fredholmrep}) of the tau function in the Fourier basis. This motivates us to introduce the following notation for the integral kernels of the $3$-point projection operators
   $\mathsf a^{[k]}$, $\mathsf b^{[k]}$, $\mathsf c^{[k]}$, $\mathsf d^{[k]}$ from
   (\ref{alphadelta}):
   \begin{subequations}\label{alphadeltaF}
   \begin{alignat}{4}
   \label{alphadeltaF01}
   &\mathsf a^{[k]}\lb z,z'\rb:=\frac{\Psi_+^{[k]}\lb z\rb{\Psi_+^{[k]}\lb z'\rb}^{-1}-\mathbb 1}{z-z'}&&=\sum_{p,q\in\mathbb Z'_+}\mathsf a^{[k]}{}^{\;p}_{\!\!\!\!\!-q}\,z^{-\frac12+p} z'^{-\frac12+q},\qquad\qquad && z,z'\in\mathcal C^{[k]}_{\mathrm{in}},\\
   &\mathsf b^{[k]}\lb z,z'\rb:=\quad-\frac{\Psi_+^{[k]}\lb z\rb {\Psi_+^{[k]}\lb z'\rb}^{-1}}{z-z'}\; &&=
   \sum_{p,q\in\mathbb Z'_+}\mathsf b^{[k]}{}^{p}_{q}\, z^{-\frac12+p} {z'}^{-\frac12-q},\qquad\qquad && z\in 
   \mathcal C^{[k]}_{\mathrm{in}},z'\in \mathcal C^{[k]}_{\mathrm{out}},\\
   &\mathsf c^{[k]}\lb z,z'\rb:=\quad\frac{\Psi_+^{[k]}\lb z\rb {\Psi_+^{[k]}\lb z'\rb}^{-1}}{z-z'}\; &&=
   \sum_{p,q\in\mathbb Z'_+}\mathsf c^{[k]}{}^{-p}_{-q}\,z^{-\frac12-p} z'^{-\frac12+q},\qquad\qquad && z\in 
   \mathcal C^{[k]}_{\mathrm{out}},z'\in \mathcal C^{[k]}_{\mathrm{in}},\\  
   &\mathsf d^{[k]}\lb z,z'\rb:=\frac{\mathbb 1-\Psi_+^{[k]}\lb z\rb {\Psi_+^{[k]}\lb z'\rb}^{-1}}{z-z'}&&=\sum_{p,q\in\mathbb Z'_+}\mathsf d^{[k]}{}^{-p}_{\;\;\;\,q}z^{-\frac12-p} z'^{-\frac12-q},\qquad\qquad && z,z'\in\mathcal C^{[k]}_{\mathrm{out}}.
   \end{alignat}
   \end{subequations}
   Just as before in (\ref{intkernelsalphadelta2}), the overall minus signs in the expressions for $\mathsf b^{[k]}\lb z,z'\rb$ and $\mathsf d^{[k]}\lb z,z'\rb$ are introduced to absorb the negative orientation of $\mathcal C^{[k]}_{\mathrm{out}}$.
   
   Our task in this subsection is to understand the dependence of matrix elements $\ds\mathsf a^{[k]}{}^{\,p}_{\!\!\!\!\!\!-q}$, $\ds\mathsf b^{[k]}{}^{p}_{q}$, $\ds\mathsf c^{[k]}{}^{-p}_{-q}$, $\ds\mathsf d^{[k]}{}^{-p}_{\;\;\;\,q}$ on their indices $p,q\in\mathbb Z'_+$. To this end recall that (cf (\ref{jump3point}))
   \beq\label{aux3prel}
   \Psi^{[k]}_+\lb z\rb =\begin{cases}
   \lb-z\rb^{-\mathfrak S_{k-1}}S_{k-1}^{-1}\Phi^{[k]}\lb z\rb ,\qquad & z\in \mathcal C^{[k]}_{\mathrm{in}},\\
   \lb-z\rb^{-\mathfrak S_{k}}S_{k}^{-1}\Phi^{[k]}\lb z\rb ,\qquad & z\in \mathcal C^{[k]}_{\mathrm{out}}.
   \end{cases}
   \eeq
   where $\Phi^{[k]}\lb z\rb $ denotes the fundamental solution of the $3$-point Fuchsian system (\ref{FS3point}). 
   \begin{theo}\label{theocauchy} Denote by $\mathfrak r^{[k]}$ the rank of the matrix $A_1^{[k]}$ which appears in the Fuchsian system (\ref{FS3point}). Let ${u_r^{[k]}, v_r^{[k]}\in\Cb^N}$ with $r=1,\ldots,\mathfrak r^{[k]}$ be the column and row vectors giving the decomposition 
     \beq\label{tensorA1k}
     a_{k}A_1^{[k]}=-\sum_{r=1}^{\mathfrak r^{[k]}}u_r^{[k]}\otimes {v_r^{[k]}}.
     \eeq
   Let $\bigl(\psi^{[k]}_{r}\bigr){\bigl.}^p, \bigl(\bar \psi^{[k]}_{r}\bigr){}_{p}, \bigl(\varphi^{[k]}_{r}\bigr){\bigl.}^{-p}, \bigl(\bar\varphi^{[k]}_{r}\bigr){}_{-p}\in \Cb^N$  be the coefficients of the Fourier expansions 
   \begin{subequations}\label{auxfactor03}
    \begin{alignat}{3}\label{auxfactor03a}
     &\begin{cases}\begin{aligned}
     \ds\frac{ {\Psi_+^{[k]}\lb z\rb}u_r^{[k]}}{z-a_{k}}\;\;\;=&\,\sum\limits_{p\in\mathbb Z'_+}\bigl(\psi^{[k]}_{r}\bigr){\bigl.}^p z^{-\frac12+p},\\
     \ds\frac{{v_r^{[k]}}{\Psi_+^{[k]}\lb z\rb}^{-1}}{z-a_{k}}=&\,\sum\limits_{p\in\mathbb Z'_+}\bigl(\bar \psi^{[k]}_{r}\bigr){\bigl.}_{p} \,z^{-\frac12+p},
     \end{aligned}
     \end{cases}\qquad && z\in 
     \mathcal C^{[k]}_{\mathrm{in}},\\
         &\begin{cases}
         \begin{aligned}
     \ds \frac{ {\Psi_+^{[k]}\lb z\rb}u_r^{[k]}}{z-a_{k}}\;\;\;
     =&\,\sum_{ p\in\mathbb Z'_+}\bigl(\varphi^{[k]}_{r}\bigr){\bigl.}^{-p} z^{-\frac12-p},\\
     \ds\frac{{v_r^{[k]}}{\Psi_+^{[k]}\lb z\rb}^{-1}}{z-a_{k}}=&\,\sum_{p\in\mathbb Z'_+}
     \bigl(\bar\varphi^{[k]}_{r}\bigr){\bigl.}_{-p} z^{-\frac12-p},
     \end{aligned}
     \end{cases}\qquad && z\in \mathcal C^{[k]}_{\mathrm{out}}.
     \end{alignat}
   \end{subequations}  
   Then  the operators $\mathsf a^{[k]}$, $\mathsf b^{[k]}$, $\mathsf c^{[k]}$, $\mathsf d^{[k]}$ can be represented as  sums of a finite number of infinite-dimensional Cauchy matrices with respect to the indices $p,q$, explicitly given by
   \begin{subequations}\label{alphadeltacauchy}
     \begin{alignat}{3}
     \label{alphadeltacauchy01}
     &\;\mathsf a^{[k]}{}^{\;\; p;\alpha}_{\!\!\!-q;\beta}&&=&&\,\sum_{r=1}^{\mathfrak r^{[k]}} \frac{\bigl(\psi^{[k]}_{r}\bigr){\bigl.}^{p;\alpha}
       {\bigl(\bar \psi^{[k]}_{r}\bigr){}_{q;\beta}}}{
       p+q+\sigma_{k-1,\alpha}-\sigma_{k-1,\beta}},\\
     \label{alphadeltacauchy02}
    &\;\; \mathsf b^{[k]}{}^{p;\alpha}_{q;\beta}&&=&&\,\sum_{r=1}^{\mathfrak r^{[k]}} \frac{\bigl(\psi^{[k]}_{r}\bigr){\bigl.}^{p;\alpha}
       \bigl(\bar\varphi^{[k]}_{r}\bigr){}_{-q;\beta}}{
       q-p-\sigma_{k-1,\alpha}+\sigma_{k,\beta}},\\
     \label{alphadeltacauchy03}
     &\mathsf c^{[k]}{}^{-p;\alpha}_{-q;\beta}&&=&&\,\sum_{r=1}^{\mathfrak r^{[k]}} \frac{\bigl(\varphi^{[k]}_{r}\bigr){\bigl.}^{-p;\alpha}
      {\bigl(\bar \psi^{[k]}_{r}\bigr){}_{q;\beta}}}{
       q-p+\sigma_{k,\alpha}-\sigma_{k-1,\beta}},\\
     \label{alphadeltacauchy04}
    & \mathsf d^{[k]}{}^{-p;\alpha}_{\;\;\; q;\beta}&&=&&\,\sum_{r=1}^{\mathfrak r^{[k]}} \frac{\bigl(\varphi^{[k]}_{r}\bigr){\bigl.}^{-p;\alpha} \bigl(\bar\varphi^{[k]}_{r}\bigr){}_{-q;\beta}}{
       p+q-\sigma_{k,\alpha}+\sigma_{k,\beta}},                  
     \end{alignat}
   \end{subequations}  
     where the color indices $\alpha,\beta=1,\ldots,N$ correspond to internal structure of the blocks $\ds\mathsf a^{[k]}{}^{\,p}_{\!\!\!\!\!\!-q}$, $\ds\mathsf b^{[k]}{}^{p}_{q}$, $\ds\mathsf c^{[k]}{}^{-p}_{-q}$, $\ds\mathsf d^{[k]}{}^{-p}_{\;\;\;\,q}$.  
   \end{theo}
   \pf
   The Fuchsian system (\ref{FS3point}) can be used to differentiate the integral kernels (\ref{alphadeltaF}) with respect to $z$ and~$z'$. Consider, for instance, the operator
   \ben
   \mathcal{L}_0=z\partial_z+z'\partial_{z'}+1.
   \ebn
   It is easy to check that\footnote{The reader with acquintance with two-dimensional conformal field theory will recognize in this equation the dilatation Ward identity for the $2$-point correlator of Dirac fermions.}
   $\mathcal L_0\frac{1}{z-z'}=0$. Combining this with (\ref{alphadeltaF01}), (\ref{aux3prel}) and (\ref{FS3point}),  one obtains e.g. that
   \ben\label{auxfactor01}
   \mathcal L_0\mathsf a^{[k]}\lb z,z'\rb=\frac{
   \lb z\partial_z+z'\partial_{z'}\rb \Psi_+^{[k]}\lb z\rb {\Psi_+^{[k]}\lb z'\rb}^{-1}}{z-z'}=\left[\mathsf a^{[k]}\lb z,z'\rb,\mathfrak S_{k-1}\right]-\frac{\Psi_+^{[k]}\lb z\rb}{z-a_{k}}a_{k}A_1^{[k]}\frac{ {\Psi_+^{[k]}\lb z'\rb}^{-1}}{z'-a_{k}},
   \ebn 
   where $z,z'\in\mathcal C^{[k]}_{\mathrm{in}}$.
   The crucial point here is that the dependence of the second term on $z$ and $z'$ is completely factorized. Indeed, 
   it follows from the last identity, the form of $\mathcal L_0$ and the notation (\ref{auxfactor03a}) that  $N\times N$ matrix $\ds\mathsf a^{[k]}{}^{\,p}_{\!\!\!\!\!\!-q}$ from (\ref{alphadeltaF01}) satisfies the equation
   \ben\label{auxfactor02}
   \lb p+q+\operatorname{ad}_{\mathfrak S_{k-1}}\rb\mathsf a^{[k]}{}^{\,p}_{\!\!\!\!\!\!-q}=
   \sum_{r=1}^{\mathfrak r^{[k]}}\bigl(\psi^{[k]}_{r}\bigr){\bigl.}^p\otimes
   {\bigl(\bar \psi^{[k]}_{r}\bigr){\bigl.}_{q}}.
   \ebn
   The formula (\ref{alphadeltacauchy01}) is nothing but a rewrite of this identity. The proof of Cauchy type representations
   (\ref{alphadeltacauchy02})--(\ref{alphadeltacauchy04}) for the other three operators is completely analogous.
   \epf
 
   \subsection{Combinatorics of determinant expansion
   \label{subseccomb}}
   This subsection develops a systematic approach to the computation of multivariate series expansion of the Fredholm determinant $\tau\lb a\rb=\operatorname{det}\lb \mathbb 1-K\rb$.
   Recall that, according to Theorem~\ref{TauF}, the isomonodromic tau function $\tau_{\mathrm{JMU}}\lb a\rb$ coincides with $\tau\lb a\rb$ up to an elementary explicit prefactor.
 
  Let $A\in \Cb^{\mathfrak X\times \mathfrak X}$ be a matrix indexed by a discrete and possibly infinite set $\mathfrak X$. Our basic tool for expanding  $\tau\lb a\rb$ is the von Koch's formula: 
  \beq\label{vonKoch}
  \operatorname{det}\lb \mathbb 1+A\rb=\sum_{\mathfrak Y\in 2^{\mathfrak X}}\operatorname{det} A_{\mathfrak Y},
  \eeq
  where  $\operatorname{det} A_{\mathfrak Y}$ denotes the $|\mathfrak Y|\times |\mathfrak Y|$ principal minor obtained by restriction of $A$ to a subset $\mathfrak Y\subseteq \mathfrak X$. Of course, the series in (\ref{vonKoch}) terminates when $\mathfrak X$ is finite.
  
  In our case, the role of the matrix $A$ is played by the operator $K$ written in the Fourier basis. The elements of $\mathfrak X$ are multi-indices which encode the following data:
  \begin{itemize}
  \item the positions of the blocks $\mathsf a^{[k]}$, $\mathsf b^{[k]}$, $\mathsf c^{[k]}$, $\mathsf d^{[k]}$ in $K$ defined by
  (\ref{UVW}); 
  \item a half-integer Fourier index of the appropriate block;
  \item a color index  taking its values in the set $\left\{1,\ldots,N\right\}$.
  \end{itemize} 
  It is useful to combine Fourier and color indices into one multi-index $\imath=\lb p,\alpha\rb\in \mathfrak N:= \mathbb Z'\times \left\{1,\ldots,N\right\}$. Unordered sets $\left\{ \imath_1,\ldots,\imath_m\right\}\in 2^{\mathfrak N}$ of such multi-indices are denoted by capital Roman letters $I$ or $J$. Given a matrix  $M\in\Cb^{\mathfrak N\times \mathfrak N}$, we denote by $M_{I}^{J}$ its $|I|\times |J|$ restriction to rows $I$ and columns $J$.
  
  Principal submatrices of $K$ may be labeled by pairs $\lb \vec I,\vec J\rb$, where  $\vec I=\lb I_1,\ldots,I_{n-3}\rb$, $\vec J=\lb J_1,\ldots, J_{n-3}\rb$ and
  $I_{1\ldots n-3},{J_{1\ldots n-3}\in 2^{\mathfrak N}}$. Namely,  define
  \ben
  \begin{gathered}
  K_{\vec I,\vec J}:=\lb
  \begin{array}{cccccccccc}
  0 & \lb\mathsf a^{[2]}\rb^{I_1}_{J_1} & 
  \lb \mathsf b^{[2]}\rb^{I_1}_{I_2} & 0  & 0 & 0 & 
  \cdot & \cdot  & 0 & 0 \\
  \lb \mathsf d^{[1]}\rb^{J_1}_{I_1} & 0 & 0 & 0   & 0 & 0 & \cdot & \cdot  & 0 & 0 \\
  0 & 0 & 0 & \lb\mathsf a^{[3]}\rb^{I_2}_{J_2} & 
  \lb \mathsf b^{[3]}\rb^{I_2}_{I_3} & 0 & \cdot & \cdot & 0 & 0 \\
  0 & \lb \mathsf c^{[2]}\rb^{J_2}_{J_1} & 
  \lb\mathsf d^{[2]}\rb^{J_2}_{I_2} & 
  0 & 0 & 0 & \cdot & \cdot & 0 & 0 \\
  0 & 0 & 0 & 0 & 0 & \lb\mathsf a^{[4]}\rb^{I_3}_{J_3} & 
  \cdot & \cdot & \cdot & \cdot \\
  0 & 0 & 0 & \lb\mathsf c^{[3]}\rb^{J_3}_{J_2} & 
  \lb\mathsf d^{[3]}\rb^{J_3}_{I_3} & 0 & \cdot & \cdot & \cdot & \cdot \\
  \cdot & \cdot & \cdot & \cdot & \cdot & \cdot & \cdot & \cdot 
  & \lb \mathsf b^{[n-3]}\rb^{I_{n-2}}_{I_{n-3}} & 0 \\
  \cdot & \cdot & \cdot & \cdot & \cdot & \cdot & \cdot & \cdot 
  & 0 & 0 \\
  0 & 0 & 0 & 0 & \cdot & \cdot & 0 & 0 & 0 & 
  \lb \mathsf a^{[n-2]}\rb^{I_{n-3}}_{J_{n-3}} \\
  0 & 0 & 0 & 0 & \cdot & \cdot & 0 & 
  \lb \mathsf c^{[n-3]} \rb^{J_{n-3}}_{J_{n-4}}
  & \lb \mathsf d^{[n-3]}\rb^{J_{n-3}}_{I_{n-3}} & 0    
  \end{array}\rb
  \end{gathered}
  \ebn
  For reasons that will become apparent below, the pairs $\lb \vec I,\vec J\rb$ will be referred to as configurations. It is useful to keep in mind that the lower index in $I_k,J_k$ corresponds to the annulus $\mathcal A_k$, and the blocks of $K$ are acting between spaces of holomorphic functions on the appropriate annuli. 

  \begin{defin}
  A configuration $\lb\vec I,\vec J\rb\in\lb 2^{\mathfrak N}\rb^{\times 2\lb n-3\rb}$ is called
  \begin{itemize} 
  \item balanced if 
    $ |I_k|=|J_k|$ for $k=1,\ldots,n-3$;
  \item proper if all elements of $I_k$ (and $J_k$)  have positive (resp. negative) Fourier indices for $k=1,\ldots,n-3$. 
  \end{itemize}
  The sets of all balanced and proper balanced configurations will be denoted by $\mathsf{Conf}$ and
        $\mathsf{Conf}_+$, respectively.
  \end{defin}
  \begin{defin}
  For $\lb \vec I,\vec J\rb\in\mathsf{Conf}$, define
  \begin{subequations}
  \label{partZ06}
    \beq\label{partZ01}
    Z^{I_{k-1},J_{k-1}}_{\;I_{k},J_{k}}\lb \mathcal T^{[k]}\rb:=
    \lb -1\rb^{\left|I_k\right|}
    \operatorname{det}\lb\begin{array}{ll}
         \lb\mathsf a^{[k]}\rb^{I_{k-1}}_{J_{k-1}} & 
           \lb \mathsf b^{[k]}\rb^{I_{k-1}}_{I_k}\vspace{0.1cm}
           \\
         \lb \mathsf c^{[k]}\rb^{J_k}_{J_{k-1}} & 
           \lb\mathsf d^{[k]}\rb^{J_k}_{I_k}  
         \end{array}\rb, \qquad k=1,\ldots,n-2.
    \eeq
    In order to have uniform notation, here we set $I_0=J_0=I_{n-2}=J_{n-2}\equiv\emptyset$, so that
    \beq\label{partZ02}
    Z^{\;\emptyset,\,\emptyset}_{I_{1},J_{1}}\lb \mathcal T^{[1]}\rb=
    \lb -1\rb^{\left|I_1\right|}\operatorname{det}\lb \mathsf d^{[1]}\rb^{J_1}_{I_1},\qquad
    Z_{\;\emptyset,\,\emptyset}^{I_{n-3},J_{n-3}}\lb \mathcal T^{[n-2]}\rb=
        \operatorname{det}\lb \mathsf a^{[n-2]}\rb^{I_{n-3}}_{J_{n-3}}.
    \eeq
  \end{subequations}  
  \end{defin}
  \begin{prop}\label{propfactor}
  The principal minor $D_{\vec I,\vec J}:=\operatorname{det} K_{\vec I,\vec J}$ vanishes unless $\lb \vec I,\vec J\rb\in \mathsf{Conf}_+$, in which case it factorizes into a product of
  $n-2$ finite $\lb\left|I_{k-1}\right|+\left|I_{k}\right|\rb\times \lb\left|I_{k-1}\right|+\left|I_{k}\right|\rb$ determinants as 
     \beq\label{factordet}
     D_{\vec I,\vec J}=\prod_{k=1}^{n-2} Z^{I_{k-1},J_{k-1}}_{\;I_{k},J_{k}}\lb \mathcal T^{[k]}\rb.
     \eeq     
  \end{prop}
  \pf For $k=1,\ldots ,n-3$, exchange the $(2k-1)$-th and $2k$-th block row of the matrix $K_{\vec I,\vec J}$. As such permutation can only change the sign of the determinant, the proposition for balanced configurations  follows immediately from the block structure of the resulting matrix. The sign change is taken into account by
  the factor $\lb -1\rb^{\left|I_k\right|}$ in~(\ref{partZ01}).
  
  The only non-zero Fourier coefficients of $\mathsf a^{[k]},\mathsf b^{[k]},\mathsf c^{[k]},\mathsf d^{[k]}$  are given by (\ref{alphadeltaF}). Therefore, if a configuration  $\lb \vec I,\vec J\rb\in\mathsf{Conf}$ is not proper, then at least one of the factors on the right of 
  (\ref{factordet}) vanishes due to the presence of zero rows or columns in the relevant matrices. \epf
  
  \begin{cor} Fredholm determinant $\tau\lb a\rb$ is given by
  \beq\label{partZ03}
  \tau\lb a\rb=\sum_{\lb\vec I,\vec J\rb\in\mathsf{Conf}_+}\; \prod_{k=1}^{n-2} Z^{I_{k-1},J_{k-1}}_{\;I_{k},J_{k}}\lb \mathcal T^{[k]}\rb.
  \eeq
  \end{cor}
  \pf Another useful consequence of the block structure of the operator $K$ is that  $\operatorname{Tr} K^{2m+1}=0$ for $m\in\mathbb Z_{\ge0}$. This implies that $\operatorname{det}\lb \mathbb 1-K\rb=\operatorname{det}\lb \mathbb 1+K\rb$. It now suffices to combine this symmetry with von Koch's formula~(\ref{vonKoch}) and Proposition~\ref{propfactor}. \epf
  
\begin{figure}
    \centering
    \includegraphics[height=4.2cm]{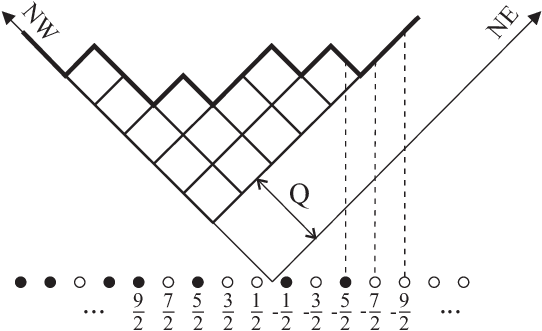}
    \begin{minipage}{0.75\textwidth}
    \caption{\label{figmaya}The correspondence between Maya diagrams and charged partitions; \hspace{\textwidth} here the charge $Q\lb \mathsf m\rb=2$ and the positions of particles and holes are given by $\mathsf p\lb \mathsf m\rb=\lb \frac{13}2,\frac72,\frac32,\frac12\rb$ and $\mathsf h\lb \mathsf m\rb=\lb -\frac52,-\frac12\rb$. }\end{minipage}
    \end{figure}

  Let us now give a combinatorial description of the
  set $\mathsf{Conf}_+$ of proper balanced configurations in terms of Maya diagrams and charged partitions.
  \begin{defin}
  A Maya diagram is a map $\mathsf m:\mathbb Z'\to \left\{-1,1\right\}$ subject to the condition that $\mathsf m\lb p\rb=\pm 1$ for all but finitely many $p\in\mathbb Z'_{\pm}$. The set of all Maya diagrams will be denoted by $\mathbb{M}$.
  \end{defin} 
  A convenient graphical representation of $\mathsf m\in \mathbb M$ is obtained by replacing $-1$'s and
    $1$'s by white and black circles located at the sites of half-integer lattice, see bottom part of Fig.~\ref{figmaya} for an example.
  The white circles in $\mathbb Z'_+$ and black circles in $\mathbb Z'_-$ are referred to as particles and holes in the Dirac sea, which itself corresponds to the  diagram $\mathsf m_0$ defined by $\mathsf m_0\lb\mathbb Z'_{\pm}\rb=\pm1$. An arbitrary diagram is completely determined by a sequence $\mathsf{p}\lb \mathsf m\rb=\lb p_1,\ldots,p_r\rb$ of strictly decreasing positive half-integers   $p_1>\ldots >p_r$ giving the positions of particles, and a sequence  $\mathsf{h}\lb \mathsf m\rb=\lb -q_1,\ldots,-q_s\rb$ of strictly increasing negative half-integers  $-q_1<\ldots <-q_s$ corresponding to the positions of holes. The integer $Q\lb \mathsf m\rb:=\left|\mathsf{p}\lb \mathsf m\rb\right|-\left|\mathsf{h}\lb \mathsf m\rb\right|$ is called the charge of $\mathsf m$. 
  
  Given a configuration $\lb \vec I,\vec J\rb\in\mathsf{Conf}_+$, consider a pair of its multi-indices  $\lb I_k,J_k\rb$ associated to the annulus~$\mathcal A_k$. Recall that the Fourier indices of elements of $I_k$ (and $J_k$) are positive (resp. negative). They can therefore be interpreted as positions of particles and holes of
  $N$ different colors. This yields a bijection between the set of pairs $\lb I_k,J_k\rb$ verifying the balance condition $\left|I_k\right|=\left|J_k\right|$ and
  the set 
  \ben
  \mathbb M^N_0=\left\{\lb \mathsf m^{(1)},\ldots,
    \mathsf m^{(N)}\rb\in\mathbb M^N\left|\,\sum\nolimits_{\alpha=1}^N
    Q\lb \mathsf m^{(\alpha)}\rb=0\right.\right\}
  \ebn
  of $N$-tuples of Maya diagrams with vanishing total charge. We thereby obtain a one-to-one correspondence 
  \ben\mathsf{Conf}_+\cong \underbrace{\mathbb M^N_0\times \ldots \times \mathbb M^N_0}_{n-3\text{ factors}}.
  \ebn
  \begin{defin}
  A charged partition is a pair $\hat Y=\lb Y,Q\rb\in
  \mathbb Y\times \mathbb Z$. The integer $Q$ is called the charge of $\hat Y$.
  \end{defin}
  
  There is a well-known bijection between Maya diagrams and charged partitions, whose construction is illustrated in Fig.~\ref{figmaya}. Given a Maya diagram $\mathsf m\in\mathbb M$, we start far on the north-west axis and draw a segment directed to the south-east above each black circle and a segment directed  north-east above each white circle. The resulting polygonal line defines the outer boundary of the Young diagram $Y$ corresponding to $\mathsf m$. The charge $Q=Q\lb \mathsf m\rb$ of $\hat Y$ is the signed distance between $Y$ and the  north-east axis. In the case $Q\lb \mathsf m\rb=0$, the sequences $\mathsf p\lb \mathsf m\rb$ and $-\mathsf h\lb \mathsf m\rb$ give the Frobenius coordinates of $Y$.
  
  Let us write $N$-tuples  $\lb\hat Y^{(1)},\ldots ,\hat Y^{(N)}\rb$ of charged partitions as
  $\lb \vec Y,\vec Q\rb$, with $\vec Y=\lb Y^{(1)},\ldots , Y^{(N)}\rb\in \mathbb Y^N $ and $\vec Q=\lb Q^{(1)},\ldots , Q^{(N)}\rb\in \mathbb Z^N$. The set of such $N$-tuples with zero total charge can be identified with
  $\mathbb M^N_0\cong  \mathbb Y^N\times\mathfrak{Q}_N$,
  where $\mathfrak Q_N$ denotes the $A_{N-1}$ root lattice:
  \ben
  \mathfrak Q_N:=\left\{\vec Q\in\mathbb Z^N\,\left|\,
  \sum\nolimits_{\alpha=1}^NQ^{(\alpha)}=0\right.\right\}.
  \ebn 
  This suggests to introduce an alternative notation for elementary finite determinant factors in (\ref{partZ03}). For $\left|I_{k-1}\right|=\left|J_{k-1}\right|$ and $\left|I_k\right|=\left|J_k\right|$, we define
  \beq\label{partZ05}
  Z^{\vec Y_{k-1},\vec Q_{k-1}}_{\;\vec Y_{k},\vec Q_{k}}\lb
  \mathcal T^{[k]}\rb:= Z^{I_{k-1},J_{k-1}}_{\;I_{k},J_{k}}\lb \mathcal T^{[k]}\rb,
  \eeq
  where $\lb  \vec Y_{k-1},\vec Q_{k-1}\rb,\lb  \vec Y_{k},\vec Q_{k}\rb \in \mathbb Y^N\times \mathfrak Q_N$ are associated to 
  $N$-tuples of Maya diagrams describing subconfigurations
  $\lb I_{k-1},J_{k-1}\rb$, $\lb I_{k},J_{k}\rb$. In what follows, the two notations are used interchangeably. 
  
  The structure of the expansion of $\tau\lb a\rb$  may now be summarized as follows.
  
  \begin{theo} Fredholm determinant $\tau\lb a\rb$ giving the isomonodromic tau function $\tau_{\mathrm{JMU}}\lb a\rb$ can be written as a combinatorial series
  \beq\label{partZ04}
  \tau\lb a\rb=\sum_{\vec Q_1,\ldots\vec Q_{n-3}\in\mathfrak Q_N}\;
  \sum_{\vec Y_1,\ldots\vec Y_{n-3}\in \mathbb Y^{N}}
  \prod_{k=1}^{n-2}Z^{\vec Y_{k-1},\vec Q_{k-1}}_{\;\vec Y_{k},\vec Q_{k}}\lb
    \mathcal T^{[k]}\rb,
  \eeq
  where  $Z^{\vec Y_{k-1},\vec Q_{k-1}}_{\;\vec Y_{k},\vec Q_{k}}\lb
      \mathcal T^{[k]}\rb$ are expressed by (\ref{partZ05}), (\ref{partZ06}) in terms of matrix elements of $\,3$-point  Plemelj operators in the Fourier basis.
  \end{theo}
  \begin{eg} Let us outline simplifications to the above scheme in the case $N=2$, $n=4$ corresponding to the Painlev\'e VI equation. Here a configuration $\lb \vec I,\vec J\rb\in\mathsf{Conf}_+$ is given by a single pair $\lb I,J\rb$ of multi-indices whose structure may be described as follows: $I$ (and $J$) encode the positions of particles (resp. holes) of two colors $\{+,-\}$, and the total number of particles in $I$ coincides with the total number of holes in $J$. Relative positions of particles and holes of each color are described by two Young diagrams $Y_+,Y_-\in\mathbb Y$. The vectors  $\lb Q_+,Q_-\rb\in\mathfrak Q_2$ of the charge lattice are labeled by a single integer $n=Q_+=-Q_-\in \mathbb Z$.
    In the notation of Subsection~\ref{subsec4pts}, the series (\ref{partZ03}) can be
  rewritten as
  \beq
  \begin{gathered}
  \begin{aligned}
  \tau\lb t\rb=&\,\sum_{n\in\mathbb Z}\;
  \sum_{\substack{\mathsf p_+,\mathsf p_-\in 2^{\mathbb Z'_+};\,
  \mathsf h_+,\mathsf h_-\in 2^{\mathbb Z'_-}\\ 
  \left|\mathsf p_+\right|-\left|\mathsf h_+\right|=\left|\mathsf h_-\right|-\left|\mathsf p_-\right|=n}}
    \lb -1\rb^{\left|\mathsf p_+\right|+
    \left|\mathsf p_-\right|}
    \operatorname{det}\,{\mathsf a}^{\mathsf p_+,\mathsf p_-}_{\mathsf h_+,\mathsf h_-}
    \operatorname{det}\,{ \mathsf d}^{\,\mathsf h_+,\mathsf h_-}_{\,\mathsf p_+,\mathsf p_-}=\\
  =&\,\sum_{n\in\mathbb Z}\;
  \sum_{Y_+,Y_-\in\mathbb Y}Z_{Y_+,Y_-,n}\lb \mathcal T^{[L]}\rb
  Z^{Y_+,Y_-,n}\lb \mathcal T^{[R]}\rb,
  \end{aligned}
  \end{gathered}
  \eeq
  where $Z_{Y_+,Y_-,n}\lb \mathcal T^{[L]}\rb= \lb -1\rb^{\left|\mathsf p_+\right|+
      \left|\mathsf p_-\right|}\operatorname{det}\,{ \mathsf d}^{\,\mathsf h_+,\mathsf h_-}_{\,\mathsf p_+,\mathsf p_-}$ and $Z^{Y_+,Y_-,n}\lb \mathcal T^{[R]}\rb =\operatorname{det}\,{\mathsf a}^{\mathsf p_+,\mathsf p_-}_{\mathsf h_+,\mathsf h_-}$. In these equations, the particle/hole positions $( \mathsf p _+,\mathsf h_+)$ and  $( \mathsf p_-,\mathsf h_-)$ for the 1st and 2nd color are identified with a pair of Maya diagrams, subsequently interpreted as charged
  partitions $\lb Y_+,n\rb$ and $( Y_-,-n)$.
  \end{eg}
  \begin{rmk}
  Describing the elements of $\mathsf{Conf}_+$ in terms of 
  $N$-tuples of Young diagrams and vectors of the $A_{N-1}$~root lattice is inspired by their appearance in the four-dimensional $\mathcal N=2$ supersymmetric linear quiver gauge theories. Combinatorial structure of the dual partition functions of such theories \cite{Nekrasov,NO} coincides with that of (\ref{partZ04}). These partition functions can in fact be obtained from our construction or its higher genus/irregu\-lar extensions by imposing additional spectral constraints on monodromy.  
  It will shortly become clear that the multiple sum over $\mathfrak Q_N$ is responsible for a Fourier transform structure of the isomonodromic tau functions. This structure was discovered in  \cite{GIL12,ILT13} for  Painlev\'e VI, understood for $N=2$ and arbitrary number of punctures within the framework of  Liouville conformal field theory \cite{ILTe}, and conjectured to appear in higher rank in~\cite{Gav}. It might be interesting to mention the appearance of a possibly related structure in the study of topological string partition functions \cite{GHM,BGT}.
  \end{rmk}
 
 \section{Rank two case\label{secranktwo}}
 For $N=2$, the elementary $3$-point RHPs can be solved in terms of Gauss hypergeometric functions so that Fredholm determinant representation
 (\ref{fredholmrep})  becomes completely explicit. Being rewritten in Fourier components, the blocks of $K$ may be reduced to single infinite Cauchy matrices acting in $\ell^2\lb \Zb\rb$. We are going to use this observation to calculate the building blocks $Z^{\vec Y_{k-1},\vec Q_{k-1}}_{\;\vec Y_{k},\vec Q_{k}}\lb
     \mathcal T^{[k]}\rb$ of principal minors of $K$ in terms of monodromy data, and derive thereby a multivariate series representation for the isomonodromic tau function of the Garnier system.

  \subsection{Gauss and Cauchy in rank $2$\label{subsecGC}}
  The form of the Fuchsian system (\ref{fuchsys}) is preserved by the following non-constant scalar gauge transformation of the fundamental solution and coefficient matrices:
  \ben
  \begin{gathered}
  \Phi\lb z\rb\mapsto \hat{\Phi}\lb z\rb
  \prod_{l=0}^{n-2}\lb z-a_l\rb^{\kappa_l},\\
    A_l\mapsto\hat A_l+\kappa_l\mathbb 1, \qquad l=0,\ldots,n-2.
    \end{gathered}
  \ebn
  Under this transformation, the monodromy matrices $M_l$ are multiplied by  $e^{-2\pi i \kappa_l}$, and the associated Jimbo-Miwa-Ueno tau function transforms as
  \ben
  \tau_{\mathrm{JMU}}\lb a\rb\mapsto \hat{\tau}_{\mathrm{JMU}}\lb a\rb 
  \prod_{0\le k<l\le n-2}\lb a_l-a_k\rb^{-N\kappa_k\kappa_l+\kappa_k\operatorname{Tr}\Theta_l+
  \kappa_l\operatorname{Tr}\Theta_k}.
  \ebn
  The freedom in the choice of $\kappa_0,\ldots,\kappa_{n-2}$ allows to make the following assumption.
  \begin{ass}\label{assrank2} One of the eigenvalues of each of the matrices $\Theta_0,\ldots,\Theta_{n-2}$ is equal to $0$.
  \end{ass} 
   \noindent This involves no loss in generality and means in particular that the ranks $\mathfrak r^{[k]}$ of the coefficient matrices $A_1^{[k]}$ in the auxiliary $3$-point Fuchsian systems (\ref{FS3point}) are at most $N-1$.
  
  For $\mathfrak r^{[k]}=1$, the factor $Z^{I_{k-1},J_{k-1}}_{\; I_{k},J_{k}}\lb \mathcal T^{[k]}\rb$
  in (\ref{partZ04}) can be computed in explicit form. In this case the sums such as (\ref{tensorA1k}) or (\ref{alphadeltacauchy}) contain only one term, and the index $r$ can therefore be omitted. The matrix $A_1^{[k]}\in \Cb^{N\times N}$  may be written as
    \ben
    a_{k}A_1^{[k]}=-u^{[k]}\otimes v^{[k]}.
    \ebn
    The crucial observation is  that the blocks (\ref{alphadeltacauchy}) are now given by single Cauchy matrices conjugated by diagonal factors (instead of being a sum of such matrices). In order to put this to a good use,  let us introduce two complex sequences $\bigl( x^{[k]}_{\imath}\bigr)_{\imath\in I_{k-1}\sqcup J_k}$, $\bigl( y^{[k]}_{\jmath}\bigr)_{\jmath\in J_{k-1}\sqcup I_k}$ of the same
    finite length $\left|I_{k-1}\right|+\left|I_{k}\right|$. Their elements are defined by shifted particle/hole positions:
      \begin{subequations}
      \label{phcauchy}
      \begin{align}
      x^{[k]}_{\imath}:=&\,\begin{cases}
      p+\sigma_{k-1,\alpha},\qquad & \imath\equiv \lb p,\alpha\rb\in I_{k-1},\\
      -p+\sigma_{k,\alpha},\qquad & \imath\equiv \lb -p,\alpha\rb\in J_k,
      \end{cases}\\
      y^{[k]}_{\jmath}:=&\,\begin{cases}
      -q+\sigma_{k-1,\beta},\quad\;\; & \jmath\equiv\lb -q,\beta\rb\in J_{k-1},\\
      q+\sigma_{k,\beta},\;\;\qquad & \jmath\equiv \lb q,\beta\rb\in I_k.
      \end{cases}
      \end{align}
      \end{subequations}

  \begin{lemma}\label{lemmafacca} If $\;\mathfrak r^{[k]}=1$, then
  $Z^{I_{k-1},J_{k-1}}_{\; I_{k},J_{k}}\lb
    \mathcal T^{[k]}\rb$ can be written as
  \beq\label{cauchyN2}
  \begin{gathered}
  \begin{aligned}
  Z^{I_{k-1},J_{k-1}}_{\; I_{k},J_{k}}\lb
  \mathcal T^{[k]}\rb=&\, \pm\!\!\!
  \prod_{\lb p,\alpha\rb\in I_{k-1}}\lb\psi^{[k]}\rb^{p;\alpha}
   \!\!\!\!\!
     \prod_{\lb -p,\alpha\rb\in J_{k-1}}\lb\bar\psi^{[k]}\rb_{p;\alpha}
  \prod_{\lb -p,\alpha\rb\in J_{k}}\lb\varphi^{[k]}\rb^{-p;\alpha}
  \!\!\!\!\!
    \prod_{\lb p,\alpha\rb\in I_{k}}\lb\bar \varphi^{[k]}\rb_{-p;\alpha}  \times\\
  &\times \frac{\ds\prod_{\imath,\jmath\in I_{k-1}\sqcup J_k; \imath<\jmath}\lb x^{[k]}_{\imath}-x^{[k]}_{\jmath}\rb
    \prod_{\imath,\jmath\in J_{k-1}\sqcup I_k;\imath<\jmath}\lb y^{[k]}_{\jmath}-y^{[k]}_{\imath}\rb}{\ds\prod_{\imath\in I_{k-1}\sqcup J_k}\prod_{\jmath \in J_{k-1}\sqcup I_k}\lb x^{[k]}_{\imath}-y^{[k]}_{\jmath}\rb}.
  \end{aligned}
  \end{gathered}
  \eeq
  \end{lemma}
  \pf The diagonal factors in (\ref{alphadeltacauchy}) produce the
  first line of (\ref{cauchyN2}). It remains to compute the determinant
  \beq\label{cauchyN02}
  \operatorname{det}\lb\begin{array}{cc}
  \left[\ds\frac{1}{p+\sigma_{k-1,\alpha}+q-\sigma_{k-1,\beta}}\right]^{\;\;\lb p,\alpha\rb\in I_{k-1}}_{\lb -q,\beta\rb\in J_{k-1}}&
  \left[\ds\frac{1}{p+\sigma_{k-1,\alpha}-q-\sigma_{k,\beta}}\right]^{
  \lb p,\alpha\rb\in I_{k-1}}_{\lb q,\beta\rb\in I_{k}}\vspace{0.2cm}\\
    \left[\ds\frac{1}{-p+\sigma_{k,\alpha}+q-
    \sigma_{k-1,\beta}}\right]^{\lb -p,\alpha\rb\in J_{k}}_{\lb -q,\beta\rb\in J_{k-1}}&
    \left[\ds\frac{1}{
    -p+\sigma_{k,\alpha}-q-\sigma_{k,\beta}}\right]^{
    \lb -p,\alpha\rb\in J_{k}}_{\;\;\lb q,\beta\rb\in I_{k}}
  \end{array}\rb,
  \eeq
  which already includes the sign $\lb -1\rb^{\left|I_k\right|}$ in (\ref{partZ01}). The $\pm$ sign in (\ref{cauchyN2}) depends on the ordering of rows and columns of the determinant (\ref{partZ01}). This ambiguity does not play any role as the relevant sign  appears twice in the full product (\ref{factordet}). 
  
  On the other hand, the notation introduced above allows to rewrite (\ref{cauchyN02}) as a $\lb\left|I_{k-1}\right|+
  \left|I_{k}\right|\rb \times \lb\left|I_{k-1}\right|+
    \left|I_{k}\right|\rb $ Cauchy determinant
    \ben\operatorname{det}\lb\frac{1}{x_{\imath}^{[k]}-y_{\jmath}^{[k]}}\rb\biggl._{\jmath \in J_{k-1}\sqcup I_k}^{\imath\in I_{k-1}\sqcup J_k},
    \ebn
  and the factorized expression (\ref{cauchyN2}) easily follows.
   \epf

  We now restrict ourselves to the case $N=2$, where the condition $\mathfrak r^{[1]}=\ldots=\mathfrak r^{[n-2]}=1$ does not lead to restrictions on monodromy. Let us start by preparing a suitable notation. 
  \begin{itemize}
  \item The color indices will take values
  in the set $\{+,-\}$ and will be denoted by $\epsilon,\epsilon'$. 
  \item Recall that the spectrum of $A_1^{[k]}$ coincides with that of $\Theta_{k}$. According to Assumption~\ref{assrank2}, the diagonal matrix 
  $\Theta_k$ has a zero eigenvalue for ${k=0,\ldots, n-2}$.  Its second eigenvalue will be denoted by
  $-2\theta_{k}$. Obviously, there is a relation
  \ben
  2\theta_{k}a_{k}=v^{[k]}\cdot u^{[k]},\qquad k=1,\ldots, n-2,
  \ebn
  where $v\cdot u=v_+u_++v_-u_-$ is the standard bilinear form 
  on $\Cb^2$. The eigenvalues of the remaining local monodromy exponent $\Theta_{n-1}$ may be parameterized as
  \ben
  \theta_{n-1,\epsilon}=\sum_{k=0}^{n-2}\theta_k+\epsilon
  \theta_{n-1},\qquad
  \epsilon=\pm.  
  \ebn
  \item Also, the spectra of $A^{[k]}_0$ and $A^{[k]}_{\infty}=-A^{[k]}_0-A^{[k]}_1$ coincide with the spectra
  of $\mathfrak S_{k-1}$ and $-\mathfrak S_{k}$. Since furthermore $\operatorname{Tr}
  \mathfrak S_{k}=\sum_{j=0}^{k}\operatorname{Tr}\Theta_j$, we may  write the eigenvalues of $\mathfrak S_k$ as
  \beq\label{sigmasN2}
  \sigma_{k,\epsilon}=
  -\sum_{j=0}^{k}\theta_j+\epsilon\sigma_k,\qquad \qquad
  \epsilon=\pm,\qquad  k=0,\ldots,n-2,
  \eeq
  where $\sigma_0\equiv\theta_0$ and $\sigma_{n-2}\equiv -\theta_{n-1}$.
  \end{itemize}
  The non-resonancy of monodromy exponents and Assumption~\ref{asssp} imply that
    \begin{alignat*}{2}
    &2\theta_k\notin \mathbb Z\backslash\left\{0\right\},\qquad && k=0,\ldots,n-1,\\
    &\left|\Re \sigma_k\right|\leq\frac12,\quad \sigma_k\neq\pm\frac12,\qquad\qquad && k=1,\ldots,n-3.
    \end{alignat*}
  To simplify the exposition, we add to this extra genericity conditions.
  \begin{ass}
  For $k=1,\ldots,n-2$, we have
  \ben
  \sigma_{k-1}+\sigma_{k}\pm\theta_{k}\notin\mathbb Z,\qquad
  \sigma_{k-1}-\sigma_{k}\pm\theta_{k}\notin \mathbb Z.
  \ebn   
  It is also assumed that $\sigma_k\ne 0$ for $k=0,\ldots,n-2$.
  \end{ass}  
  Let us introduce the space
   \ben
    \mathcal M_{\Theta}=\left\{\left[M_0,\ldots, M_{n-1}\right]\in
    \lb\mathrm{GL}\lb N,\Cb\rb\rb^n/\sim\,\bigl|\,
    M_0\ldots M_{n-1}=\mathbb 1,\; M_k\in[e^{2\pi i\Theta_k}]\text{ for }k=0,\ldots,n-1 \right\}
    \ebn
  of  conjugacy classes of monodromy representations of the fundamental group with fixed local exponents.
  The parameters $\sigma_1,\ldots,\sigma_{n-3}$ are associated to annuli
  $\mathcal A_1,\ldots,\mathcal A_{n-3}$ and provide $n-3$ local coordinates on $\mathcal M_{\Theta}$  (that is, exactly one half of 
  ${\operatorname{dim}\mathcal M_{\Theta}=2n-6}$).  The remaining $n-3$ coordinates will be defined below.
  
  Our task is now to find the $3$-point solution $\Psi^{[k]}$ explicitly. The freedom in the choice of its normalization allows to pick any representative in the conjugacy class
  $\bigl[A_0^{[k]},A_1^{[k]}\bigr]$ for the construction of the $3$-point Fuchsian system (\ref{FS3point}). An important feature of the $N=2$ case is that this conjugacy class is completely fixed by local monodromy exponents $\mathfrak S_{k-1}$, $\Theta_{k}$ and $-\mathfrak S_{k}$. We can set in particular
  \ben
  A_0^{[k]}=\operatorname{diag}\left\{\sigma_{k-1,+},
  \sigma_{k-1,-}\right\},\qquad
  a_{k}A_1^{[k]}=-u^{[k]}\otimes v^{[k]},
  \ebn
   with $\sigma_{k-1,\pm}$ parameterized as in (\ref{sigmasN2}) and
   \ben
   u^{[k]}_{\pm}=\frac{\lb \sigma_{k-1}\pm\theta_{k}\rb^2-\sigma_{k}^2}{2\sigma_{k-1}}\,a_k,\qquad v^{[k]}_{\pm}=\pm1.
   \ebn
   
   As in Subsection~\ref{subsec4pts}, one may first construct the solution $\tilde{\Phi}^{[k]}$ of the rescaled system
   \beq\label{AUX00}
   \partial_z \tilde{\Phi}^{[k]}=\tilde{\Phi}^{[k]}\lb\frac{A_0^{[k]}}{z}+
   \frac{A_1^{[k]}}{z-1}\rb,
   \eeq
   having the same monodromy around $0$, $1$, $\infty$ as the solution ${\Phi}^{[k]}$ of the original
   system (\ref{FS3point}) has around $0$, $a_{k}$ and $\infty$. To write it explicitly in terms of the Gauss hypergeometric function ${}_2F_1\biggl[\begin{array}{c} 
         a,b \vspace{-0.1cm} \\ c \end{array};z\biggr]$, we introduce a convenient notation,
   \beq\label{chiphi}      
   \begin{aligned}
   \chi\biggl[\begin{array}{c} 
   \theta_2 \vspace{-0.05cm} \\ \theta_1\quad\theta_3 \end{array};z\biggr]:=&\,{}_2F_1\biggl[\begin{array}{c}
   \theta_1+\theta_2+\theta_3,\theta_1+\theta_2-\theta_3 \\
   2\theta_1 
   \end{array};z\biggr],\\
      \phi\biggl[\begin{array}{c} 
      \theta_2 \vspace{-0.05cm} \\ \theta_1\quad\theta_3 \end{array};z\biggr]:=&\,\frac{\theta_3^2-\lb \theta_1+\theta_2\rb^2}{2\theta_1\lb
      1+2\theta_1\rb}\,z\,{}_2F_1\biggl[\begin{array}{c}
      1+\theta_1+\theta_2+\theta_3,
      1+\theta_1+\theta_2-\theta_3 \\
      2+2\theta_1 
      \end{array};z\biggr].
   \end{aligned}
   \eeq      
   The solution of (\ref{AUX00}) can then be written as
   \beq\label{3pRHP01}
   \tilde{\Phi}^{[k]}\lb z\rb=S_{k-1}\lb -z\rb^{\mathfrak S_{k-1}}
      \tilde{\Psi}^{[k]}_{\mathrm{in}}\lb z\rb,
   \eeq
    where $S_{k-1}$ is a constant connection matrix encoding the monodromy (cf (\ref{jump3point})), and $\tilde{\Psi}^{[k]}_{\mathrm{in}}$ is given by
   \beq\label{3pRHP02} 
   \begin{aligned}
   \lb\tilde{\Psi}^{[k]}_{\mathrm{in}}\rb_{\pm\pm}\lb z\rb
   =&\,\chi\biggl[\begin{array}{c} 
   \quad\theta_k \vspace{-0.05cm} \\ \pm\sigma_{k-1}\quad\sigma_k \end{array};z\biggr], \\
   \lb\tilde{\Psi}^{[k]}_{\mathrm{in}}\rb_{\pm\mp}\lb z\rb
   =&\,\phi\biggl[\begin{array}{c} 
   \quad\theta_k \vspace{-0.05cm} \\ \pm\sigma_{k-1}\quad\sigma_k \end{array};z\biggr].   
   \end{aligned}    
   \eeq     
   It follows that $\Phi^{[k]}\lb z\rb=\tilde{\Phi}^{[k]}\lb \frac{z}{a_{k}}\rb$ and
   \begin{subequations}
   \label{sol2pRHPin}
   \beq\label{sol2pRHPinA}
   \Psi^{[k]}_+\lb z\rb=a_{k}^{-\mathfrak S_{k-1}}\,\tilde{\Psi}^{[k]}_{\mathrm{in}}\lb \frac{z}{a_{k}}\rb, \qquad\qquad z\in \mathcal C^{[k]}_{\mathrm{in}}.
   \eeq
   Let us also note that $\operatorname{det}\tilde{\Phi}^{[k]}\lb z\rb=
   \operatorname{const}\cdot \lb -z\rb^{\operatorname{Tr} A_0^{[k]}}\lb 1-z\rb^{\operatorname{Tr} A_1^{[k]}}$ implies that $\operatorname{det}\tilde{\Psi}^{[k]}_{\mathrm{in}}\lb z\rb=\lb 1-z\rb^{-2\theta_{k}}$, which in turn yields a simple representation for the inverse matrix
     \beq\label{sol2pRHPinB} 
    {\Psi^{[k]}_+\lb z\rb}^{-1}=\lb 1-\frac{z}{a_{k}}\rb^{2\theta_{k}}
    \lb\begin{array}{rr}
    \lb\tilde{\Psi}^{[k]}_{\mathrm{in}}\rb_{--}\lb\frac{z}{a_{k}}\rb
    & -\lb\tilde{\Psi}^{[k]}_{\mathrm{in}}\rb_{+-}
    \lb\frac{z}{a_{k}}\rb\vspace{0.1cm}\\
    -\lb\tilde{\Psi}^{[k]}_{\mathrm{in}}\rb_{-+}
       \lb\frac{z}{a_{k}}\rb &
    \lb\tilde{\Psi}^{[k]}_{\mathrm{in}}\rb_{++}\lb\frac{z}{a_{k}}\rb
    \end{array}\rb
    a_{k}^{\mathfrak S_{k-1}},\qquad
    z\in \mathcal C^{[k]}_{\mathrm{in}}.  
      \eeq    
   \end{subequations}
   
   The equations (\ref{3pRHP01})--(\ref{3pRHP02}) are adapted for the description of local behavior of $\Psi^{[k]}\lb z\rb$   inside the disk around~$0$ bounded by the circle $\mathcal C^{[k]}_{\mathrm{in}}$, cf left part of Fig.~\ref{figgammahat}. To calculate $\Psi^{[k]}_+\lb z\rb$ inside the disk around $\infty$ bounded by $\mathcal C^{[k]}_{\mathrm{out}}$, let us first rewrite (\ref{3pRHP01}) using the well-known $_2F_1$ transformation formulas. One can show that
   \beq\label{3pRHP03}
   \tilde\Phi^{[k]}\lb z\rb=S_{k-1} C^{[k]}_{\infty}\lb -z\rb^{\mathfrak S_{k}}
   \tilde\Psi^{[k]}_{\mathrm{out}}\lb z\rb G_{\infty}^{[k]},
   \eeq
   where
   \beq\label{3pRHP03bis}   
   \begin{gathered}   
      \begin{aligned}
      \lb\tilde{\Psi}^{[k]}_{\mathrm{out}}\rb_{\pm\pm}\lb z\rb=&\,\chi\biggl[\begin{array}{c} 
         \quad\theta_k \vspace{-0.05cm} \\ 
     \mp\sigma_{k}\quad\sigma_{k-1} \end{array};z^{-1}\biggr],\\
      \lb\tilde{\Psi}^{[k]}_{\mathrm{out}}\rb_{\pm\mp}\lb z\rb=&\,
    \phi\biggl[\begin{array}{c} 
             \quad\theta_k \vspace{-0.05cm} \\ 
         \mp\sigma_{k}\quad\sigma_{k-1} \end{array};z^{-1}\biggr],        
      \end{aligned}
      \end{gathered}
      \eeq    
   and
   \begin{gather}
   \label{3pRHP04}  
   G_{\infty}^{[k]}=\frac{1}{2\sigma_{k}}
   \lb\begin{array}{rr}
   -\theta_{k}+\sigma_{k-1}+\sigma_{k} &
   \theta_{k}+\sigma_{k-1}-\sigma_{k} \\
   -\theta_{k}+\sigma_{k-1}-\sigma_{k} &
   \theta_{k}+\sigma_{k-1}+\sigma_{k}
   \end{array}\rb,\\
   \label{3pRHP05}  
   C^{[k]}_{\infty}=\lb\begin{array}{rr}
   \ds\frac{\Gamma\lb2\sigma_{k-1}\rb\Gamma\lb 1+2\sigma_{k}\rb}{
   \Gamma\lb1+\sigma_{k-1}+\sigma_{k}-\theta_{k}\rb
   \Gamma\lb \sigma_{k-1}+\sigma_{k}+\theta_{k}\rb} & 
    \ds -\frac{\Gamma\lb2\sigma_{k-1}\rb\Gamma\lb 1-2\sigma_{k}\rb}{
      \Gamma\lb1+\sigma_{k-1}-\sigma_{k}-\theta_{k}\rb
      \Gamma\lb \sigma_{k-1}-\sigma_{k}+\theta_{k}\rb}
      \vspace{0.2cm} \\
   \ds -\frac{\Gamma\lb -2\sigma_{k-1}\rb\Gamma\lb1+2\sigma_{k}\rb}{
   \Gamma\lb1-\sigma_{k-1}+\sigma_{k}-\theta_{k}\rb
   \Gamma\lb \theta_{k}-\sigma_{k-1}+\sigma_{k}\rb} & 
   \ds\frac{\Gamma\lb -2\sigma_{k-1}\rb\Gamma\lb 1-2\sigma_{k}\rb}{
      \Gamma\lb1-\sigma_{k-1}-\sigma_{k}-\theta_{k}\rb
      \Gamma\lb \theta_{k}-\sigma_{k-1}-\sigma_{k}\rb}        
   \end{array}\rb.
   \end{gather}
   As a consequence,
   \begin{subequations}
   \beq
   \Psi_+^{[k]}\lb z\rb=D^{[k]}_{\infty}a_{k}^{-\mathfrak S_{k}}\,
      \tilde\Psi^{[k]}_{\mathrm{out}}\lb \frac{z}{a_{k}}\rb G_{\infty}^{[k]},\qquad\qquad z\in \mathcal C^{[k]}_{\mathrm{out}},
   \eeq
   where $D^{[k]}_{\infty}=\operatorname{diag}\left\{
   d^{[k]}_{\infty,+},d^{[k]}_{\infty,-}\right\}$ is a diagonal matrix expressed in terms of monodromy as 
   \ben D^{[k]}_{\infty}=
   S_{k}^{-1}S_{k-1} C^{[k]}_{\infty}.
   \ebn
   Analogously to (\ref{sol2pRHPinB}), it may be shown that for
   $z\in \mathcal C^{[k]}_{\mathrm{out}}$
   \beq
      {\Psi^{[k]}_+\lb z\rb}^{-1}=\lb 1-\frac{a_{k}}z\rb^{2\theta_{k}}\!\!
      {G^{[k]}_{\infty}}^{-1}
      \lb\begin{array}{rr}
      \lb\tilde{\Psi}^{[k]}_{\mathrm{out}}\rb_{--}
      \lb\frac{z}{a_{k}}\rb
      & -\lb\tilde{\Psi}^{[k]}_{\mathrm{out}}\rb_{+-}
      \lb\frac{z}{a_{k}}\rb\vspace{0.1cm}\\
      -\lb\tilde{\Psi}^{[k]}_{\mathrm{out}}\rb_{-+}
         \lb\frac{z}{a_{k}}\rb &
      \lb\tilde{\Psi}^{[k]}_{\mathrm{out}}\rb_{++}
      \lb\frac{z}{a_{k}}\rb
      \end{array}\rb
      a_{k}^{\mathfrak S_{k}}{D^{[k]}_{\infty}}^{-1} .
   \eeq
   \end{subequations}
  
  We now have at our disposal all quantities that are necessary to 
  compute the explicit form of the integral kernels of $\mathsf a^{[k]}$,
  $\mathsf b^{[k]}$, $\mathsf c^{[k]}$, $\mathsf d^{[k]}$ in the Fredholm determinant representation (\ref{fredholmrep}) of the Jimbo-Miwa-Ueno tau function, as well as of diagonal factors 
  $\psi^{[k]}$, $\varphi^{[k]}$, $\bar\psi^{[k]}$, $\bar\varphi^{[k]}$
  in the building blocks (\ref{cauchyN2}) of its combinatorial expansion (\ref{partZ04}). 
  
    \begin{lemma}\label{abcd2F1}
    For $N=2$, the integral kernels (\ref{alphadeltaF}) can be expressed as
    \begin{subequations}
    \begin{alignat}{3}
    \mathsf{a}^{[k]}\lb z,z'\rb=&\quad
    a_{k}^{-\mathfrak S_{k-1}}&&
     \frac{\lb 1-\frac{z'}{a_{k}}\rb^{2\theta_{k}}
      \lb\begin{array}{cc}
      K_{++}\lb z\rb & K_{+-}\lb z\rb \\
      K_{-+}\lb z\rb & K_{--}\lb z\rb
      \end{array}\rb
       \lb\begin{array}{rr}
          K_{--}\lb z'\rb & \!\!\!-K_{+-}\lb z'\rb \\
          \!\!\!-K_{-+}\lb z'\rb & K_{++}\lb z'\rb
          \end{array}\rb
          -\mathbb{1}}{z-z'}\,a_{k}^{\mathfrak S_{k-1}},\\
    \mathsf{b}^{[k]}\lb z,z'\rb=&\, -      
    a_{k}^{-\mathfrak S_{k-1}}\,&&
         \frac{\lb 1-\frac{a_{k}}{z'}\rb^{2\theta_{k}}
          \lb\begin{array}{cc}
          K_{++}\lb z\rb & K_{+-}\lb z\rb \\
          K_{-+}\lb z\rb & K_{--}\lb z\rb
          \end{array}\rb {G_{\infty}^{[k]}}^{-1}
           \lb\begin{array}{rr}
              \bar K_{--}\lb z'\rb & \!\!\!-\bar K_{+-}\lb z'\rb \\
              \!\!\!-\bar K_{-+}\lb z'\rb & \bar K_{++}\lb z'\rb
              \end{array}\rb}{z-z'}\,a_{k}^{\mathfrak S_{k}}{D_{\infty}^{[k]}}^{-1},\\
    \mathsf{c}^{[k]}\lb z,z'\rb=&\; {D_{\infty}^{[k]}} a_{k}^{-\mathfrak S_{k}}\,
         &&\frac{\lb 1-\frac{z'}{a_{k}}\rb^{2\theta_{k}}
          \lb\begin{array}{cc}
          \bar K_{++}\lb z\rb & \bar K_{+-}\lb z\rb \\
          \bar K_{-+}\lb z\rb & \bar K_{--}\lb z\rb
          \end{array}\rb {G_{\infty}^{[k]}}
           \lb\begin{array}{rr}
              K_{--}\lb z'\rb & \!\!\!-K_{+-}\lb z'\rb \\
              \!\!\!-K_{-+}\lb z'\rb & K_{++}\lb z'\rb
              \end{array}\rb
              }{z-z'}\,a_{k}^{\mathfrak S_{k-1}} ,\\
   \mathsf{d}^{[k]}\lb z,z'\rb=&\; {D_{\infty}^{[k]}} a_{k}^{-\mathfrak S_{k}}\,  &&    \frac{
   \mathbb{1}-\lb 1-\frac{a_{k}}{z'}\rb^{2\theta_{k}}
         \lb\begin{array}{cc}
         \bar K_{++}\lb z\rb & \bar K_{+-}\lb z\rb \\
         \bar K_{-+}\lb z\rb & \bar K_{--}\lb z\rb
         \end{array}\rb
          \lb\begin{array}{rr}
             \bar K_{--}\lb z'\rb & \!\!\!-\bar K_{+-}\lb z'\rb \\
             \!\!\!-\bar K_{-+}\lb z'\rb & \bar K_{++}\lb z'\rb
             \end{array}\rb
             }{z-z'} \,a_{k}^{\mathfrak S_{k}}{D_{\infty}^{[k]}}^{-1},                   
    \end{alignat}
    \end{subequations}
    where we introduced a shorthand notation
    $K\lb z\rb =\tilde\Psi^{[k]}_{\mathrm{in}}\lb\frac{z}{a_{k}}\rb$,
    $\bar K\lb z\rb =\tilde\Psi^{[k]}_{\mathrm{out}}\lb\frac{z}{a_{k}}\rb$; the matrices
    $\tilde\Psi^{[k]}_{\mathrm{in,out}}\lb z\rb$  and ${G_{\infty}^{[k]}}$  are defined by
    (\ref{3pRHP02}), (\ref{3pRHP03}) and (\ref{3pRHP04}).
    \end{lemma}
    \pf Straightforward substitution. \epf
  
  \begin{lemma}
  Under genericity assumptions on parameters formulated above, the  Fourier coefficients which appear in (\ref{auxfactor03}) are given by
  \begin{subequations}
  \label{diagfactors}
  \begin{align}
  \label{diagfactors01}
  \bigl(\psi^{[k]}\bigr){\bigl.}^{p;\epsilon}=&\, \frac{\prod_{\epsilon'=\pm}\lb \theta_{k}+\epsilon\sigma_{k-1}+\epsilon'\sigma_{k} \rb_{p+\frac12}
    }{\lb p-\frac12\rb! \,\lb 2\epsilon\sigma_{k-1}\rb_{p+\frac12}}\,a_{k}^{\sum_{j=0}^{k-1}
    \theta_k-\epsilon\sigma_{k-1}-\lb p-\frac12\rb}\lb
    -\epsilon\rb,\\
    \bigl(\bar \psi^{[k]}\bigr){\bigl.}_{p;\epsilon}=&\,
    \frac{\prod_{\epsilon'=\pm}\lb1- \theta_{k}-\epsilon\sigma_{k-1}+\epsilon'\sigma_{k} \rb_{p-\frac12}}{\lb p-\frac12\rb!\,\lb 1-2\epsilon\sigma_{k-1}\rb_{p-\frac12}}\,
    a_{k}^{-\sum_{j=0}^{k-1}
        \theta_k+\epsilon\sigma_{k-1}-\lb p+\frac12\rb} \lb -\epsilon\rb,\\
  \label{diag14c}
  \bigl(\varphi^{[k]}\bigr){\bigl.}^{-p;\epsilon}=
  &\,
  \frac{\prod_{\epsilon'=\pm}
  \lb\theta_{k}+\epsilon'\sigma_{k-1}-\epsilon
  \sigma_{k}\rb_{p+\frac12}
   }{\lb p-\frac12\rb!\,\lb -2\epsilon\sigma_{k}\rb_{p+\frac12}}
    \,a_{k}^{\sum_{j=0}^{k}
            \theta_k-\epsilon\sigma_{k}+\lb p+\frac12\rb}
    d_{\infty,\epsilon}^{[k]}\epsilon,\\
    \label{diag14d}
    \bigl(\bar \varphi^{[k]}\bigr){\bigl.}_{-p;\epsilon}=&\,
    \frac{\prod_{\epsilon'=\pm}\lb1- \theta_{k}+\epsilon'\sigma_{k-1}+\epsilon\sigma_{k} \rb_{p-\frac12}}{\lb p-\frac12\rb!\,\lb 1+2\epsilon\sigma_{k}\rb_{p-\frac12}}\,
    a_{k}^{-\sum_{j=0}^{k}
                \theta_k+\epsilon\sigma_{k}+\lb p-\frac12\rb}
        {d_{\infty,\epsilon}^{[k]}}^{-1}\epsilon ,
  \end{align}
  \end{subequations}
  where $\epsilon=\pm$ and $\lb c\rb_l:=\ds\frac{\Gamma\lb c+l\rb}{\Gamma\lb c\rb}$ denotes the Pochhammer symbol.
  \end{lemma}
  \pf
  From the first equation in (\ref{auxfactor03a}), the representation (\ref{sol2pRHPinA}) for $\Psi_+^{[k]}\lb z\rb$ on $\mathcal C^{[k]}_{\mathrm{in}}$, and hypergeometric contiguity relations such as
  \ben
  _2F_1\biggl[\begin{array}{c}
  a,b \\ c\end{array};z\biggr]+\lb z-1\rb\,{}
  _2F_1\biggl[\begin{array}{c}
    a+1,b+1 \\ c+1\end{array};z\biggr]=
    \frac{\lb c-a\rb\lb c-b\rb}{c\lb c+1\rb}\, z\,{}
     _2F_1\biggl[\begin{array}{c}
        a+1,b+1 \\ c+2\end{array};z\biggr],
  \ebn
  it follows that
  \ben
  \sum_{p\in\mathbb Z_+'}\bigl(\psi^{[k]}\bigr){\bigl.}^p z^{-\frac12+p}=-a_{k}^{-\mathfrak S_{k-1}}\lb
  \begin{array}{c}
  \ds \frac{\lb\theta_{k}+\sigma_{k-1}\rb^2-
  \sigma_{k}^2}{2\sigma_{k-1}}\,
  {}_2F_1\biggl[\begin{array}{c}
  1+\theta_{k}+\sigma_{k-1}+\sigma_{k},
  1+\theta_{k}+\sigma_{k-1}-\sigma_{k}\\
  1+2\sigma_{k-1}
  \end{array};\frac{z}{a_{k}}\biggr]\vspace{0.1cm}\\ \ds
  \frac{\lb\theta_{k}-\sigma_{k-1}\rb^2-
  \sigma_{k}^2}{2\sigma_{k-1}}\,
    {}_2F_1\biggl[\begin{array}{c}
    1+\theta_{k}-\sigma_{k-1}+\sigma_{k},
    1+\theta_{k}-\sigma_{k-1}-\sigma_{k}\\
    1-2\sigma_{k-1}
    \end{array};\frac{z}{a_{k}}\biggr]
  \end{array} \rb.
  \ebn
  This in turn implies the equation (\ref{diagfactors01}). The proof of three other identities is similar.
  \epf
  
  The Cauchy determinant in (\ref{cauchyN2}) remains invariant upon
  simultaneous translation of all $x^{[k]}_{\imath}$ and $y^{[k]}_{\jmath}$ by the same amount. Let us use this to replace the notation (\ref{phcauchy}) in the case $N=2$ by
       \begin{subequations}
        \label{phcauchyv2}
        \begin{align}
        x^{[k]}_{\imath}:=&\,\begin{cases}
        p+\epsilon\sigma_{k-1},\qquad & \imath\equiv \lb p,\epsilon\rb\in I_{k-1},\\
        -p-\theta_{k}+\epsilon\sigma_{k},\qquad & \imath\equiv \lb -p,\epsilon\rb\in J_k,
        \end{cases}\\
        y^{[k]}_{\jmath}:=&\,\begin{cases}
        -q+\epsilon\sigma_{k-1},\qquad & \jmath\equiv\lb -q,\epsilon\rb\in J_{k-1},\\
        q-\theta_{k}+\epsilon\sigma_{k},\;\;\;\;\qquad & \jmath\equiv \lb q,\epsilon\rb\in I_k.
        \end{cases}
        \end{align}
        \end{subequations}
  Define a notation for the charges
  \ben
  m_{k}:=\left|\lb \,\cdot\,,+\rb\in I_k\right|-
  \left|\lb \,\cdot\,,+\rb\in J_k\right|=
  \left|\lb \,\cdot\,,-\rb\in J_k\right|-
    \left|\lb \,\cdot\,,-\rb\in I_k\right|,\qquad k=1,\ldots,n-3,
  \ebn
  and combine them into a vector $\mathbf m:=\lb m_1,\ldots,m_{n-3}\rb\in\mathbb Z^{n-3}$. We will also write $\boldsymbol \sigma:=\lb \sigma_1,\ldots,\sigma_{n-3}\rb\in \mathbb C^{n-3}$ and further define
  \beq\label{defeta}
  \boldsymbol\eta:=\lb \eta_1,\ldots,\eta_{n-3}\rb,\qquad 
  e^{i\eta_{k}}:=\frac{d^{[k]}_{\infty,-}}{d^{[k]}_{\infty,+}}.
  \eeq
  The parameters $\boldsymbol \eta$ provide the remaining $n-3$ local coordinates on the space $\mathcal M_{\Theta}$ of monodromy data.
    The main result of this section may now be formulated as follows.
  
  \begin{theo}\label{theogarnier} The isomonodromic tau function of the Garnier system admits the following multivariate combinatorial expansion:
  \begin{gather}
  \label{tauGar}
  \begin{aligned}
  \tau_{\mathrm{Garnier}}\lb a\rb=&\;\mathrm{const}\cdot a_1^{-\theta_0^2}\prod_{k=1}^{n-3}
  a_k^{-\theta_k^2}\prod_{1\leq k<l\leq n-2}\lb 1-\frac{a_k}{a_l}\rb^{-2\theta_k\theta_l}\times\\
  \times &\,\sum_{\mathbf{m}\in
  \mathbb Z^{n-3}}e^{i\mathbf m\cdot \boldsymbol{\eta}}\!\!\!\!\!\!\sum_{\vec Y_1,\ldots,\vec Y_{n-3}\in\mathbb Y^2}
  \prod_{k=1}^{n-3}\lb\frac{a_{k}}{a_{k+1}}\rb^{\lb\sigma_{k}+m_{k}\rb^2+\left|\vec Y_{k}\right|}
  \prod_{k=1}^{n-2}Z^{\vec Y_{k-1},m_{k-1}}_{\;\vec Y_{k},m_{k}}\lb
      \mathcal T^{[k]}\rb,
  \end{aligned}
  \end{gather}
  where $\vec Y_{k}$ stands for the pair of charged Young diagrams associated to  $\lb I_k,J_k\rb$,  the total number of boxes in $\vec Y_{k}$ is denoted by $\left|\vec Y_{k}\right|$,  and
  \begin{gather}
  \label{formula418}
  \begin{aligned}
  Z^{\vec Y_{k-1},m_{k-1}}_{\;\vec Y_{k},m_{k}}\lb
        \mathcal T^{[k]}\rb=\!\!
   &\prod_{\lb p,\epsilon\rb\in I_{k-1}}\!\!\frac{\prod_{\epsilon'=\pm}\lb \theta_{k}+\epsilon\sigma_{k-1}+\epsilon'\sigma_{k} \rb_{p+\frac12}}{\lb p-\frac12\rb! \,\lb 2\epsilon\sigma_{k-1}\rb_{p+\frac12}}\!\!\!\!\!\!
         \prod_{\lb -p,\epsilon\rb\in J_{k-1}}\!\!\!\!\!\! \frac{
         \prod_{\epsilon'=\pm}\lb1- \theta_{k}-\epsilon\sigma_{k-1}+\epsilon'\sigma_{k} \rb_{p-\frac12}}{\lb p-\frac12\rb!\,\lb 1-2\epsilon\sigma_{k-1}\rb_{p-\frac12}}\times\\
        \times\, & \,  \prod_{\lb -p,\epsilon\rb\in J_{k}}\!\!\frac{\prod_{\epsilon'=\pm}\lb\theta_{k}+
         \epsilon'\sigma_{k-1}-\epsilon
           \sigma_{k}\rb_{p+\frac12}
             }{\lb p-\frac12\rb!\,\lb -2\epsilon\sigma_{k}\rb_{p+\frac12}}\;
            \prod_{\lb p,\epsilon\rb\in I_{k}}\frac{\prod_{\epsilon'=\pm}\lb1- \theta_{k}+\epsilon'\sigma_{k-1}+\epsilon\sigma_{k} \rb_{p-\frac12}}{\lb p-\frac12\rb!\,\lb 1+2\epsilon\sigma_{k}\rb_{p-\frac12}}\times\\
    \times &\,\frac{\ds\prod_{\imath,\jmath\in I_{k-1}\sqcup J_k; \imath<\jmath}\lb x^{[k]}_{\imath}-x^{[k]}_{\jmath}\rb
        \prod_{\imath,\jmath\in J_{k-1}\sqcup I_k;\imath<\jmath}\lb y^{[k]}_{\jmath}-y^{[k]}_{\imath}\rb}{\ds\prod_{\imath\in I_{k-1}\sqcup J_k}\prod_{\jmath \in J_{k-1}\sqcup I_k}\lb x^{[k]}_{\imath}-y^{[k]}_{\jmath}\rb}.       
  \end{aligned}
  \end{gather}
  \end{theo}
  \pf Consider the product in the first line of (\ref{cauchyN2}). The balance conditions $|I_k|=|J_k|$ imply that the factors such as $e^{\sum_{j=0}^{k-1}\theta_k}$ in (\ref{diagfactors}) cancel out from
  $Z^{I_{k-1},J_{k-1}}_{\; I_{k},J_{k}}\lb
      \mathcal T^{[k]}\rb$. The factors of the form $\pm \epsilon$ also compensate each other in the product of elementary determinants in (\ref{partZ03}). The factors ${d^{[k]}_{\infty,\epsilon}}^{\pm 1}$ in  (\ref{diag14c}) and
      (\ref{diag14d}) produce the exponential
      $e^{i\mathbf m\cdot\boldsymbol \eta}$ in (\ref{tauGar}).
  
\begin{figure}
        \centering
        \includegraphics[height=3.cm]{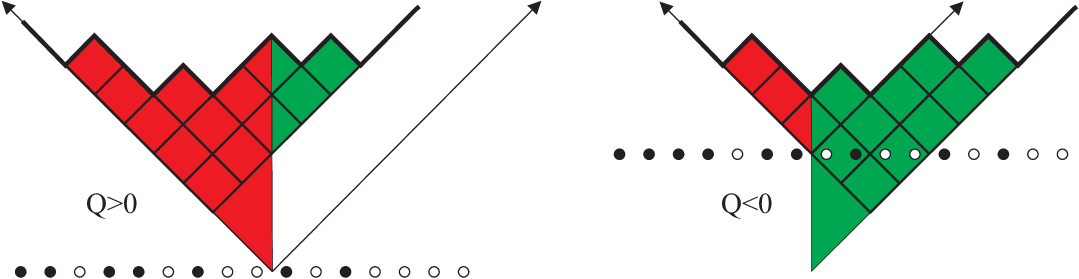}
        \begin{minipage}{0.75\textwidth}
        \caption{\label{figsumofps} A charged Maya diagram $\mathsf m$ and the associated partition $Y\lb \mathsf m\rb$ for
            positive and negative charges
            $Q\lb \mathsf m\rb$.
          Given the positions  $\mathsf p\lb \mathsf m\rb=\lb p_1,\ldots, p_r\rb$ and 
            $\mathsf q\lb \mathsf m\rb=\lb -q_1,\ldots, -q_s\rb$ of particles and holes, the red and green 
            areas represent the sums 
            $\sum p_k$ and $\sum q_k$. We clearly have
            $\sum p_k+\sum q_k=\frac{{Q\lb \mathsf m\rb}^2}{2}+
            |Y\lb\mathsf m\rb |$ in both cases. }\end{minipage}
        \end{figure}

  The total power in which the coordinate $a_{k}$ appears in (\ref{cauchyN2}) is equal to 
  \begin{gather*}
  2m_{k}\sigma_{k}
  -2m_{k-1}\sigma_{k-1}-\sum_{\lb p,\epsilon\rb\in I_{k-1}}\!\!\!p
  -\sum_{\lb -p,\epsilon\rb\in J_{k-1}}\!\!\!\!\! p+
  \sum_{\lb -p,\epsilon\rb\in J_{k}}\!\!\! \!\!p 
  +\sum_{\lb p,\epsilon\rb\in I_{k}}\!\!\! \!\! p
  =\\
  =\lb 2m_{k}\sigma_{k}+m_{k}^2+\left|\vec Y_{k}\right|\rb-\lb 2m_{k-1}\sigma_{k-1}+m_{k-1}^2+\left|\vec Y_{k-1}\right|\rb.
  \end{gather*}
 The last equality is demonstrated graphically in Fig.~\ref{figsumofps}. The prefactor in the first line of (\ref{tauGar}) comes from two sources: i) the shifts of (initially traceless) Garnier monodromy exponents $\Theta_k$ by $-\theta_k\mathbb 1$ making one of their eigenvalues equal to $0$ and ii) the prefactor
 $\Upsilon \lb a\rb$ from Theorem~\ref{TauF}.
  \epf
  
   In the Appendix, we show that the formula (\ref{formula418}) can be rewritten in terms of Nekrasov functions. In the Painlevé VI case ($n=4$), this transforms Theorem~\ref{theogarnier} into Theorem~B of the Introduction.

  
  \subsection{Hypergeometric kernel}
  Recall that the matrices $\Theta_0,\ldots,\Theta_{n-1}$ are by convention diagonal with eigenvalues distinct modulo non-zero integers. However, all of the results of Section~\ref{sectionfred} remain valid if the diagonal parts corresponding to the degenerate eigenvalues are replaced by appropriate Jordan blocks. 
  
  In this subsection we will consider in more detail a specific example of this type by revisiting the  $4$-point tau function. We will thus follow the notational conventions of Subsection~\ref{subsec4pts}. Fix $n=4$, $N=2$ and assume furthermore that the monodromy representation $\rho^{[L]}:\pi_1\lb \Pb^1\backslash\left\{0,t,\infty\right\}\rb\to \mathrm{GL}\lb 2,\Cb\rb$ associated to the  internal trinion $\mathcal T^{[L]}$ is reducible, whereas its counterpart $\rho^{[R]}:\pi_1\lb \Pb^1\backslash\left\{0,1,\infty\right\}\rb\to\mathrm{GL}\lb 2,\Cb\rb $ for the external trinion $\mathcal T^{[R]}$ remains generic.  
  For instance, one may set
  \ben
  \begin{gathered}
  \Theta_0=\mathfrak S=\lb \begin{array}{cc}
  0 & 0 \\
  0 & -2\sigma
  \end{array} \rb,\qquad
  \Theta_t=\lb\begin{array}{cc}0 & 0 \\ 1 & 0\end{array}\rb,
  \end{gathered}
  \ebn
  so that the monodromy matrices $M_0$, $M_t$ can be assumed to have the lower triangular form
  \beq\label{monodromyFL} 
    M_0=\lb\begin{array}{cc}1 & 0 \\ 
    -2\pi i\kappa e^{-2\pi i \sigma} & e^{-4\pi i \sigma}\end{array}\rb,\qquad 
    M_t=\lb\begin{array}{cc}1 & 0 \\ 2\pi i\kappa e^{2\pi i 
    \sigma} & 1\end{array}\rb,
    \qquad M_0M_t=e^{2\pi i \mathfrak S}.
  \eeq
  
  The solution $\Psi^{[L]}\lb z\rb$ of the appropriate internal $3$-point RHP  may be constructed from the fundamental solution of a Fuchsian system
  \beq\label{fuchsf21}
  \partial_z\Phi^{[L]}=\Phi^{[L]}\lb
  \begin{array}{cc}
  0 &  0 \\ \ds\frac{\varrho t}{ z\lb z-t\rb} & \ds-\frac{2\sigma}{z}
  \end{array}\rb,
  \eeq
  with a suitably chosen value of the parameter $\varrho$. Taking into account the diagonal monodromy around $\infty$, such a solution $\Phi^{[L]}\lb z\rb$ on $\mathbb C\backslash \mathbb R_{\ge 0}$ can be written~as
  \begin{align}\label{fuchsf21aux01}
  \Phi^{[L]}\lb z\rb=&\,\lb\begin{array}{cc}
    1 & 0 \\
    {\ds\frac{\varrho t \lb - z\rb^{-2\sigma-1}}{ 1+2\sigma}}\,l_{2\sigma}\lb\frac tz\rb & \lb - z\rb^{-2\sigma}
    \end{array}\rb=\tilde C_0 \lb\begin{array}{cc}
            1 & 0 \\
            {\ds\frac{\varrho  \lb - z\rb^{-2\sigma}}{ 2\sigma}}\,l_{-1-2\sigma}\lb\frac zt\rb & \lb - z\rb^{-2\sigma}
            \end{array}\rb  ,            
  \end{align}
  where $l_{a}\lb z\rb:={}_2F_1\biggl[\begin{array}{c} 
  1+a,1 \vspace{-0.1cm} \\ 2+a \end{array};z\biggr]$, and the modified connection matrix $\tilde C_0$ is lower-triangular: 
  \ben
  \tilde C_0=\lb\begin{array}{cc}
                          1 & 0 \\
                          \ds -\frac{\pi\varrho t^{-2\sigma}}{
                                                    \sin2\pi\sigma} & 1
                          \end{array}\rb.
  \ebn
  The monodromy matrix around $0$ is clearly equal to $M_0=\tilde C_0 e^{2\pi i \Theta_0}{\tilde C_0}^{-1}$. This allows to relate the mono\-dromy parameter $\kappa$ to the coefficient $\rho$ of the Fuchsian system (\ref{fuchsf21}) as 
  \beq\label{monlambda}
  \kappa=
  \varrho t^{-2\sigma}.
  \eeq
  
  The $3$-point RHP solution $\Psi^{[L]}\lb z\rb$ inside the annulus $\mathcal A$ is thus explicitly given by
  \beq\label{uppertraux}
  \Psi^{[L]}\lb z\rb\Bigl|_{\mathcal A}=
  \lb\begin{array}{cc}
    1 & 0 \\
    0 & \lb -z\rb^{2\sigma}
    \end{array}\rb \Phi^{[L]}\lb z\rb
  =\lb\begin{array}{cc}
  1 & 0 \\
  {\ds-\frac{\varrho t}{\lb 2\sigma+1\rb z}}\, l_{2\sigma}\lb \frac tz\rb & 1
  \end{array}\rb.
  \eeq
  This formula leads to substantial simplifications in the Fredholm determinant representation (\ref{det4pttau}) of the tau function $\tau_{\mathrm{JMU}}\lb t\rb$.
  It follows from from the structure of (\ref{uppertraux}) and (\ref{intkernelsalphadelta2}) that 
  \beq\label{F21detaux00}
  \mathsf d_{-+}\lb z,z'\rb =\frac{\varrho}{1+2\sigma}\frac{
  \frac tz  l_{2\sigma}\lb \frac tz\rb - 
  \frac t{z'}  l_{2\sigma}\lb \frac t{z'}\rb }{z-z'}
  \eeq
  is the only non-zero element of the $2\times 2$ matrix integral kernel $\mathsf d\lb z,z'\rb$  (note that the lower indices here are color and should not be confused with half-integer Fourier modes). This in turn implies that the only entry of $\mathsf a\lb z,z'\rb$ contributing to the determinant is 
  \beq\label{hyperapm}
  \mathsf a_{+-}\lb z,z'\rb=\frac{1}{\operatorname{det}\Psi^{[R]}\lb z'\rb}\frac{\Psi^{[R]}_{+-}\lb z\rb \Psi^{[R]}_{++}\lb z'\rb
  -\Psi^{[R]}_{++}\lb z\rb \Psi^{[R]}_{+-}\lb z'\rb}{z-z'}.
  \eeq
  Therefore, (\ref{det4pttau}) reduces to
  \beq\label{F21detaux01}
  \tau_{\mathrm{JMU}}\lb t\rb=\operatorname{det}\lb\mathbb{1}-
  \mathsf a_{+-}\mathsf d_{-+}\rb.
  \eeq

  The action of the operators $\mathsf a_{+-}$, $\mathsf d_{-+}$ involves integration along a circle $\mathcal C\subset \mathcal A$. The kernel
  $\mathsf a_{+-}\lb z,z'\rb$ extends to a function holomorphic in both arguments inside $\mathcal C$. Therefore in the computation of contributions 
 of different exterior powers to the determinant one may try to shrink all integration contours to the branch cut
  $\mathcal B:=[0,t]\subset \mathbb R$. The latter comes from two branch points $0,t$  of $\mathsf d_{-+}\lb z,z'\rb$ defined by (\ref{F21detaux00}). 
  \begin{lemma}\label{lemmaF21}
  Let $\left|\Re \sigma\right|<\frac12$. 
  For $m\in\mathbb Z_{\ge0}$, denote $X_m=\operatorname{Tr}\lb\mathsf a_{+-}\mathsf d_{-+}\rb^m$. We have
  \ben
  X_m=\operatorname{Tr}  K_{F}^m,
  \ebn
  where $ K_F$ denotes an integral operator on $L^2\lb \mathcal B\rb$ with the kernel
  \beq \label{kernelKF}
   K_F\lb z,z'\rb=-\kappa \lb zz'\rb^{\sigma}  \mathsf a_{+-}\lb z,z'\rb .
  \eeq
  \end{lemma}
  \pf Let us denote by $\mathcal B_{\text{up}}$ and $\mathcal B_{\text{down}}$ the upper and lower edge of the branch cut $\mathcal B$. After shrinking of the integration contours in the multiple integral $I_k$ to  $\mathcal B$,  the operators 
  $\mathsf a_{+-}$, $\mathsf d_{-+}$ should be interpreted as acting on $\mathcal W=L^2\lb \mathcal B_{\text{up}}\rb\oplus
  L^2\lb \mathcal B_{\text{down}}\rb$ instead of  $L^2\lb\mathcal C\rb$. Here $L^2\lb \mathcal B_{\text{up,down}}\rb$ arise as appropriate completions of spaces of boundary values of functions holomorphic inside $D_{\mathcal C}\backslash\mathcal B$, where $D_{\mathcal C}$ denotes the disk bounded by $\mathcal C$. 
  The space $\mathcal W$ can be decomposed as $\mathcal W=\mathcal W_{+}\oplus
  \mathcal W_-$, where the elements of $\mathcal W_+$ are continuous across the branch cut, whereas the elements of $\mathcal W_-$ have opposite signs on its two sides:
  \ben
  \mathcal W_{\pm}=\left\{f\in\mathcal W\,:\,f\lb z+i0\rb=\pm f\lb z-i0\rb,z\in\mathcal B\right\}.
  \ebn
  We will denote by $\operatorname{pr}_{\pm}$ the projections on $\mathcal W_{\pm}$ along $\mathcal W_{\mp}$.
  
  Since $\mathsf a_{+-}\lb z,z'\rb$ is holomorphic in $z,z'$ inside $\mathcal C$, it follows that $\operatorname{im}\mathsf a_{+-}\subseteq\mathcal W_+\subseteq \operatorname{ker}\mathsf a_{+-}$. Therefore $X_k$ remains unchanged if  $\mathsf a_{+-}$ is replaced by $\operatorname{pr}_{+}\circ\, \mathsf a_{+-}\circ \operatorname{pr}_{-}$. This is in turn equivalent to replacing $\mathsf d_{-+}$ by $\operatorname{pr}_{-}\circ\, \mathsf d_{-+}\circ \operatorname{pr}_{+}$. Given $f=g\oplus g\in\mathcal W_+$ with $g\in L^2\lb \mathcal B\rb$, the action of 
  $\mathsf d_{-+}$ on $f$ is given by
  \begin{align*}
  \lb  \mathsf d_{-+} f\rb\lb z\rb=&\,\frac{1}{2\pi i}
  \int_0^t\left[\mathsf d_{-+}\lb z,z'-i0\rb-\mathsf d_{-+}\lb z,z'+i0\rb\right]g\lb z'\rb dz'=\\
  =&\,\frac{\varrho t}{2\pi i \lb 1+2\sigma\rb}
  \int_0^t\frac{l_{2\sigma}\lb\frac t{z'+i0}\rb-
    l_{2\sigma}\lb\frac t{z'-i0}\rb}{z'\lb z-z'\rb}\,g\lb z'\rb dz'.
  \end{align*}
  An important consequence of the lower triangular monodromy is that the jump of $l_{2\sigma}\lb \frac t{z'}\rb$ on $\mathcal B$ yields an elementary function, cf (\ref{fuchsf21aux01}):
  \ben
  l_{2\sigma}\lb\frac t{z'+i0}\rb-
      l_{2\sigma}\lb\frac t{z'-i0}\rb=-2\pi i \lb2\sigma+1\rb\lb\frac  {z'}t\rb^{2\sigma+1}.
  \ebn 
  Substituting this jump back into the previous formula and using (\ref{monlambda}), one obtains
  \ben
  \lb  \mathsf d_{-+} f\rb\lb z\rb=\kappa \int_0^t  \frac{z'^{2\sigma}g\lb z'\rb dz'}{z'-z},\qquad\qquad
  z\in D_{\mathcal C}\backslash \mathcal B.
  \ebn   
  Next we have to compute the projection $\operatorname{pr}_-$ of this expression onto $\mathcal W_-$. Write $\operatorname{pr}_-\circ\,\mathsf d_{-+} f=h\oplus\lb -h\rb$, with ${h\in L^2\lb \mathcal B\rb}$. Then
  \ben
  h\lb z\rb=\frac12\left[
  \lb  \mathsf d_{-+} f\rb\lb z+i0\rb-\lb  \mathsf d_{-+} f\rb\lb z-i0\rb\right]=\pi i \kappa
   z^{2\sigma}g\lb z\rb,\qquad z\in \mathcal B.
  \ebn
  Finally, write $\mathsf a_{+-}\circ\operatorname{pr}_-\circ\,\mathsf d_{-+} f$ as $\tilde g\oplus \tilde g\in\mathcal W_+$. It follows from the previous expression for $h\lb z\rb$ that 
  \ben
  \tilde g\lb z\rb=-\kappa \int_0^t \mathsf a_{+-}\lb z,z'\rb {z'}^{2\sigma}g\lb z'\rb dz',\qquad\qquad
  z\in\mathcal B.
  \ebn
  The minus sign in front of the integral is related to orientation of the contour $\mathcal C$ in the definition of $\mathsf a$.
  We have thereby computed the action of $\mathsf a_{+-}\circ\operatorname{pr}_-\circ\,\mathsf d_{-+}$ on $\mathcal W_+$. Raising this operator to an arbitrary power $k\in\mathbb Z_{\ge0}$ and symmetrizing the factors ${z'}^{2\sigma}$ under the trace immediately yields the statement of the lemma.
  \epf
  \begin{theo}\label{theorem2F1}
  Given complex parameters $\theta_1,\theta_{\infty},\sigma$ satisfying previous genericity assumptions, let
  \begin{subequations}
  \begin{align}
  \varphi\lb x\rb:=&\,x^{\sigma}\lb 1-x\rb^{\theta_1}{}_2F_1\biggl[
  \begin{array}{c}
  \sigma+\theta_1+\theta_{\infty},
  \sigma+\theta_1-\theta_{\infty}\\ 2\sigma
  \end{array};x\biggr],\\
  \psi\lb x\rb:=&\,x^{1+\sigma}\lb 1-x\rb^{\theta_1}{}_2F_1\biggl[
    \begin{array}{c}
    1+\sigma+\theta_1+\theta_{\infty},
    1+\sigma+\theta_1-\theta_{\infty}\\ 2+2\sigma
    \end{array};x\biggr].
  \end{align}
  \end{subequations}
  Define the continuous $_2F_1$ kernel by 
  \beq\label{kernelKFtilde}
  \tilde K_F\lb x,y\rb:= \frac{
  \psi\lb x\rb\varphi\lb y\rb-\varphi\lb x\rb \psi\lb y\rb
  }{x-y},
  \eeq
  and consider Fredholm determinant 
  \beq\label{detKF}
  D\lb t\rb:=\operatorname{det}\lb\mathbf 1-\lambda\tilde K_F\bigl|_{(0,t)}\bigr.\rb,\qquad \lambda\in\Cb.
  \eeq
  Then $D\lb t\rb$ is a tau function of the Painlev\'e VI equation with parameters $\vec{\theta}=\lb\theta_0=\sigma,\theta_t=0,\theta_1,\theta_{\infty}\rb$. The conjugacy class of monodromy representation for the associated $4$-point Fuchsian system is generated by the matrices (\ref{monodromyFL}) and
  \begin{subequations}
  \begin{align} 
  \label{monodromyKFa} M_1=&\,
    \frac{e^{-2\pi i \theta_{1}}}{i\sin2\pi \sigma}\lb\begin{array}{cc}
      \cos2\pi\theta_{\infty}-e^{-2\pi i \sigma}\cos2\pi\theta_1 &
    s^{-1}e^{-2\pi i \sigma}\left[
    \cos2\pi\theta_{\infty}-\cos2\pi\lb\theta_1-\sigma\rb\right]   
      \\
    s e^{2\pi i \sigma}\left[\cos2\pi\lb\theta_1+\sigma\rb-
          \cos2\pi\theta_{\infty}\right]  
      &  e^{2\pi i \sigma}\cos2\pi\theta_1-\cos2\pi\theta_{\infty}
      \end{array}\rb,\\
      \label{monodromyKFb}
  M_{\infty}=&\,\frac{e^{-2\pi i \theta_{\infty}}}{i\sin2\pi \sigma}\lb\begin{array}{cc}
  \cos2\pi\theta_1-e^{-2\pi i \sigma}\cos2\pi\theta_{\infty} & 
  s^{-1}\left[\cos2\pi\lb\theta_1-\sigma\rb-
    \cos2\pi\theta_{\infty}\right]  \\
  s\left[\cos2\pi\theta_{\infty}-
    \cos2\pi\lb\theta_1+\sigma\rb\right]
  &  e^{2\pi i \sigma}\cos2\pi\theta_{\infty}-\cos2\pi\theta_1
  \end{array}\rb=M_{1}^{-1}e^{-2\pi i \mathfrak S}.
  \end{align}
  \end{subequations}
  where 
  \begin{gather}
  \label{lambdakappa}
  \lambda=\kappa\frac{\lb\theta_1+\sigma\rb^2-\theta_{\infty}^2}{
  2\sigma\lb 2\sigma+1\rb},\\
  s=-\frac{\Gamma\lb 1-2\sigma\rb
  \Gamma\lb \theta_1+\sigma+\theta_{\infty}\rb
  \Gamma\lb \theta_1+\sigma-\theta_{\infty}\rb}{\Gamma\lb 1+2\sigma\rb
  \Gamma\lb \theta_1-\sigma+\theta_{\infty}\rb
    \Gamma\lb \theta_1-\sigma-\theta_{\infty}\rb}.
  \end{gather}
  \end{theo}
  \pf To prove that $D\lb t\rb$ is a Painlev\'e VI tau function with
  $\lambda$ and $\kappa$ related by (\ref{lambdakappa}), it suffices to combine the determinant representation (\ref{F21detaux01}) with Lemma~\ref{lemmaF21}, and substitute into the formula (\ref{hyperapm}) for $\mathsf a_{+-}\lb z,z'\rb$ explicit hypergeometric expressions (\ref{3pRHP02}).
  
  The formula (\ref{monodromyKFb}) follows from $M_{\infty}=C_{\infty} e^{2\pi i\Theta_{\infty}}C_{\infty}^{-1}$, where $C_{\infty}$ is obtained from the connection matrix (\ref{3pRHP05}) by replacements
  $\lb \theta_{k},\sigma_{k-1},\sigma_{k}\rb\to 
  \lb \theta_1,\sigma,-\theta_{\infty}\rb$. The expression (\ref{monodromyKFa}) for $M_1$ is then most easily deduced from the diagonal form of the product $M_1M_{\infty}=e^{-2\pi i \mathfrak S}$.  \epf
  \begin{rmk}
  The $_2F_1$ kernel is related to the so-called $ZW$-measures \cite{BorodinAnnals} arising in the representation theory of the infinite-dimensional unitary group $U\lb\infty\rb$.  It produces various other classical integrable kernels (such as sine and Whittaker) as limiting cases.
  The first part of Theorem~\ref{theorem2F1}, namely
  the Painlev\'e VI equation for $D\lb t\rb$, was proved
  by Borodin and Deift in \cite{BorodinDeift}. Monodromy data for the associated Fuchsian system have been identified in \cite{dyson2F1}.
  To facilitate the comparison, let us note that indroducing instead of
  $\lambda$ and $\kappa$  a new parameter $\bar\sigma$ de\-fi\-ned by
  \ben
  \lambda=\frac{\sin\pi\lb\bar\sigma-\theta_1\rb
  \sin\pi\lb\bar\sigma+\theta_1\rb}{\pi^2}\frac{\prod_{\epsilon,\epsilon'=\pm}\Gamma\lb 1+\sigma+\epsilon\theta_1+\epsilon'\theta_{\infty}\rb}{
  \Gamma\lb 1+2\sigma\rb \Gamma\lb 2+2\sigma\rb},
  \ebn
  we have in particular that $\operatorname{Tr} M_{\infty}M_0=2e^{-2\pi i \lb\sigma+\theta_{\infty}\rb}\cos2\pi\bar\sigma$ and
  $\operatorname{Tr} M_tM_1=2e^{2\pi i \lb\sigma+\theta_{\infty}\rb}\cos2\pi\bar\sigma$. The relation
  between parameters $z,z',w,w'$  of \cite{BorodinDeift} and ours is
  \ben
  \lb z,z',w,w'\rb_{\text{\cite{BorodinDeift}}}=\lb \bar\sigma+\theta_1,\bar\sigma-\theta_1,
  \sigma-\bar\sigma+\theta_{\infty},
  \sigma-\bar\sigma-\theta_{\infty}\rb.
  \ebn
  \end{rmk}
 
 
   \appendix
   \section{Relation to Nekrasov functions}

  Here we demonstrate that the formula (\ref{formula418}) can be rewritten in terms of Nekrasov functions. This rewrite is conceptually important for identification of isomonodromic
  tau functions with dual partition functions of quiver gauge theories \cite{NO}. It is also useful from a
  computational point of view: naively, the formula (\ref{formula418}) may produce poles in the tau function expansion coefficients when
  $\theta_k\pm\sigma_k\pm'\sigma_{k-1}\in\mathbb Z$. Our calculation shows that these poles actually cancel.
  
  The statement we are going to prove\footnote{In the present paper we do it only for $N=2$ but the generalization is relatively straightforward.} is the  relation
  \eq{
  Z^{\vec Y',m'}_{\vec Y,m}\lb\mc T\rb=(-1)^{{\rm lsgn}\lb\vec Y',m'\rb+{\rm lsgn}\lb\vec Y,m\rb}\hat Z^{\vec Y',\vec Q'}_{\vec Y,\vec Q}\lb\mc T\rb,
  }
  where
  \begin{gather}
  \label{Zhat}
  \begin{aligned}
  \hat Z^{\vec Y',\vec Q'}_{\vec Y,\vec Q}(\mc T)=&\,
  \frac{\prod_{\alpha,\beta}^N
  C\lb\left.\sigma'_\alpha-\sigma_\beta\right|
    Q'_\alpha,Q_\beta\rb}{\prod_{\alpha<\beta}^N
    C\bigl(\bigl.\sigma'_\alpha-\sigma'{}_\beta\bigr|
    Q'_\alpha,Q'_\beta\bigr)\,
    C\bigl(\sigma_\alpha-
    \sigma_\beta|Q_\alpha,Q_\beta\bigr)}
    \frac{\lb -1\rb^{N |\vec Y| + \frac N2 |\vec Q|^2} e^{i\delta\vec\eta'\cdot\vec Q'+
    i\delta\vec\eta\cdot\vec Q}}{
    \prod_{\alpha}^N \left|Z_{\,\mathsf{bif}}\lb 0\bigl|\bigr. Y_\alpha,Y_\alpha\rb\right|^{\frac12}
    \left|Z_{\,\mathsf{bif}}\lb 0\bigl|\bigr. Y'_\alpha,Y'_\alpha\rb\right|^{\frac12}}\times\\
   \times & \,
  \frac{\prod_{\alpha,\beta}^N Z_{\,\mathsf{bif}}\lb\left.
  \sigma'_\alpha+Q'_\alpha-\sigma_\beta-Q_\beta
    \right|Y'_\alpha,Y_\beta\rb}{
    \prod_{\alpha<\beta}^N
    Z_{\,\mathsf{bif}}\bigl(
    \bigl.\sigma'_\alpha+Q'_\alpha-\sigma'_\beta-Q'_\beta
    \bigr|Y'_\alpha,Y'_\beta\bigr)\, Z_{\,\mathsf{bif}}\lb\left.\sigma_\alpha+Q_\alpha-
    \sigma_\beta-Q_\beta\right|Y_\alpha,Y_\beta\rb }.
  \end{aligned}
  \end{gather}
  The notation used in these formulas means the following:
  \begin{itemize}
  \item $\vec Q=\lb m,-m\rb$, $\vec Q'=\lb m',-m'\rb$,
  though the right side of (\ref{Zhat}) is defined even without this specialization.
  \item $Y'$ and $Y$ are identified, respectively, with $Y_{k-1}$ and $Y_k$ in (\ref{formula418}). Similar conventions will be used for all other quantities. We denote, however,  $\sigma'_\pm=\pm\sigma_{k-1}$ and $\sigma_\pm=-\theta_{k}\pm\sigma_{k}$;
  $\mc T$ stands for $\mc T^{[k]}$.
  \item  ${\rm lsgn}\lb\vec Y,m\rb\in\mathbb Z/2\mathbb Z$  means the ``logarithmic $\rm sign$'',
  \eq{
  {\rm lsgn}\lb\vec Y,m\rb:=|\mathsf q_+|\cdot |\mathsf p_+|+\Sum_i\lb q_{+,i}+\frac12\rb+\Sum_i\lb p_{-,i}+\frac12\rb.
  }
  Here, for example, $|\mathsf p_+|$ denotes the number of  coordinates $p_{+,i}$ of particles in the Maya diagram corresponding to the charged partition $\lb Y_+,m\rb$. The logarithmic signs cancel in the product
  $\prod_{k=1}^{n-2}Z^{\vec Y_{k-1},m_{k-1}}_{\;\vec Y_{k},m_{k}}$  which appears in the representation (\ref{tauGar}) for the Garnier tau function.
  \item $\delta\vec\eta$ and $\delta\vec\eta'$ are some explicit functions which are computed below. They just shift Fourier transformation parameters and their relevant combinations are explicitly given by
  \begin{gather}
  \label{apeq4}
  \begin{aligned}
  e^{i\delta\eta'_+-i\delta\eta'_-}&=\frac{1}{2\sigma_{k-1}}
  \frac{\lb\theta_{k}+\sigma_{k-1}\rb^2-\sigma_{k}^2}{
  \lb\theta_{k}-\sigma_{k-1}\rb^2-\sigma_{k}^2},\\
   e^{i\delta\eta_+-i\delta\eta_-}&=\;\;\;\frac{-1}{2\,
   \sigma_{k}}\;\,
     \frac{\lb\theta_{k}+\sigma_{k}\rb^2-\sigma_{k-1}^2}{
     \lb\theta_{k}-\sigma_{k}\rb^2-\sigma_{k-1}^2}.
  \end{aligned}
  \end{gather}
  \item $Z_{\,\mathsf{bif}}\lb\nu|Y',Y\rb$ is the Nekrasov bifundamental contribution
  \beq
  Z_{\,\mathsf{bif}}\lb\nu|Y',Y\rb:=\prod_{\square\in Y'}\bigl(\nu+1+a_{Y'}\lb\square\rb+l_{Y}\lb\square\rb\bigr)
  \prod_{\square\in Y}\bigl(\nu-1-a_{Y}\lb\square\rb-l_{Y'}\lb\square\rb\bigr).
  \eeq
  In particular, we have $\left|Z_{\,\mathsf{bif}}\lb 0|Y,Y\rb\right|^{\frac12}=\prod_{\square\in Y}h_Y\lb \square\rb$.
  \item The three-point function $C\lb\nu|Q',Q\rb$ is defined by
  \eq{
  C\lb\nu|Q',Q\rb\equiv  C\lb\nu|Q'-Q\rb=\frac{G\lb 1+\nu+Q'-Q\rb}{G\lb 1+\nu\rb\Gamma\lb 1+\nu\rb^{Q'-Q}},
  }
  where $G\lb x\rb$ is the Barnes $G$-function. The only property of this function  essential for our purposes is the recurrence relation $G\lb x+1\rb=\Gamma\lb x\rb G\lb x\rb$.
  \item Using the formula (\ref{formula418}), we assume a concrete ordering: $p_+', p_-', q_+, q_-$, $p_1>p_2>\ldots$, and in \rf{Zhat} we suppose that $+<-$.
  \end{itemize}
  An important feature of the product \rf{Zhat} is that the combinatorial part in the 2nd line depends only on combinations such as $\sigma_{\alpha}+Q_{\alpha}$, $\sigma'_{\alpha}+Q'_{\alpha}$. This is most crucial for the Fourier transform structure of the full aswer for the tau function $\tau_{\mathrm{Garnier}}\lb a\rb$.  
  
  Let us now present the plan of the proof, which will be divided into several self-contained parts. Most computations will be done up to an overall sign, and sometimes we will omit to indicate this. In the end we will consider the limit $\theta_k\to+\infty$, $\sigma_k,\sigma_{k-1}\ll\theta_k$,
  $\sigma_k,\sigma_{k-1}\to+\infty$
  to recover the correct sign.

  \begin{enumerate}
  \item First we will rewrite the formula (\ref{formula418}) as
  $$Z^{\vec Y',m'}_{\vec Y,m}\lb\mc T\rb=\pm e^{i\delta_1\vec\eta'\cdot\vec Q'+i\delta_1\vec\eta\cdot\vec Q}\hat{\hat{Z}}^{\vec Y',\vec Q'}_{\vec Y,\vec Q}\lb\mc T\rb,$$ 
  where $\hat{\hat{Z}}^{\vec Y',\vec Q'}_{\vec Y,\vec Q}\lb\mc T\rb$ is expressed in terms of yet another function $\tilde Z_{\,\mathsf{bif}}\lb\nu\bigl|Q',Y';Q,Y\rb$,
  \begin{gather}
  \begin{aligned}
  \hat{\hat{Z}}^{\vec Y',\vec Q'}_{\vec Y,\vec Q}\lb\mc T\rb=&\;
  \prod_{\alpha}^N\,\bigl|\tilde Z_{\mathsf{bif}}\lb 0\bigl|Q_\alpha,Y_\alpha,Q_\alpha,Y_\alpha\rb\bigr|^{-\frac12}
    \bigl|
    \tilde Z_{\mathsf{bif}}\lb 0\bigl|Q'_\alpha,Y'_\alpha,
    Q'_\alpha,Y'_\alpha\rb\bigr|^{-\frac12}\;\times\\
    \times&\,
  \frac{
  \prod_{\alpha,\beta}^N \tilde Z_{\mathsf{bif}}\lb\sigma'_\alpha-\sigma_\beta\bigl|Q'_\alpha,
  Y'_\alpha;Q_\beta,Y_\beta\rb}{
  \prod_{\alpha<\beta}^N
  \tilde Z_{\,\mathsf{bif}}\bigl(\sigma'_\alpha-\sigma'_\beta\bigl|
  Q'_\alpha,Y'_\alpha;
  Q'_\beta,Y'_\beta\bigr)
  \tilde Z_{\,\mathsf{bif}}\lb\sigma_\alpha-\sigma_\beta\bigl|
  Q_\alpha,Y_\alpha;Q_\beta,Y_\beta\rb},
  \label{Zhhat}
  \end{aligned}
  \end{gather}
  which is defined as
  \begin{gather}
  \label{apeq2}
  \begin{aligned}
  \tilde Z_{\,\mathsf{bif}}\lb\nu\bigl|Q',Y';Q,Y\rb=&\,\prod_i\lb-\nu\rb_{q_i'+\frac12}
  \prod_i
  \lb\nu+1\rb_{q_i-\frac12}
  \prod_i\lb -\nu\rb_{p_i+\frac12}
  \prod_i\lb\nu+1\rb_{p_i'-\frac12}\times \\
  \times &\,
  \frac{\prod_{i,j}\lb\nu-q_i'-p_j\rb\prod_{i,j}\lb\nu+p_i'+q_j\rb}{
  \prod_{i,j}\lb\nu-q_i'+q_j\rb\prod_{i,j}\lb\nu+p_i'-p_j\rb}.
  \end{aligned}
  \end{gather}
  \item At the second step, we prove that $\tilde Z_{\,\mathsf{bif}}\lb\nu\bigl|0,Y';0,Y\rb\equiv\tilde Z_{\,\mathsf{bif}}\lb\nu\bigl|Y',Y\rb=\pm Z_{\,\mathsf{bif}}\lb\nu\bigl|Y',Y\rb$.
  \item Next it will be shown that 
  \beq\tilde Z_{\,\mathsf{bif}}\lb\nu\bigl|Q',Y';Q,Y\rb=C\lb\nu\bigl|Q',Q\rb
  Z_{\,\mathsf{bif}}\lb\nu+Q'-Q\bigl|Y',Y\rb.
  \eeq
  \item Finally, we check the overall sign and compute extra contribution to $\vec\eta$ to absorb it.  
  \end{enumerate}
  A realization of this plan is presented below.
  \subsection*{Step 1}
  It is useful to decompose the product (\ref{formula418}) into two different parts: a ``diagonal'' one, containing the products of functions of one particle/hole coordinate, and a ``non-diagonal'' part containing the products of pairwise sums/differences. Careful comparison of the formulas (\ref{formula418}) and \rf{Zhhat} shows that their non-diagonal parts actually
  coincide. Further analysis of \rf{Zhhat} shows that its diagonal part is given by 
  \ben
  \prod_{\lb p',\epsilon\rb\in I'}\psi_{p',\epsilon}\prod_{\lb -q',\epsilon\rb\in J'}\bar\psi_{q',\epsilon}
  \prod_{\lb -q,\epsilon\rb\in J}\varphi_{q,\epsilon}\prod_{\lb p,\epsilon\rb\in I}\bar\varphi_{p,\epsilon}.
  \ebn
  with
  \begin{gather}
  \begin{aligned}
  \psi_{p',\epsilon}&=\frac{
  \lb 1+\epsilon\sigma_{k-1}+\theta_{k}-\sigma_{k}\rb_{p'-\frac12}
  \lb 1+\epsilon\sigma_{k-1}+\theta_{k}+\sigma_{k}\rb_{p'-\frac12}
  }{\bigl[\epsilon=+ : \lb 1+2\sigma_{k-1}\rb_{p'-\frac12};\,\epsilon=- : \lb -2\sigma_{k-1}\rb_{p'+\frac12}\bigr]\lb p'-\frac12\rb!},\\
  \bar\psi_{q',\epsilon}&=\frac{\lb-\epsilon\sigma_{k-1}-\theta_{k}+\sigma_{k}\rb_{q'+\frac12}\lb-\epsilon\sigma_{k-1}-\theta_{k}-\sigma_{k}\rb_{q'+\frac12}}{
  \bigl[\epsilon=+ : \lb-2\sigma_{k-1}\rb_{q'+\frac12};\, \epsilon=- : \lb 1+2\sigma_{k-1}\rb_{q'-\frac12}\bigr]\lb q'-\frac12\rb!},\\
  \varphi_{q,\epsilon}&=\frac{\lb\sigma_{k-1}+\theta_{k}-\epsilon\sigma_{k}+1\rb_{q-\frac12}\lb-\sigma_{k-1}+\theta_{k}-\epsilon\sigma_{k}+1\rb_{q-\frac12}}{
  \bigl[\epsilon=+ : \lb-2\sigma_{k}\rb_{q+\frac12};\, \epsilon=- : \lb 1+2\sigma_{k}\rb_{q-\frac12}\bigr]\lb q-\frac12\rb!},\\
  \bar\varphi_{p,\epsilon}&=\frac{\lb-\sigma_{k-1}-\theta_{k}+\epsilon\sigma_{k}\rb_{p+\frac12}\lb\sigma_{k-1}-\theta_{k}+\epsilon\sigma_{k}\rb_{p+\frac12}}{
  \bigl[\epsilon=+ : \lb 1+2\sigma_{k}\rb_{p-\frac12};\, \epsilon=- : \lb-2\sigma_{k}\rb_{p+\frac12}\bigr]\lb p-\frac12\rb!}.
  \end{aligned}
  \end{gather}
  The notation $[\epsilon=+: X; \epsilon=-: Y]$ means that we should substitute this construction by $X$ when $\epsilon=+$ and
  by~$Y$ when $\epsilon=-$.  
  Comparing these expressions  with 
  (\ref{formula418}), we may compute the ratios of diagonal factors which appear {in~$Z^{\vec Y',m'}_{\;\vec Y,m}\bigl/
  \hat{\hat{Z}}^{\vec Y',\vec Q'}_{\vec Y,\vec Q}\bigr.$}:
  
  \eqs{
  \delta\psi_{p',\epsilon}&=\frac{
  \lb\epsilon\sigma_{k-1}+\theta_{k}-\sigma_{k}\rb
  \lb\epsilon\sigma_{k-1}+\theta_{k}+\sigma_{k}\rb
  }{[\epsilon=+ : 2\sigma_{k-1};\,\epsilon=- : 1]},\\
  \delta\bar\psi_{q',\epsilon}&=\frac{
  \lb-\epsilon\sigma_{k-1}-\theta_{k}+\sigma_{k}\rb^{-1}
  \lb-\epsilon\sigma_{k-1}-\theta_{k}-\sigma_{k}\rb^{-1}}{
  [\epsilon=+ : \lb-2\sigma_{k-1}\rb^{-1};\, \epsilon=- : 1]},\\
  \delta\varphi_{q,\epsilon}&=\frac{
  \lb\sigma_{k-1}+\theta_{k}-\epsilon\sigma_{k}\rb
  \lb-\sigma_{k-1}+\theta_{k}-\epsilon\sigma_{k}\rb}{
  [\epsilon=+ : 1;\, \epsilon=- :2\sigma_k ]},\\
  \delta\bar\varphi_{p,\epsilon}&=\frac{
  \lb-\sigma_{k-1}-\theta_{k}+\epsilon\sigma_{k}\rb^{-1}
  \lb\sigma_{k-1}-\theta_{k}+\epsilon\sigma_{k}\rb^{-1}}{
  [\epsilon=+ : 1;\, \epsilon=- : \lb-2\sigma_k\rb^{-1}]}.
  }
  Since $|\mathsf p_\pm|-|\mathsf q_\pm|=Q_\pm$, these formulas allow to determine the corrections $\delta_1\eta_\pm$:
  \begin{gather}
  \label{apeq3}
  \begin{aligned}
  e^{i\delta_1\eta'_+}&=\frac{
  \lb\theta_{k}+\sigma_{k-1}\rb^2-\sigma_{k}^2
  }{2\sigma_{k-1}},\qquad e^{-i\delta_1\eta_+}=\,\lb\theta_k-\sigma_k\rb^2-\sigma_{k-1}^2,\\
  e^{i\delta_1\eta'_-}&=\,
  \lb\theta_{k}-\sigma_{k-1}\rb^2-\sigma_{k}^2,\qquad
  e^{-i\delta_1\eta_-}=\frac{\lb\theta_k+\sigma_k\rb^2
  -\sigma_{k-1}^2}{2\sigma_k}.
  \end{aligned}
  \end{gather}
  One could notice that some minus signs should also be taken into account, so that
  \ben
  Z^{\vec Y',m'}_{\vec Y,m}\lb\mc T\rb=\lb-1\rb^{|\mathsf q_+'|+
  |\mathsf p_-|}
  e^{i\delta_1\vec\eta'\cdot\vec Q'+i\delta_1\vec\eta\cdot\vec Q}\hat{\hat{Z}}^{\vec Y',\vec Q'}_{\vec Y,\vec Q}\lb\mc T\rb.
  \ebn
  This is however not essential, as these  signs will be recovered at the last step. A more important thing to note is that in the reference limit described by $\theta_k\to+\infty$, $\sigma_k,\sigma_{k-1}\ll\theta$, $\sigma_k,\sigma_{k-1}\to+\infty$
  one has 
  \ben
  {\rm sgn}\bigl( e^{i\delta_1\eta_\pm}\bigr)={\rm sgn}\bigl( e^{i\delta_1\eta'_\pm}\bigr)=1.
  \ebn
  \subsection*{Step 2}
\begin{figure}[h!]
      \centering
      \includegraphics[height=3cm]{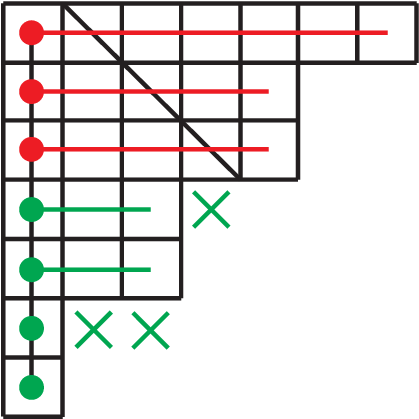}\\
       \begin{minipage}{0.7\textwidth}
      \caption{\label{appfig1}A Young diagram $Y'^*$ obtained from $Y'=\left\{6,4,4,2,2\right\}$ by addition of a column of length
      $L=7$.}
      \end{minipage}
      \end{figure}

  Let us now formulate and prove combinatorial
  \begin{theo}\label{appth1} $\tilde Z_{\,\mathsf{bif}}\lb\nu\bigl|0,Y';0,Y\rb\equiv\tilde Z_{\,\mathsf{bif}}\lb\nu\bigl|Y',Y\rb=\pm Z_{\,\mathsf{bif}}\lb\nu\bigl|Y',Y\rb$.
  \end{theo}
  \noindent This statement follows from the following two  lemmas.
  \begin{lemma} Equality $Z_{\,\mathsf{bif}}=\pm\tilde Z_{\,\mathsf{bif}}$ holds for the diagrams $Y', Y\in\mathbb Y$ iff it holds for $Y', Y$  with added one column of admissible height $L$.
  \end{lemma}
  \pf  Let us denote the  new value of $Z_{\,\mathsf{bif}}$ by $Z_{\,\mathsf{bif}}^*$ \footnote{Everywhere in this appendix $X^*$ denotes the value of a quantity
  $X$ after appropriate transformation.}, then
  \eq{\label{apeq1}
  Z_{\,\mathsf{bif}}^*=\frac{\lb1+\nu\rb_{L}\prod_i\lb L+p_i'+\frac12+\nu\rb}{\prod_i\lb L-q_i'+\frac12+\nu\rb}\frac{\lb1-\nu\rb_{L}\prod_i\lb L+p_i+\frac12-\nu\rb}{\prod_i\lb L-q_i+\frac12-\nu\rb}
  Z_{\,\mathsf{bif}}.
  }
  The extra factor comes only from the product over $2L$ new boxes. To explain how its expression is obtained, we will use the conventions of Fig.~\ref{appfig1}. 
  
  To compute the contribution from the red boxes it is enough just to multiply the corresponding shifted hook lengths, which yields $\prod_i\lb L+p_i'+\frac12+\nu\rb$.
  To compute the contribution from the green boxes one has to first write down the product of numbers from $\nu+L$ to $\nu+1$ (i.e. the Pochhammer symbol $\lb 1+\nu\rb_{L}$ in the numerator), keeping in mind that each
  step down by one box decreases the leg-length of the box by at least one. Then one has to take into account that some jumps in this sequence are greater than one: this
  happens exactly when we meet some rows of the transposed diagram. We mark with the green crosses the boxes whose contributions should be cancelled from the initial product: they produce the denominator.
  
  Next let us check what happens with $\tilde Z_{\,\mathsf{bif}}$. We have 
  \ben
  \tilde Z_{\,\mathsf{bif}}^*\lb\nu|Y',Y\rb=
  \prod_i\lb-\nu\rb_{q_i'^*+\frac12}
  \prod_i\nu^{-1}\lb\nu\rb_{q_i^*+\frac12}
  \prod_i\lb-\nu\rb_{p_i^*+\frac12}
  \prod_i\nu^{-1}\lb\nu\rb_{p_i'^*+\frac12}
  \frac{\prod_{i,j}\lb\nu-q_i'^*-p_j^*\rb
  \prod_{i,j}\lb\nu+p_i'^*+q_j^*\rb}{\prod_{i,j}
  \lb q_i'^*-q_j^*-\nu\rb\prod_{i,j}\lb p_i'^*-p_j^*+\nu\rb},
  \ebn
  where
  \begin{gather*}
  \left\{q_i^*\right\}=\left\{\lb L-1/2\rb,\lb q_1-1\rb,\ldots,\lb q_{d-1}-1\rb,\widetilde{(q_d-1)}\right\},\\
  \left\{p_i^*\right\}=\left\{\lb p_1+1\rb,\ldots,\lb p_{d}+1\rb,\widetilde{1/2}\right\},\\
  \left\{q_i'^*\right\}=\bigl\{\lb L-1/2\rb,\lb q'_1-1\rb,\ldots,\lb q'_{d'-1}-1\rb,\widetilde{\widetilde{(q'_{d'}-1)}}\bigr\},\\
  \left\{p_i'^*\right\}=\bigl\{\lb p'_1+1\rb,\ldots,\lb p'_{d'}+1\rb,\widetilde{\widetilde{1/2}}\bigr\},
  \end{gather*}
  and $d,d'$ denote the number of boxes on the main diagonals of $Y,Y'$.
  The above notation means that one has either to simultaneously include or not to include the coordinates tilded in the same way. These numbers are included in the case
  when both of them are positive (it implies that $q_d\neq\frac12$ or
  $q'_{d'}\ne \frac12$). Fig.~\ref{appfig2} below illustrates the difference between these two cases.
  
  \begin{figure}[h!]
     \centering
     \includegraphics[height=3cm]{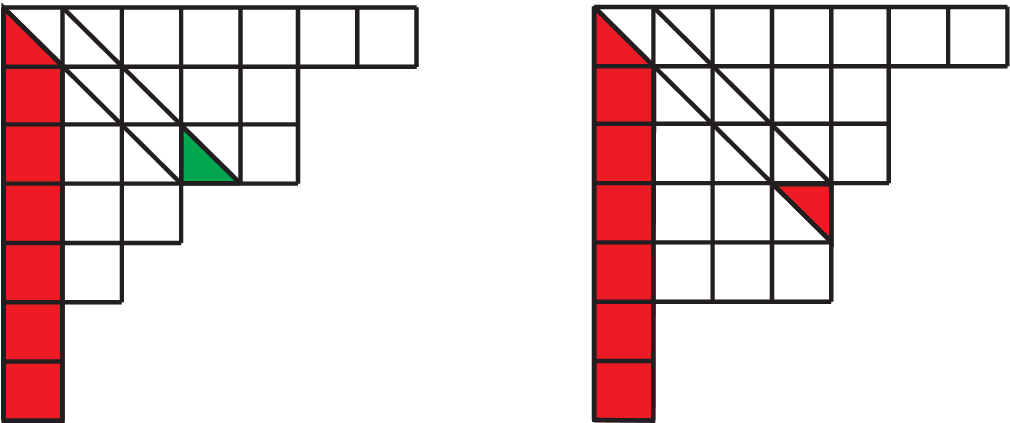}
     \begin{minipage}{0.65\textwidth}
     \caption{\label{appfig2}Possible mutual configurations of main diagonals of $Y$, $Y^*$; $q_d=\frac12$ (left) and $q_d\ne\frac12$ (right).}
     \end{minipage}
     \end{figure}
  
  We may now consider one by one four possible options, namely:
  i) $q_d\neq \frac12$, $q'_{d'}\neq\frac12$; ii) $q_d=q'_{d'}=\frac12$; iii)~${q_d\ne\frac12}$, $q'_{d'}=\frac12$;
  iv) $q_d=\frac12$, $q'_{d'}\ne\frac12$. For instance, for $q_d\neq \frac12$, $q'_{d'}\neq\frac12$ after massive cancellations one obtains
  \begin{gather*}
  \begin{aligned}
  \frac{\tilde Z_{\,\mathsf{bif}}^*}{\tilde Z_{\,\mathsf{bif}}}=&\,\prod_{i=1}^{d'}\frac1{-\nu+q_i'-\frac12}
  \lb-\nu\rb_L\prod_{i=1}^{d}\frac1{\nu+q_i-\frac12}\nu^{-1}
  \lb\nu\rb_L
  \prod_{i=1}^d\lb-\nu+p_i+\frac12\rb\lb-\nu\rb_1\prod_{i=1}^{d'}
  \lb\nu+p_i'+\frac12\rb\times\\
  \times&\,\frac{\prod_i\lb\nu-L-\frac12-p_i\rb
  \prod_i\lb\nu-q_i'+\frac12\rb
  \prod_i\lb\nu-\frac12+q_i\rb\prod_i\lb\nu+p_i'+L+\frac12\rb}{
  \prod_i\lb L+\frac12-q_i-\nu\rb \prod_i\lb q_i'-L-\frac12-\nu\rb\prod_i\lb p_i'+\frac12+\nu\rb
  \prod_i\lb-\frac12-p_i+\nu\rb}\frac{\lb\nu-L\rb\lb\nu+L\rb}{\nu^2}=
  \\=&\
  \lb 1-\nu\rb_{L}\lb 1+\nu\rb_{L}\frac{\prod_i(\nu-L-\frac12-p_i)\prod_i(\nu+p'_i+L+\frac12)}{\prod_i(L+\frac12-q_i-\nu)\prod_i(q_i'-L-\frac12-\nu)}=
  \frac{Z_{\,\mathsf{bif}}^*}{Z_{\,\mathsf{bif}}},
  \end{aligned}
  \end{gather*}
  where the first line of the first equality corresponds to the ratio of diagonal parts and the second to non-diagonal ones.
  The proof in the other three cases is analogous. \epf

 \begin{cor} $Z_{\,\mathsf{bif}}=\tilde Z_{\,\mathsf{bif}}$ for arbitrary $Y, Y'\in\mathbb Y$ iff $Z_{\,\mathsf{bif}}=\pm Z_{\,\mathsf{bif}}^*$ for diagrams with $\left\{q_i\right\}=\left\{\frac12,\ldots,L-\frac12\right\}$
  (that is, for the diagrams containing a large square on the left).
  \end{cor}
  
  \begin{lemma} The equality $Z_{\,\mathsf{bif}}=\tilde Z_{\,\mathsf{bif}}$ holds for given diagrams $Y, Y'\in\mathbb Y$ with a large square iff it holds for the diagrams with a large square and
  one deleted box.
  \end{lemma}
  \pf
     \begin{figure}[h!]
         \centering
         \includegraphics[height=2.5cm]{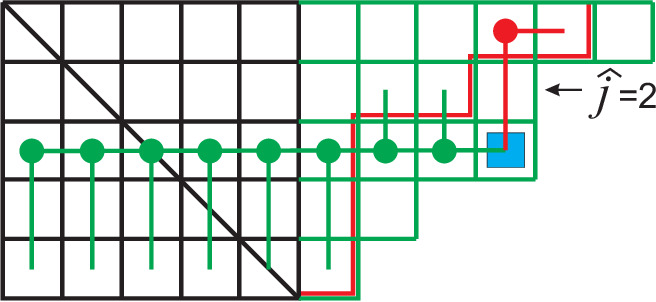}\\
         \begin{minipage}{0.6\textwidth}
         \caption{\label{appfig3}A pair of Young diagrams (red and green) with a large
         square (black) and added box (blue square).}
         \end{minipage}
         \end{figure}
  Suppose that we have added one box to the $i$th row of $Y'$. The only boxes whose contribution to $Z_{\,\mathsf{bif}}$ depends on the added box lie on its left  in the diagram $Y'$ and above it
  in the diagram $Y$, see Fig.~\ref{appfig3}.    
  The contribution from the boxes on the left (green circles) was initially given by
  \ben
  Z_{\,\mathsf{bif}}^{\mathrm{left}}=\frac{\lb\nu\rb_{p'_i+L+\frac12}}{\prod_{j\geq\hat j}\lb p'_i-p_j+\nu\rb\cdot 
  \lb\nu\rb_{\hat j-i+1}},
  \ebn
  where $\hat j=\min\left[\left\{j|p_j+j\leq p'_i+i+1\right\}\cup \left\{L\right\}\right]$ (notice that we can move $\hat j$ in the range where $p_j+j=p'_i+i+1$). The contribution from the boxes on the top (red circles) was
  $ Z_{\, \mathsf{bif}}^{\mathrm{top}}=\prod\nolimits_{j<\hat j}\lb-\nu+p_j-p_i'-1\rb$.
  After addition of one box (blue square) it transforms into
  $Z_{\,\mathsf{bif}}^{*\mathrm{top}}=\prod\nolimits_{j<\hat j}\lb-\nu+p_j-p_i'\rb$,
  whereas the previous part becomes
  \ben
  Z_{\,\mathsf{bif}}^{*\mathrm{left}}=\frac{\lb\nu\rb_{p'_i+L+\frac32}}{\prod_{j\geq\hat j}\lb p'_i-p_j+1+\nu\rb\cdot\lb \nu\rb_{\hat j-i+1}}.
  \ebn
  The ratio of the transformed and initial functions is then given by
  \ben
  \frac{Z_{\,\mathsf{bif}}^*}{Z_{\,\mathsf{bif}}}=\frac{\lb p_i'+L+\frac12+\nu\rb\prod_{j<\hat j}\lb p_i'-p_j+\nu\rb\prod_{j\geq\hat j}\lb p'_i-p_j+\nu\rb}{\prod_{j\geq\hat j}\lb p'_i-p_j+1+\nu\rb
  \prod_{j<\hat j}\lb-\nu+p_j-p_i'-1\rb}=\frac{\lb p_i'+L+\frac12\rb\prod_{j}\lb p_i'-p_j+\nu\rb}{\prod_{j}\lb p'_i-p_j+1+\nu\rb}.
  \ebn
  
  On the other hand, the ratio $\tilde Z_{\,\mathsf{bif}}^*/\tilde Z_{\,\mathsf{bif}}$ is easier to compute since the addition of one box to the $i$th row of~$Y'$ simply shifts one coordinate, $p_i'\mapsto p_i'+1$. From (\ref{apeq2}) and the large square condition $\left\{q_i\right\}=\left\{\frac12,\ldots,L-\frac12\right\}$ it follows that 
  \ben
  \frac{\tilde Z_{\,\mathsf{bif}}^*}{\tilde Z_{\,\mathsf{bif}}}=
  \lb p'_i+\frac12+\nu\rb\times
  \frac{\lb p'_i+\frac12+L+\nu\rb\prod_{j}\lb p_i'-p_j+\nu\rb}{\lb p_i'+\frac12+\nu\rb\prod_{j}\lb p'_i-p_j+1+\nu\rb}=
  \frac{Z_{\,\mathsf{bif}}^*}{Z_{\,\mathsf{bif}}},
  \ebn
  which finishes the proof. \epf
  
  Using  two inductive procedures described above, any pair of diagrams $Y,Y'\in\mathbb Y$ can be reduced to equal squares, in which case the statement of Theorem~\ref{appth1} can be checked directly.
  
  \subsection*{Step 3}
  Let us move to the third part of our plan and prove 
  
  \begin{theo}  
  $Z_{\,\mathsf{bif}}\lb\nu\bigl|Q',Y';Q,Y\rb=C\lb\nu\bigl|Q'-Q\rb
    Z_{\,\mathsf{bif}}\lb\nu+Q'-Q\bigl|Y',Y\rb$.
  \end{theo}
  \pf
  It is useful to start by computing $Z_{\,\mathsf{bif}}$ for the ``vacuum state''
  \begin{alignat*}{2}
  &\mathsf p_{\alpha}=\mathsf p_{\alpha}^Q:=
  \Bigl\{\frac12,\frac32,\ldots,Q^{(\alpha)}-\frac12\Bigr\},
  \quad
  \mathsf q_{\alpha}=\emptyset\qquad
   && \text{for }Q^{(\alpha)}>0, \\
  &\mathsf p_{\alpha}=\emptyset,\quad
  \mathsf q_{\alpha}=\mathsf q_{\alpha}^Q:=\Bigl\{\frac12,\frac32,\ldots,-Q^{(\alpha)}-
  \frac12\Bigr\}
  \qquad && \text{for } Q^{(\alpha)}<0.
  \end{alignat*}
  One obtains
  \begin{gather*}
  \begin{aligned}
  \tilde Z_{\,\mathsf{bif}}\lb\nu\bigl|\mathsf p^{Q'},\emptyset;\mathsf p^Q,\emptyset\rb=&\,\lb-1\rb^{Q\lb Q+1\rb/2}\Prod_{i=1}^{Q'}\nu^{-1}\lb\nu\rb_{i}
  \Prod_{i=1}^{Q}\lb-\nu\rb_{i}{\Prod_{i=1}^{Q'}\Prod_{j=1}^Q
  \lb\nu+i-j\rb^{-1}}=\\=&\,
  \lb-1\rb^{Q\lb Q+1\rb/2}\Prod_{i=1}^{Q'}\frac{\Gamma\lb\nu+i\rb}{\Gamma\lb\nu+1\rb}\Prod_{i=1}^Q\frac{\Gamma\lb i-\nu\rb}{\Gamma\lb-\nu\rb}\Prod_{j=1}^{Q}
  \frac{\Gamma\lb \nu-j+1\rb}{\Gamma\lb\nu-j+Q'+1\rb}=\\=&\,\frac{G\lb 1+\nu+Q'\rb}{
  G\lb 1+\nu\rb}\frac{G\lb 1-\nu+Q\rb}{G\lb 1-\nu\rb}
  \frac{\lb-1\rb^{Q\lb Q+1\rb/2}}{\Gamma\lb\nu+1\rb^{Q'}\Gamma\lb-\nu\rb^Q}
  \frac{G\lb\nu+1\rb}{G\lb\nu+1-Q\rb}\frac{G\lb\nu+Q'+1-Q\rb}{G\lb\nu+Q'+1\rb}=\\=&\,\lb-1\rb^{Q\lb Q+1\rb/2}\frac{G\lb 1-\nu+Q\rb}{G\lb1+\nu-Q\rb}\frac{G\lb 1+\nu+Q'-Q\rb
  }{G\lb 1-\nu\rb\Gamma\lb 1+\nu\rb^{Q'}\Gamma\lb-\nu\rb^Q}.
  \end{aligned}
  \end{gather*}
  Using the recurrence relation
    \begin{gather*}
  \frac{G\lb1-\nu+Q\rb}{G\lb1+\nu-Q\rb}=\lb-1\rb^{Q\lb Q-1\rb/2}\frac{G\lb 1-\nu\rb}{G\lb 1+\nu\rb}\lb\frac{\pi}{\sin\pi\nu}\rb^Q,
    \end{gather*}
  and the reflection formula $
    \Gamma\lb-\nu\rb\Gamma\lb 1+\nu\rb=-\frac{\pi}{\sin\pi\nu}$, the last expression can be rewritten as
  \begin{gather*}
  C\lb\nu|Q'-Q\rb:=\tilde Z_{\,\mathsf{bif}}\lb\nu\bigl|\mathsf p^{Q'},\emptyset;\mathsf p^Q,\emptyset\rb=
  \frac{G\lb 1+\nu+Q'-Q\rb}{G\lb 1+\nu\rb\Gamma\lb 1+\nu\rb^{Q'-Q}}.
  \end{gather*}

  Next let us rewrite the expression for $\tilde Z_{\,\mathsf{bif}}\lb\nu|Y',Q';Y,Q\rb$ for charged Young diagrams in terms of uncharged ones. To do this, we will try to
  understand how this expression changes under the following transformation, shifting in particular all particle/hole coordinates associated to $Y'$:
  \ben
  p'_i\mapsto p'_i+1,\qquad
  q'_i\mapsto q'_i-1,\qquad
  \nu\mapsto\nu-1.
  \ebn
  It should also be specified that if we had $q'=\frac12$, then this value should be dropped from the new set of hole coordinates; if not, we should add a new particle at $p'=\frac12$. Looking at Fig.~\ref{figmaya}, one may
  understand that this transformation is exactly the shift $Q'\mapsto Q'+1$ preserving the form of the Young diagram.
  
  Now compute what happens with 
  $\tilde Z_{\,\mathsf{bif}}\lb\nu|Y',Q';Y,Q\rb$. One should distinguish two cases:
  \begin{enumerate}
  \item If there is no hole at $q'=\frac12$ in $\lb Y',Q'\rb$, then it follows from (\ref{apeq2}) that
  \ben
  \frac{\tilde Z_{\,\mathsf{bif}}
  \lb \nu-1|Q'+1,Y';Q,Y\rb}{\tilde Z_{\,\mathsf{bif}}\lb\nu|Q',Y';Q,Y\rb}=
  \frac{\prod_i\lb\nu-\frac12+q_i\rb}{\prod_i\lb\nu-\frac12-p_i\rb}
  \prod_i\frac\nu{\nu+q_i-\frac12}
  \prod_i\frac{-\nu+p_i+\frac12}\nu\times\nu^{|\mathsf p'|}\nu^{-| \mathsf q'|}=\nu^{Q'-Q}.
  \ebn
  \item Similarly, if there is a hole at $q'=\frac12$ to be removed, then
  \ben
  \frac{\tilde Z_{\,\mathsf{bif}}(\nu-1|Q'+1,Y';Q,Y)}{\tilde Z_{\,\mathsf{bif}}\lb\nu|Q',Y';Q,Y\rb}=\nu^{-1}\frac{
  \prod_i\lb\nu-\frac12+q_i\rb}{\prod_j\lb\nu-\frac12-p_j\rb}\times
  \nu^{|\mathsf p'|}
  \nu^{-|\mathsf q'|+1}\prod_i\frac{\nu-p_i-\frac12}\nu\prod_i
  \frac\nu{\nu-\frac12+q_i}=\nu^{Q'-Q}.
  \ebn
  \end{enumerate}
  The computation of the shift of $Q$ is absolutely analogous thanks to the symmetry properties of $\tilde Z_{\,\mathsf{bif}}$.
  
  Introducing
  \ben
  \tilde Z^?_{\,\mathsf{bif}}\lb\nu|Q',Y';Q,Y\rb=\frac{Z_{\,\mathsf{bif}}
  \lb\nu|Q',Y';Q,Y\rb}{C\lb\nu|Q'-Q\rb},
  \ebn
  it is now straightforward to check that
  \ben
  \frac{\tilde Z^?_{\,\mathsf{bif}}\lb\nu-1|Q'+1,Y';Q,Y\rb}{\tilde Z^?_{\,\mathsf{bif}}\lb\nu|Q',Y';Q,Y\rb}=
  \frac{\tilde Z^?_{\,\mathsf{bif}}\lb\nu+1|Q',Y';Q+1,Y\rb}{\tilde Z^?_{\,\mathsf{bif}}\lb\nu|Q',Y';Q,Y\rb}=1,
  \ebn
  and therefore $\tilde Z^?_{\,\mathsf{bif}}\lb\nu|Q',Y';Q,Y\rb=
  \tilde Z^?_{\,\mathsf{bif}}\lb\nu+Q'-Q|0,Y';0,Y\rb$.
  Finally, combining this recurrence relation with $C\lb\nu|0\rb=1$, one obtains the identity
  \ben
  \frac{\tilde Z_{\,\mathsf{bif}}\lb\nu|Q',Y';Q,Y\rb}{C\lb\nu|Q'-Q\rb}=\tilde Z_{\,\mathsf{bif}}\lb\nu+Q'-Q|Y',Y\rb,
  \ebn
  which is equivalent to the statement of the theorem.
  \epf
  
  \subsection*{Step 4}
  At this point, we have already shown that
  \ben
  Z^{\vec Y',m'}_{\vec Y,m}\lb\mc T\rb=\pm
  e^{i\lb\delta_1\eta'_+-\delta_1\eta'_-\rb m'+
  i\lb\delta_1\eta_+-\delta_1\eta_-\rb m}\hat Z^{\vec Y',\vec Q'}_{\vec Y,\vec Q}\lb\mc T\rb.
  \ebn
 It remains to check the signs in the reference limit described above. Note that ${\rm sgn}\lb\hat Z\rb=1$, since ${\rm sgn}\lb C\lb\nu|Q',Q\rb\rb=1$  and ${\rm sgn}\lb Z_{\,\mathsf{bif}}\lb\nu|Y',Y\rb\rb=1$ as $\nu\to\infty$. Everywhere in this subsection the calculations are done modulo 2.
  
  First let us compute the sign of the non-diagonal part of $Z$. To do this, one has to fix the ordering as
  \begin{align*}
  x_I\colon& p_+'+\sigma_{k-1},\: p_{-}'-\sigma_{k-1},\: -q_+-\theta_k+\sigma_k,\: -q_--\theta_k-\sigma_k,\\
  y_I\colon&-q_+'+\sigma_{k-1},\: -q_-'-\sigma_{k-1},\: p_+-\theta_k+\sigma_k,\: p_--\theta_k-\sigma_k.
  \end{align*}
  The variables in each of these groups are ordered as
  $p_1,p_2,\ldots$ where $p_1>p_2>\ldots$ This gives
    \begin{gather*}
    {\rm lsgn}\lb\left.Z\right|_{\mathrm{non-diag}}\rb=|\mathsf p_-'|\cdot |\mathsf q_+'|+|\mathsf q_+|\cdot\lb|\mathsf q_+'|+|\mathsf q_-'|+|\mathsf p_+|\rb+|\mathsf q_-|\cdot\lb|\mathsf q_+'|+|\mathsf q_-'|+|\mathsf p_+|+|\mathsf p_-|\rb+\\+
    \frac{|\mathsf q_+|\lb|\mathsf q_+|-1\rb}{2}+\frac{|\mathsf q_-|\lb|\mathsf q_-|-1\rb}{2}+\frac{|\mathsf p_+|\lb|\mathsf p_+|-1\rb}{2}+
  \frac{|\mathsf p_-|\lb|\mathsf p_-|-1\rb}{2}+\\+|\mathsf q_+'|\cdot(|\mathsf q_-'|+|\mathsf p_+|+|\mathsf p_-|)+|\mathsf q_-'|\cdot(|\mathsf p_+|+|
  \mathsf p_-|)+|\mathsf p_+|\cdot|\mathsf p_-|.
    \end{gather*}
    Using the charge balance conditions
    \begin{align*}
    |\mathsf p_+|-|\mathsf q_+|=&\,|\mathsf q_-|-|\mathsf p_-|=m,\\
    |\mathsf p_+'|-|\mathsf q_+'|=&\,|\mathsf q_-'|-|\mathsf p_-'|=m',
    \end{align*}
    the above expression can be simplified to
    \begin{gather*}
    {\rm lsgn}\lb\left.Z\right|_{\mathrm{non-diag}}\rb=m+m'+m|\mathsf p_+|+m'|\mathsf p_+'|+|\mathsf p_+|+|\mathsf p_-|.
    \end{gather*}
  
    Next compute the sign of the diagonal part,
    \ben
    {\rm lsgn}\lb\left.Z\right|_{\mathrm{diag}}\rb=\Sum(p_-'+q_+'+q_++p_-)+
    \frac{|\mathsf p_-'|-|\mathsf q_+'|+|\mathsf q_+|-|\mathsf p_-|}{2}.
    \ebn
  Combining these two expressions, after some simplification we get
  \ben
    {\rm lsgn}\lb Z\rb=|\mathsf p_+|\cdot|\mathsf q_+|+|\mathsf p_+'|\cdot|\mathsf q_+'|+\sum\lb q_++\frac12\rb+
  \sum\lb q_+'+\frac12\rb+\sum\lb p_-+\frac12\rb+\sum\lb p_-'+\frac12\rb+m.
  \ebn
  This expression can be represented as
  \ben
  {\rm lsgn}\lb Z\rb=:{\rm lsgn}\lb 
  \mathsf p,\mathsf q\rb+{\rm lsgn}\lb \mathsf p',\mathsf q'\rb+m.
  \ebn
  To get the desired formula, one has to absorb $m$ by adding extra shift
    $ e^{i\delta_2\eta_+}=-1$.
    Combining this shift with the previous formulas (\ref{apeq3}), we deduce the full shift (\ref{apeq4}) of the Fourier transformation parameters.
    The final formula for the relative sign is
    \ben
    {\rm lsgn}\lb Z/\hat Z\rb={\rm lsgn}\lb \mathsf p,\mathsf q\rb+{\rm lsgn}\lb\mathsf p',\mathsf q'\rb,
    \ebn
    which completes our calculation.


\newpage
  
 \begin{thebibliography}{1000}
  
  \bibitem[AFLT]{AFLT}
  V. A. Alba, V. A. Fateev, A. V. Litvinov, G. M. Tarnopolsky,
  \textit{On combinatorial expansion of the con\-for\-mal blocks arising from
  AGT conjecture}, Lett. Math. Phys.~\textbf{98}, (2011), 33--64;  arXiv:1012.1312 [hep-th].
  
  \bibitem[AGT]{AGT}
  L. F. Alday, D. Gaiotto, Y. Tachikawa, \textit{Liouville correlation functions from
  four-dimensional gauge theories}, Lett. Math. Phys.~\textbf{91}, (2010), 167--197; arXiv:0906.3219 [hep-th].  
  
  \bibitem[Bal]{Balogh} F. Balogh, \textit{Discrete matrix models for partial sums of conformal blocks associated to Painlev\'e transcendents},  Nonlinearity~\textbf{28}, (2014), 43--56;	arXiv:1405.1871 [math-ph].
 
  \bibitem[BMPTY]{BMPTY}
  L. Bao, V. Mitev, E. Pomoni, M. Taki, F. Yagi,
  {\it Non-lagrangian theories from brane junctions},
  J. High Energ. Phys. (2014) 2014: 175;
   	arXiv:1310.3841 [hep-th]. 
 
  \bibitem[BSh]{BSh}
 M. Bershtein, A. Shchechkin, {\it Bilinear equations on Painlev\'e tau functions from CFT}, Comm. Math. Phys.~\textbf{339}, (2015), 1021--1061; arXiv:1406.3008v5 [math-ph].
 
 \bibitem[Bol]{Bolibrukh}
 A. A. Bolibrukh, \textit{On Fuchsian systems with given asymptotics and monodromy},
 Proc. Steklov Inst. Math.~\textbf{224}, (1999), 98--106; translation from Tr. Mat. Inst. Steklova \textbf{224}, (1999), 112--121.
 
 
  \bibitem[BGT]{BGT}
  G. Bonelli, A. Grassi, A. Tanzini, {\it Seiberg-Witten theory as a Fermi gas}, Lett. Math. Phys. \textbf{107}, (2017), 1--30; 	arXiv:1603.01174 [hep-th].
  
   \bibitem[BMT]{Bonelli} G. Bonelli, K. Maruyoshi, A. Tanzini,
   {\it Wild quiver gauge theories},  
   J. High Energ. Phys. 2012:31, (2012);	arXiv:1112.1691 [hep-th].  
 
  \bibitem[BO1]{BorodinAnnals}
  A. Borodin, G. Olshanski, {\it Harmonic analysis on the infinite-dimensional
  unitary group and determinantal point processes}, Ann. Math.~\textbf{161}, (2005),
  1319--1422; math/0109194 [math.RT].
  
    \bibitem[BO2]{BO} A. Borodin, G. Olshanski, \textit{$Z$-measures on partitions, Robinson-Schensted-Knuth
    correspondence, and $\beta=2$ random matrix ensembles}, in ``Random Matrix Models and their Applications'',
     (eds. P.~M.~Bleher, A.~R.~Its), Cambridge Univ. Press, (2001), 71--94; arXiv:math/9905189v1 [math.CO].
  
  \bibitem[BD]{BorodinDeift}
  A. Borodin, P. Deift, {\it Fredholm determinants, Jimbo-Miwa-Ueno tau-functions, and representation theory}, Comm. Pure Appl. Math.~\textbf{55}, (2002),
  1160--1230; math-ph/0111007.
  
  \bibitem[Bul]{Bullimore}
  M. Bullimore, {\it Defect Networks and Supersymmetric Loop
  Operators}, J. High Energ. Phys. (2015) 2015: 66; arXiv:1312.5001v1 [hep-th].
  
  \bibitem[CM]{CM} L. Chekhov, M. Mazzocco, \textit{Colliding holes in Riemann surfaces and quantum cluster algebras}, Nonlinearity~\textbf{31}, (2018), 54; arXiv:1509.07044 [math-ph].
  
  \bibitem[CMR]{CMR} L. Chekhov, M. Mazzocco, V. Rubtsov,
  {\it Painlevé monodromy manifolds, decorated character varieties and
  cluster algebras}, Int. Math. Res. Not. \textbf{2017}, (2017), 7639--7691; arXiv:1511.03851v1 [math-ph].
  
  \bibitem[FL]{FL}
  V. A. Fateev, A. V. Litvinov, {\it Integrable structure, W-symmetry and AGT relation}, J. High Energ. Phys. (2012) 2012: 51;  arXiv:1109.4042v2 [hep-th].
 
 \bibitem[FIKN]{FIKN} A. S. Fokas, A. R. Its, A. A. Kapaev, V. Yu. Novokshenov, \textit{Painlev\'e transcendents:
  the Riemann-Hilbert approach}, Mathematical Surveys and Monographs~\textbf{128}, AMS, Providence,
  RI, (2006). 
  
      \bibitem[G]{Gaiotto1}
      D. Gaiotto, \textit{Asymptotically free $\mathcal{N}=2$ theories and irregular conformal blocks},
      arXiv:0908.0307 [hep-th].
      
     \bibitem[GT]{GT}
     D. Gaiotto, J. Teschner, \textit{Irregular singularities in Liouville theory and Argyres-Douglas type
     gauge theories, I}, J. High Energ. Phys. (2012) 2012: 50; arXiv:1203.1052 [hep-th].
  
   \bibitem[Gav]{Gav}
   P.~Gavrylenko, \textit{Isomonodromic $\tau$-functions and $W_N$
   conformal blocks}, J. High Energ. Phys. (2015) 2015: 167;  arXiv:1505.00259v1 [hep-th].
   
    \bibitem[GM1]{GM1}
    P. Gavrylenko, A. Marshakov, \textit{Exact conformal blocks for the W-algebras, twist fields and isomonodromic deformations}, J. High Energ. Phys. (2016) 2016: 181;  arXiv:1507.08794 [hep-th].
    
   \bibitem[GM2]{GM2}
    P. Gavrylenko, A. Marshakov, \textit{Free fermions, W-algebras and isomonodromic deformations}, 
    Theor. Math. Phys.~{\bf 187}, (2016), 649--677;
    arXiv:1605.04554 [hep-th]. 

  \bibitem[GIL12]{GIL12}
   O. Gamayun, N. Iorgov, O. Lisovyy,  \textit{Conformal field theory of Painlev\'e~VI},
   J. High Energ. Phys. (2012) 2012: 38; arXiv:1207.0787 [hep-th].
   
    \bibitem[GIL13]{GIL13}
    O. Gamayun, N. Iorgov, O. Lisovyy,  \textit{How instanton combinatorics solves Painlevé~VI, V and III's},
     J.~Phys.~\textbf{A46}, (2013), 335203;
      {arXiv:1302.1832 [hep-th]}.
      
  \bibitem[GHM]{GHM}
  A. Grassi, Y. Hatsuda, M. Marino, {\it Topological strings from quantum mechanics}, arXiv:1410.3382 [hep-th].    
     
  \bibitem[HI]{HI}
  J. Harnad, A. R. Its, {\it Integrable Fredholm operators and dual isomonodromic deformations}, Comm. Math. Phys. {\bf 226}, (2002), 497--530; arXiv:solv-int/9706002.
 
  \bibitem[HKS]{HKS}
  L. Hollands, C. A. Keller, J. Song,
  {\it Towards a 4d/2d correspondence for Sicilian quivers},
  J. High Energ. Phys. (2011), 1110:100;
  arXiv:1107.0973v1 [hep-th].

  
   \bibitem[ILTe]{ILTe}
    N. Iorgov, O. Lisovyy, J. Teschner,  \textit{Isomonodromic tau-functions from Liouville conformal blocks}, Comm. Math. Phys. \textbf{336}, (2015),  671--694; arXiv:1401.6104 [hep-th].
    
  \bibitem[ILT13]{ILT13} N. Iorgov, O. Lisovyy, Yu. Tykhyy, \textit{Painlevé VI connection problem and monodromy of
  $c = 1$ conformal blocks}, J. High Energ. Phys. (2013) 2013: 29;
  arXiv:1308.4092v1 [hep-th].   
  
   \bibitem[IIKS]{IIKS}
   A. R. Its, A. G. Izergin, V. E. Korepin, N. A. Slavnov,
   {\it Differential equations for quantum correlation functions},
   Int. J. Mod. Phys. {\bf B4}, (1990), 1003--1037.

  \bibitem[ILP]{ILP} A. R. Its, O. Lisovyy, A. Prokhorov,
  \textit{Monodromy dependence and connection formulae for isomonodromic tau functions}, Duke Math. J.~\textbf{167}, (2018), 1347--1432; arXiv:1604.03082 [math-ph].

 \bibitem[ILT14]{ILT14} 
 A. Its, O. Lisovyy, Yu. Tykhyy, \textit{Connection problem for the sine-Gordon/Painlev\'e III tau function
  and irregular conformal blocks}, Int. Math. Res. Not. \textbf{2015}, (2015), 8903--8924; arXiv: 1403.1235 [math-ph].    
  
 \bibitem[IKSY]{IKSY}
 K. Iwasaki, H. Kimura, S. Shimomura, M. Yoshida,
 {\it From Gauss to Painlevé: A modern theory of special functions},
 Aspect of Math. {\bf E16}, Braunschweig: Vieweg, (1991). 
  
 
  \bibitem[Jim]{Jimbo} M. Jimbo,
  {\it Monodromy problem and the boundary condition for some Painlev\'e
  equations}, Publ. Res. Inst. Math. Sci.~\textbf{18}, (1982),  1137--1161.
  
   \bibitem[JMMS]{JMMS}
    M. Jimbo, T. Miwa, Y. M\^ori, M. Sato, \textit{Density matrix of an impenetrable Bose gas
    and the fifth Painlev\'e transcendent}, Physica~\textbf{1D}, (1980), 80--158.
  
  \bibitem[JMU]{JMU} M. Jimbo, T. Miwa, K. Ueno,  {\it Monodromy preserving deformation of linear ordinary differential equations with rational coefficients. I},
   Physica {\bf D2}, (1981), 306--352.
   
  \bibitem[JR]{JR}
  N. Joshi, P. Roffelsen, {\it Analytic solutions of $q$-$P\lb A_1\rb $ near its critical points}, Nonlinearity \textbf{29}, (2016), 3696; arXiv:1510.07433 [nlin.SI]. 
  
  \bibitem[Kor]{Korotkin_elliptic}
  D. A. Korotkin, {\it Isomonodromic deformations in genus zero and one: algebrogeometric solutions and Schlesinger transformations},
  in ``Integrable systems: from classical to quantum'', ed. by
  J.Harnad, G.Sabidussi and P.Winternitz, CRM Proceedings and Lecture Notes, American Mathematical Society, (2000);
  arXiv:math-ph/0003016v1.
    	
   
   \bibitem[Lis]{dyson2F1}
 O. Lisovyy, {\it Dyson's constant for the hypergeometric kernel}, in
  ``New trends in quantum integrable systems'' (eds. B. Feigin, M. Jimbo, M. Okado),
  World Scientific, (2011), 243--267; arXiv:0910.1914 [math-ph].    
   
  \bibitem[Mal]{Malgrange}
  B. Malgrange, \textit{Sur les déformations isomonodromiques, I. Singularités régulières}, in
  ``Mathematics and Physics'', (Paris, 1979/1982); Prog. Math.~\textbf{37}, Birkh\"auser, Boston, MA, (1983), 401--426.
  
  \bibitem[Ma]{Mano}
  T. Mano, {\it Asymptotic behaviour around a boundary point of the
  $q$-Painlev\'e VI equation and its connection problem}, Nonlinearity
  {\bf 23}, (2010), 1585--1608.
  
 \bibitem[Nag]{Nagoya}
 H. Nagoya, \textit{Irregular conformal blocks, with an application to the fifth and fourth Painlev\'e equations}, J.~Math. Phys.~\textbf{56}, (2015), 123505; arXiv:1505.02398v3 [math-ph].  
 
 \bibitem[Nek]{Nekrasov} N.~A.~Nekrasov,
 {\em Seiberg-Witten prepotential from instanton counting},
 Adv. Theor. Math. Phys. {\bf 7}, (2003), 831--864; 
  	arXiv:hep-th/0206161.
 
 \bibitem[NO]{NO} N. Nekrasov, A. Okounkov,
 {\it Seiberg-Witten theory and random partitions}, in ``The unity of mathematics'', pp. 525--596, Progr. Math.~\textbf{244},
 Birkh\"auser Boston, Boston, MA, (2006);  arXiv:hep-th/0306238.
 
  \bibitem[Pal1]{Palmer90}
  J. Palmer, {\it Determinants of Cauchy-Riemann operators as 
  $\tau$-functions}, Acta Appl. Math.~\textbf{18}, (1990), {199--223}.
  
  \bibitem[Pal2]{PalmerTW}
  J. Palmer, {\it Deformation analysis of matrix models}, 
  Physica {\bf D78}, (1994), 166--185;
  arXiv:hep-th/9403023v1.
  
  \bibitem[Pal3]{Palmer93}
  J. Palmer, {\it Tau functions for the Dirac operator in the Euclidean plane}, Pacific J. Math.~\textbf{160}, (1993), 259--342.
  
 
  \bibitem[Sato]{Sato}
     			M. Sato, {\it Soliton equations as dynamical systems on infinite dimensional Grassmann manifold}, North-Holland Math. Studies \textbf{81}, (1983), 259--271.
 
 \bibitem[SMJ]{SMJ}
   M. Sato, T. Miwa, M. Jimbo,\textit{Holonomic quantum fields III--IV}, Publ. RIMS Kyoto
   Univ.~\textbf{15}, (1979), 577--629;
   \textbf{15}, (1979), 871--972.
 
     \bibitem[SW]{SW} 
     G. Segal, G. Wilson, {\it Loop groups and equations of KdV type}, Publ. Math. IHES \textbf{61}, (1985), 5--65.   
 
 \bibitem[SKAO]{qVir}
 J. Shiraishi, H. Kubo, H. Awata, S. Odake,
 {\it A quantum deformation of the Virasoro algebra and the Macdonald symmetric functions},
 Lett. Math. Phys. {\bf 38}, (1996), 33--51;
 arXiv:q-alg/9507034.
   
  \bibitem[TW1]{TW1}
  C. A. Tracy, H. Widom, \textit{Level-spacing distributions and the Airy kernel},
  Comm. Math. Phys.~\textbf{159}, (1994), 151--174;  hep-th/9211141.
 
  \bibitem[TW2]{TW2}
  C. A. Tracy, H. Widom, \textit{Fredholm determinants, differential equations and matrix models},
  Comm. Math. Phys.~\textbf{163}, (1994), 33--72; hep-th/9306042.
  
  \bibitem[Tsu]{Tsuda} 
  T. Tsuda, \textit{UC hierarchy and monodromy preserving deformation}, J. Reine Angew. Math. \textbf{690}, (2014), 1--34;
  arXiv:1007.3450v2  [math.CA].
  
    \bibitem[WMTB]{WMTB}
    T. T. Wu, B. M. McCoy, C. A. Tracy, E. Barouch, \textit{Spin-spin correlation functions for
    the two-dimensional Ising model: exact theory in the scaling region}, Phys. Rev.~\textbf{B13},
    (1976), 316--374.
  
 \end{thebibliography}
\end{document}